\documentclass[10pt, letter, onecolumn]{arxiv}

\usepackage{kantlipsum, lipsum}
\usepackage{dm-colors}
\usepackage{amsmath}
\usepackage{pstricks, pst-node}
\usepackage{verbatim}
\usepackage{multirow}
\usepackage{scalerel}
\usepackage{booktabs}
\usepackage{enumitem}
\usepackage{xspace}
\usepackage{bm}
\usepackage{bbm}
\usepackage{mathtools}
\usepackage{soul}
\usepackage{epsfig}
\usepackage{graphicx}
\usepackage{amssymb}
\usepackage{colortbl}
\usepackage{arydshln}
\usepackage{mdframed}
\usepackage{csquotes}
\usepackage{setspace}
\usepackage{colortbl}
\usepackage{tabularx,ragged2e}
\usepackage{placeins}
\usepackage[dvipsnames]{xcolor}
\usepackage{bbm}
\usepackage{longtable}
\usepackage{lineno} 
\usepackage[hang,flushmargin]{footmisc}
\usepackage{nameref}
\usepackage{varioref}
\usepackage[pagebackref=false,breaklinks=false,%
            colorlinks=true,bookmarks=true,citecolor=ourdarkblue,%
            urlcolor=ourdarkblue,linkcolor=ourdarkblue]{hyperref}
\usepackage[noabbrev,capitalize]{cleveref}
\usepackage{etoc}
\usepackage{soul}

\usepackage{ifthen}

\graphicspath{{figures/}}

\newcommand{\model}[1]{{SAT}}
\newcommand{\samdataset}[1]{{SAT-DS}}

\definecolor{lightlightgray}{rgb}{200,200,200}

\title{\Large{Large-Vocabulary Segmentation for Medical Images with Text Prompts}}

\vspace{1cm}
\author[1,2]{Ziheng Zhao}
\author[2]{Yao Zhang}
\author[1,2]{Chaoyi Wu}
\author[1,2]{Xiaoman Zhang}
\author[2]{\\ \vspace{0.1cm} Xiao Zhou}
\author[1,2]{Ya Zhang}
\author[1,2,$*$]{Yanfeng Wang} 
\author[1,2,$*$]{Weidi Xie}

\affil[1]{\normalsize Shanghai Jiao Tong University, Shanghai, China \authorcr \vspace{0.1cm}}

\affil[2]{\normalsize Shanghai AI Laboratory, Shanghai, China \authorcr \vspace{0.1cm}}

\affil[$*$]{\normalsize Corresponding author \authorcr \vspace{0.1cm} Yanfeng Wang: wangyanfeng622@sjtu.edu.cn; Weidi Xie: weidi@sjtu.edu.cn}

\renewcommand{\correspondingauthor}[1]{$\dag$~Corresponding Authors.}

\begin{document}
\begin{abstract}
\textbf{Abstract.} 
This paper aims to build a model that can \textbf{S}egment \textbf{A}nything in 3D medical images, driven by medical terminologies as \textbf{T}ext prompts, termed as \textbf{\model{}}. 
Our main contributions are three-fold: 
(i) We construct the first multimodal knowledge tree on human anatomy, including \textbf{6502 anatomical terminologies}; Then, we build the largest and most comprehensive segmentation dataset for training, collecting over \textbf{22K 3D scans} from \textbf{72 datasets}, across \textbf{497 classes}, with careful standardization on both image and label space;
(ii) We propose to inject medical knowledge into a text encoder via contrastive learning and formulate a large-vocabulary segmentation model that can be prompted by medical terminologies in text form.
\textcolor{black}{(iii) We train \textbf{\model{}-Nano} (110M parameters) and \textbf{\model{}-Pro} (447M parameters). 
\textbf{\model{}-Pro} achieves comparable performance to 72 nnU-Nets---the strongest specialist models trained on each dataset (over 2.2B parameters combined)---over 497 categories.
%
Compared with the interactive approach MedSAM, \model{}-Pro consistently outperforms across all 7 human body regions with +7.1\% average Dice Similarity Coefficient (DSC) improvement, while showing enhanced scalability and robustness.
%
%
On 2 external~(cross-center) datasets, \model{}-Pro achieves higher performance than all baselines (+3.7\% average DSC), demonstrating superior generalization ability.
%
%
%
}
\end{abstract}
\maketitle


\section{Introduction}

Medical image segmentation aims to identify and delineate regions of interest (ROIs) like organs, lesions, and tissues in diverse medical images, which plays a crucial role in numerous clinical applications, 
such as disease diagnosis, treatment planning, and disease progression tracking~\cite{wang2019organ,yan2018deeplesion,BraTS2021,nouranian2015learning,jaffray2023harnessing}, as well as in medical research~\cite{nauffal2024noninvasive,bai2020population}. Traditionally, radiologists perform manual segmentation to measure volume, shape, and location in a slice-wise manner, which is both time-consuming and challenging to scale with the growing volume of medical data. Consequently, there is a pressing need for automated and robust medical image segmentation methods in clinical settings, to enhance efficiency and scalability.

Recent advancements in medical image analysis have been marked by a surge in deep learning. These developments have yielded a spectrum of segmentation models, each trained for specific tasks~\cite{AbdomenCT1K,FLARE22,MSD,MRSpineSeg,BraTS2021,KiTS23,COVID19}, often referred to as `specialist' models. 
While these models demonstrate impressive segmentation capabilities, 
their major drawback lies in their narrow specialization. 
Designed and optimized for distinct ROIs and imaging modalities, 
these models~\cite{ma2024u,UNetr,nnUNET,VNet,UNet,SwinUNetr} require distinct preprocessing methods for each dataset.
As a result, they often fall short in diverse and dynamic clinical environments, where adaptability to new conditions and imaging techniques is essential.

\begin{figure}[!t]
    \centering
    \includegraphics[width = \textwidth]{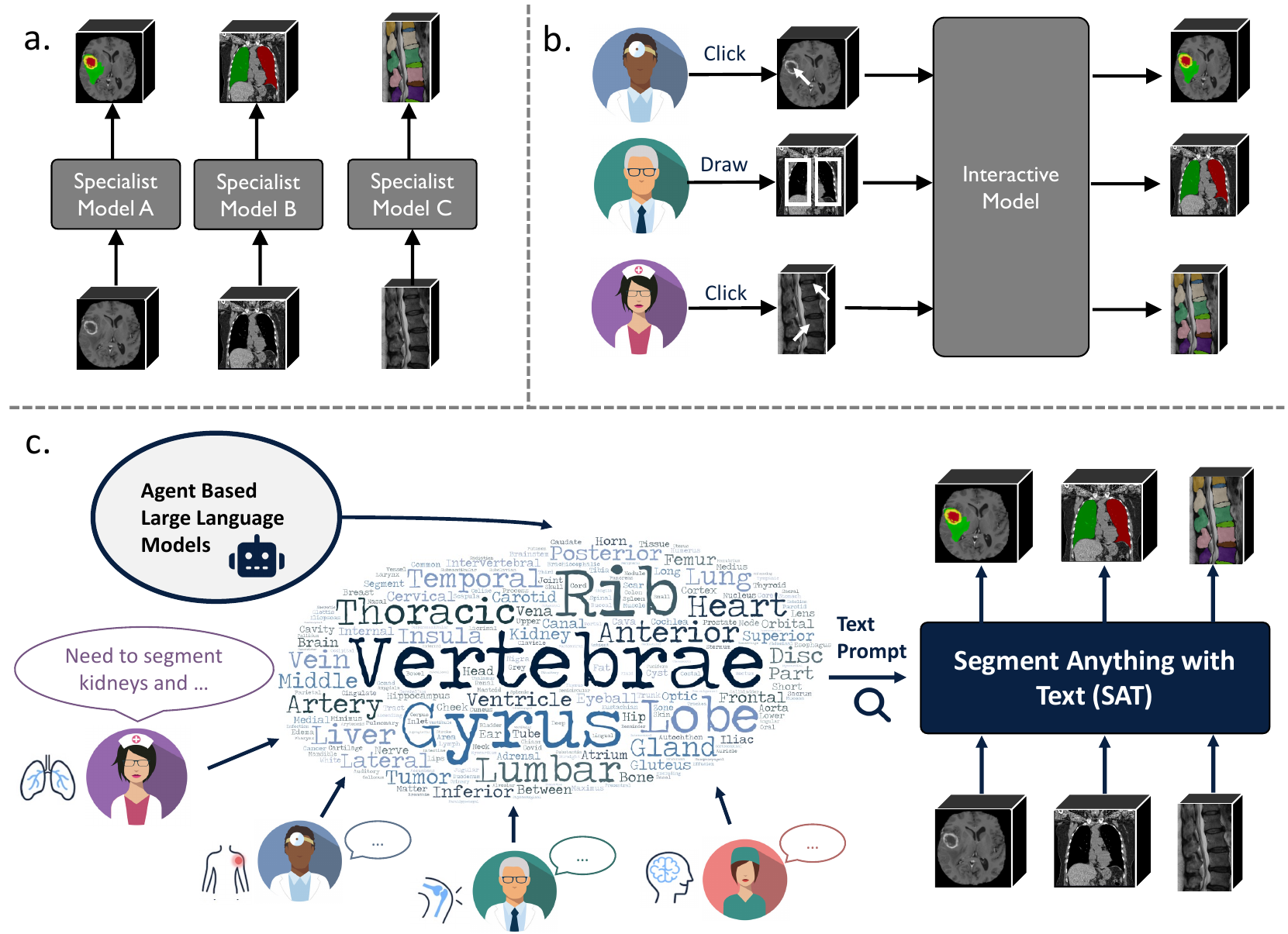}
    \caption{\textbf{Segment Anything in 3D medical images with Text.} In contrast to conventional specialist models (a) that develop specialized solution for each task, or recently proposed interactive segmentation foundation models (b) relying on real-time human interventions, Segment Anything by Text~(\textbf{SAT}) 
    directly takes 3D volumes as inputs, and use text as prompts to perform a wide array of medical image segmentation tasks across different modalities, anatomies, and body regions (c). It can be easily applied to clinics or seamlessly integrated with any agent-based large language model.}
    \label{fig:new_highlevel_idea}
    \vspace{-10pt}
\end{figure}

There is a growing interest in developing foundation models for medical image segmentation~\cite{MedSAM, SAM4Med?}, by adapting the pre-trained Segment Anything Model (SAM)~\cite{SAM} models from the computer vision community. However, while transferring to medical scenarios, these models trained on natural images suffer from fundamental limitations: 
(i) models typically perform 2D slice segmentation, 
which is later fused into 3D volumes through post-processing. 
This approach overlooks the crucial contextual information in 3D radiological imaging; 
(ii) models often require point or box inputs as prompts, thus are interactive segmentation models, requiring considerable manual effort for use in practice;
(iii) models suffer from significant domain gaps, 
from image statistics to domain-specific medical knowledge. 

In this paper, we present the \textbf{first knowledge-enhanced} foundation model for 3D medical volume segmentation, \textbf{with medical terminology as text prompt, termed as \model{}}. 
In practice, our model can effectively take 3D volumes as visual inputs along with text prompts, to seamlessly tackle various medical image segmentation tasks, across modalities, anatomies, and body regions. 
As illustrated in Figure 1, 
our proposed method distinguishes itself from previous medical segmentation paradigms, that can be seamlessly applied to clinical practice or integrated with any large language model. Specifically, we make the following contributions: 

\textbf{On dataset construction,} we construct a knowledge tree on anatomy concepts and definitions throughout the human body. 
On the visual side, we curate over 22K 3D medical image scans with 302K anatomical segmentation annotations, covering 497 categories from 72 publicly available medical segmentation datasets, termed as \textbf{\samdataset}. 
To the best of our knowledge, \samdataset{} represents the largest and most comprehensive collection of public 3D medical segmentation datasets. 
To achieve this goal, we have invested significant effort in standardizing datasets and unifying annotation labels, paving the way for training a large-vocabulary segmentation foundation model. For a complete list of datasets and download links, we refer readers to Table~\ref{tab:dataset_links}. 

\begin{figure}[t]
    \centering
    \includegraphics[width = \textwidth]{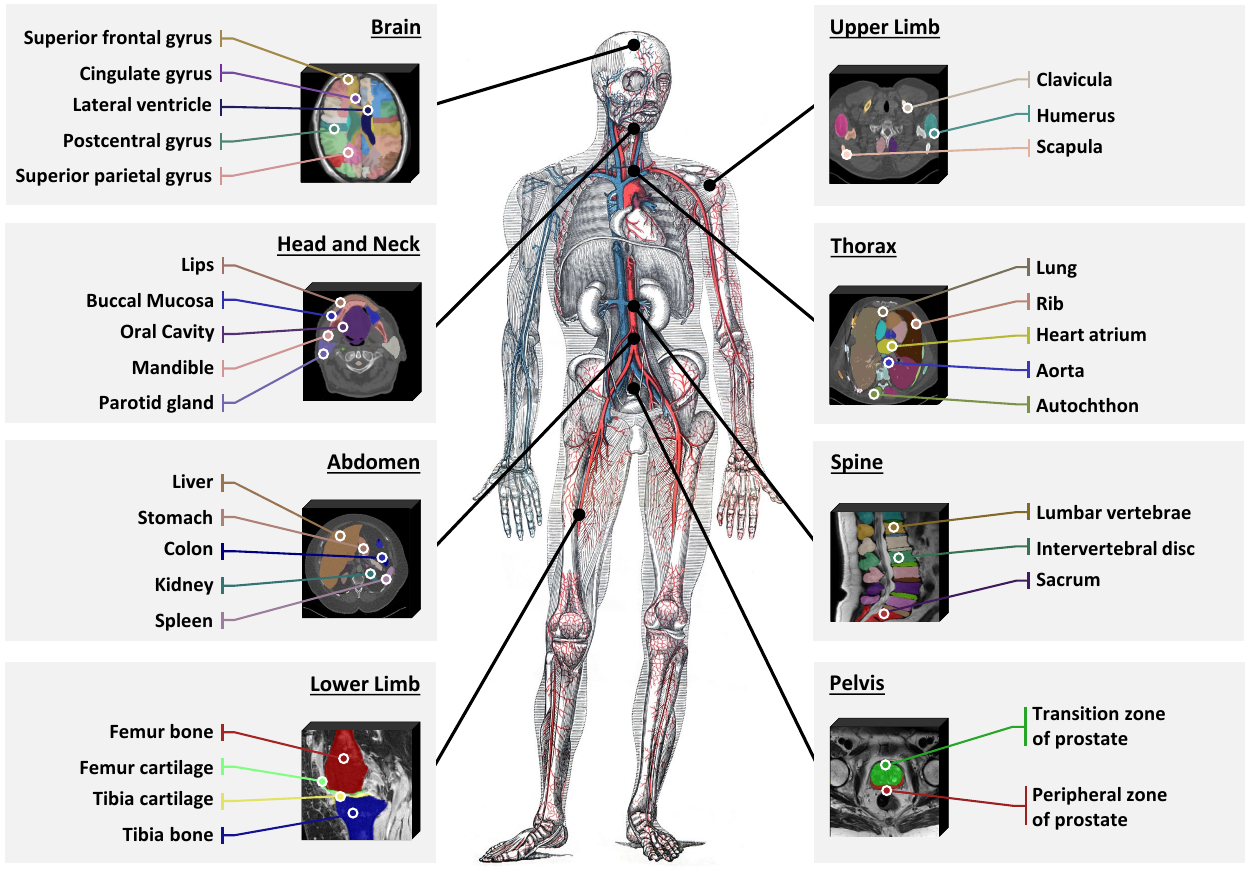}
    \caption{\textbf{Overview of \samdataset{}}, comprising diverse segmentation tasks spanning multiple imaging modalities and anatomical regions, including the brain, head and neck, thorax, spine, abdomen, upper limbs, lower limbs, and pelvis. This comprehensive dataset enables the training of a large-vocabulary segmentation foundation model.}
    \label{fig:data_overview}
    \vspace{-10pt}
\end{figure}

\textbf{On architecture design and training strategy,}
we build a large-vocabulary segmentation foundation model, that enables flexible segmentation across a spectrum of medical imaging modalities and anatomies, with text prompts. 
Specifically, we adopt knowledge-enhanced representation learning, leveraging textual anatomical knowledge and atlas segmentation of specific anatomical structures to train the visual-language encoders. 
Through this training process, the visual features of these anatomical structures are aligned with their corresponding text descriptions in the latent space, which is validated to boost the segmentation performance, especially in a long-tail distribution.
Subsequently, the text embeddings of anatomy/abnormality are treated as queries in a Transformer-based architecture, 
iteratively attending to the visual features to update queries for precise segmentation of the queried target. 
To meet requirements from different computational resources, 
we train two models of varying sizes, 
namely, \textbf{\model{}-Nano} and \textbf{\model{}-Pro}, 
and validate the effectiveness of scaling model sizes.

\textbf{On experiment evaluation}, 
we devise comprehensive metrics for large-vocabulary medical segmentation across various aspects, including region-wise average, organ-wise average, and dataset-wise average. Through extensive internal and external experiments, we demonstrate that: 

\vspace{-0.2cm}
\begin{itemize}
  \setlength\itemsep{0.5em}
    \item Building on the unprecedented dataset collection, 
    \model{} is able to handle a wide range of downstream segmentation tasks with medical terminologies as text prompts, simplifying the training and deployment procedure for conventional specialist models.
    \textcolor{black}{
    On internal evaluation, \textbf{\model{}-Pro} shows comparable overall performance to 72 nnU-Net models---the strongest specialist models that are specialized and trained individually on each dataset---over 497 categories, while using only 20\% of their combined model parameters (447M vs. 2.2B+).
    }
    
    \item Driven by text prompts, \model{} outlines a novel paradigm for segmentation foundation model, as opposed to previous interactive approaches that rely on spatial prompts. 
    This could save tremendous manual efforts from prompting in clinical applications. 
    \textcolor{black}{
    On performance, \model{}-Pro consistently outperforms the state-of-the-art interactive model MedSAM across 7 human body regions, while being robust to targets with ambiguous spatial relationships.
    }
    
    \item Compared to BiomedParse~\cite{zhao2024foundation}, a concurrent model on text-prompted biomedical image segmentation, 
    \textcolor{black}{
    \model{}-Pro not only exhibits superior performance on 29 out of 30 categories, but also showcases a significantly broader capability on radiology images.
    }

    \item On external evaluation, 
    \textcolor{black}{
    \model{}-Pro delivers the best results across both external validation datasets, and surpasses all baselines including specialist and generalist models, highlighting its strong generalization capabilities as a foundation model.
    }

    \item The text-prompted feature and large vocabulary of \model{} makes it a powerful out-of-box agents for language model. We show \model{} can be seamlessly integrated with any large language models such as GPT-4~\cite{GPT4}, automatically providing grounding ability in diverse clinical scenarios. This potentially extends the application diagram of medical segmentation models, and advance generalist medical artificial intelligence.
  
\end{itemize}
\section{Results}
\label{sec:results}

We propose \textbf{S}egment \textbf{A}nything with \textbf{T}ext~(\textbf{SAT}), a large-vocabulary segmentation foundation model for 3D medical images. 
The objective is to handle a wide range of heterogeneous tasks using text prompts.
It includes 497 anatomical targets across 8 regions and various lesions of the human body, assembled from 72 distinct datasets. 
To balance the computational cost and performance, we train and evaluate two variants \textbf{\model{}-Pro} and \textbf{\model{}-Nano}.

In this section, we detail the experiment results, 
where \textbf{\model{}} is comprehensively evaluated against three categories of methods: 
(i) \textbf{specialist models}, which are optimized and trained individually for each dataset, following the conventional mainstream practice in medical image segmentation.
We choose nnU-Nets~\cite{nnUNET}, SwinUNETR~\cite{SwinUNetr} and U-Mamba~\cite{ma2024u} for comparison, as they are widely adopted representatives for CNN-based, Transformer-based and Mamba-based architecture respectively; 
(ii) \textbf{interactive segmentation models}, which have been recently investigated to provide semi-automatic segmentation with spatial prompts. 
We choose MedSAM~\cite{MedSAM} as a typical and state-of-the-art baseline;
(iii) \textbf{text-prompted segmentation models}, which represent a paradigm shift from the previous two, capable of performing automatic segmentation across a wide range of tasks with text prompts. 
BiomedParse~\cite{zhao2024foundation} is a concurrent work to ours and compared in this study.

The evaluations are conducted on both \textbf{internal} and \textbf{external} datasets. Specifically, we split each dataset in \samdataset{} into train and test splits in 8:2 ratio, a combination of these test splits is used for internal evaluation, {\em i.e.}, in-domain data. 
When comparing to off-the-shelf models, we tailor the scope of datasets to accommodate their varying capabilities, 
to avoid overlapping the train and test data. 
The external evaluation is conducted on two very recently published datasets, namely, AbdomenAtlas 1.1~\cite{abdomenatlas} and LiQA~\cite{LiQA}, 
as they are excluded from \samdataset{} and not used in training any of these methods. 
This simulates the scenario where the models are tested on multi-center images. \textbf{Note that}, this does not involve new classes, as the segmentation targets in human body are relatively limited and fixed.

We present evaluation results from various aspects, 
including \textbf{region-wise}, \textbf{class-wise}, and \textbf{dataset-wise}, 
to give a deep understanding of the models' performance on large-scale segmentation. 
Note that, class-wise and region-wise evaluations are computed by averaging the results from different datasets. For instance, the performance metrics for the `brainstem' in CT images represent the macro average from models trained on datasets, like `HAN Seg', `PDDCA', and `SegRap2023 Task 1', 
that all include annotations for this anatomical class. 
Detailed experiment settings can be found in Section~\ref{sec:experiment_settings}.

The following sections start with experiments on internal datasets in Section~\ref{sec:specialist_internal}, \ref{sec:medsam_internal} and \ref{sec:biomedparse_internal}, with more detailed results available in the ``Detailed Internal Evaluation Results'' Section in Supplementary.
Then, we present the results of different methods on external datasets in Section~\ref{sec:external}, with more detailed results available in the ``Detailed External Evaluation Results'' Section in Supplementary.
Finally, we demonstrate the impact of knowledge injection in Section~\ref{sec:ab_study_text}, and \model{}'s potential application scenarios in Section~\ref{sec:clinic_demo}.
\textcolor{black}{
Additional ablation experiments are provided in the ``Extended Ablation Studies'' Section in Supplementary; 
Model calibration analysis is presented in the ``Calibration Analysis'' Section in Supplementary;  
}

\subsection{Comparison with Specialist Models on Automatic Segmentation} 
\label{sec:specialist_internal}

\begin{figure}[!t]
    \centering
    \includegraphics[width = \textwidth]{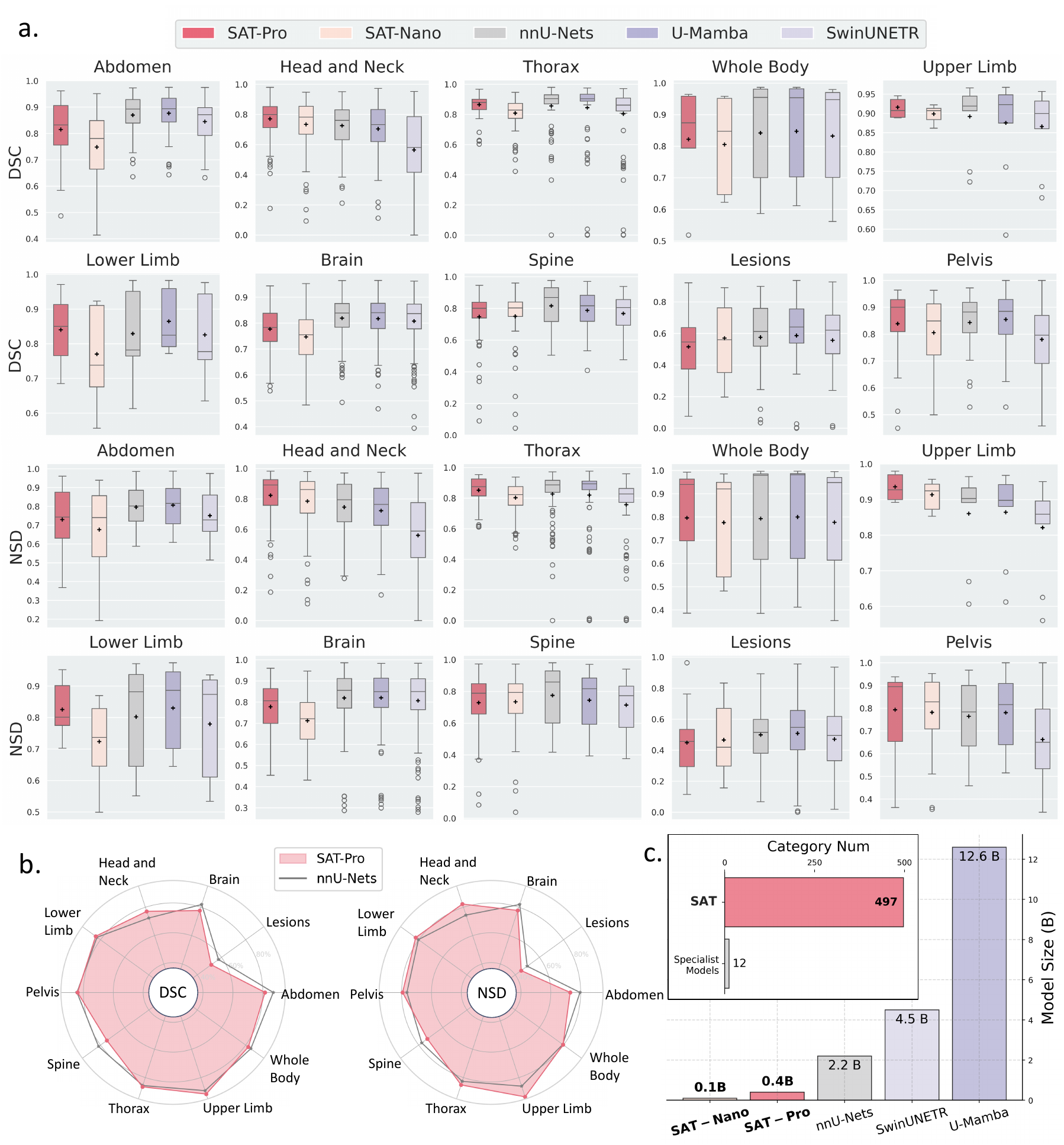}
    \vspace{-0.1cm}
    \caption{\textcolor{black}{\textbf{Internal evaluation between \model{}-Pro, \model{}-Nano, and three specialist models on 72 datasets from \samdataset{}.} Results are merged by different human body regions and lesions. \textbf{a}, Box plots on DSC and NSD results. The center line within each box indicates the median value; the bottom and top bound indicate the 25th and 75th percentiles respectively. The mean value is marked with a plus sign. The whiskers extend to 1.5 times the interquartile range. Outlier classes are plotted as individual dots. \textbf{b}, Comparison between \model{}-Pro and the most competitive specialist models nnU-Nets on performance. \textbf{c}, Comparison between \model{} and specialist models on model size and capability range. \model{} has much smaller model size compared to the ensemble of specialist models, while capable of segmenting 497 targets in one model. By comparison, each specialist model can only segment 12 targets on average.}
    }
    \label{fig:specialist_internal}
\end{figure}

In this experiment setting, we compare with specialist models~(nnU-Nets, U-Mamba, SwinUNETR) on all the 72 datasets in \textbf{\samdataset{}} as internal evaluation. All specialist models are trained with optimized configuration on each dataset with official codes. While both \textbf{\model{}-Pro} and \textbf{\model{}-Nano} are trained and evaluated on all datasets as one model. \textbf{Note that}, unless otherwise stated, \model{}-Pro and \model{}-Nano are trained on all 72 datasets of \samdataset{} throughout the following text.

Figure 3 (a) and Supplementary Table 3 shows the \textbf{region-wise results} on 8 regions of human body, including `Brain', `Head and Neck', `Thorax', `Abdomen', `Pelvis', `Spine', `Upper Limb', and `Lower Limb', as well as `Lesion', in terms of Dice Similarity Coefficient (DSC) and Normalized Surface Distance (NSD) respectively. 
Classes existing in multiple regions are specifically grouped as `Whole Body'. 

Despite having been proposed for a few years, nnU-Nets remains the best-performing specialist model overall. As a generalist model, \textbf{\model{}-Pro} consistently outperforms the most competitive baseline nnU-Nets in four regions: Head and Neck, Thorax, Upper Limb and Lower Limb. 
\textcolor{black}{
On average DSC of all 497 categories, \model{}-Pro shows comparable performance to nnU-Nets (paired $t$-test $p > 0.09$) and U-Mamba ($p > 0.13$), while surpass SwinUNETR significantly ($p < 2 \times 10^{-5}$).
}

Figure 3 (b) and (c) provide another view on the above results, where it can be seen that \model{}-Pro shows comparable segmentation performance to the 72 nnU-Nets, while being significantly smaller in size and more capable; for example, \model{}-Pro is approximately \textbf{1/5} of the ensemble of nnU-Nets, and is able to handle 497 classes, in contrast to each specialist model handling an average of only 12 classes.

We further finetune \textbf{\model{}-Pro} on each dataset, 
and report the \textbf{region-wise} results in Supplementary Table 3, denoted as \textbf{\model{}-Ft}. 
\textcolor{black}{
\model{}-Ft shows notable improvement over \model{}-Pro on all the regions and lesions.
On average performance over all categories, it outperforms U-Mamba on both DSC ($p < 2 \times 10^{-9}$) and NSD ($p < 0.01$), and nnU-Nets on NSD ($p < 6 \times 10^{-9}$).
}
This indicates that \model{} can serve as a strong pre-trained model for further adaptation.

We present dataset-wise results in Supplementary Table 5, 6, 7 and 8, and more detailed class-wise results in Supplementary Table 9, 10, 11 and 12; 

\subsection{Comparison with Interactive Segmentation Foundation Model}
\label{sec:medsam_internal}

In this section, we compare with MedSAM, an out-of-the-box interactive segmentation method trained on large-scale data. 
\textcolor{black}{
Due to inconsistent training data, we focus the internal evaluation on all the 32 datasets (out of 72) that were involved in training MedSAM for fair comparison.
}
\textbf{Note that} even though these datasets are included in MedSAM's training, we are unable to align the train-test splits. This means our test set might have been leaked in MedSAM's training.
We report \textbf{three results}: 
(i) simulate box prompts based on ground truth segmentation, using the minimum rectangle covering the ground truth (denoted as Tight), {\em i.e.}, the most accurate prompts; 
(ii) randomly shift each box corner by up to 8\% of the image resolution~(denoted as Loose), {\em i.e.}, allowing errors to some extent; 
(iii) directly use the tight box prompts as prediction~(denoted as Oracle Box), {\em i.e.}, the input baseline for MedSAM.

Figure 4 (a) and Supplementary Table 4 show the \textbf{region-wise results} for all methods. 
\textcolor{black}{
Notably, \textbf{\model{}-Pro} consistently outperforms MedSAM across all human body regions, even when MedSAM is prompted with the most accurate box (Tight), and achieve significantly superior average performance over all categories (paired $t$-test $p < 2 \times 10^{-9}$).
}
For lesion segmentation, \textbf{\model{}-Pro} underperforms MedSAM (Tight) due to the small lesion size, where the box prompts provide very strong priors to MedSAM, as evidenced by the oracle box even outperforming MedSAM's output on DSC score.
When perturbing the box prompts, MedSAM (Loose) shows significant performance drops across all regions and metrics.

\begin{figure}[!t]
    \centering
    \includegraphics[width = \textwidth]{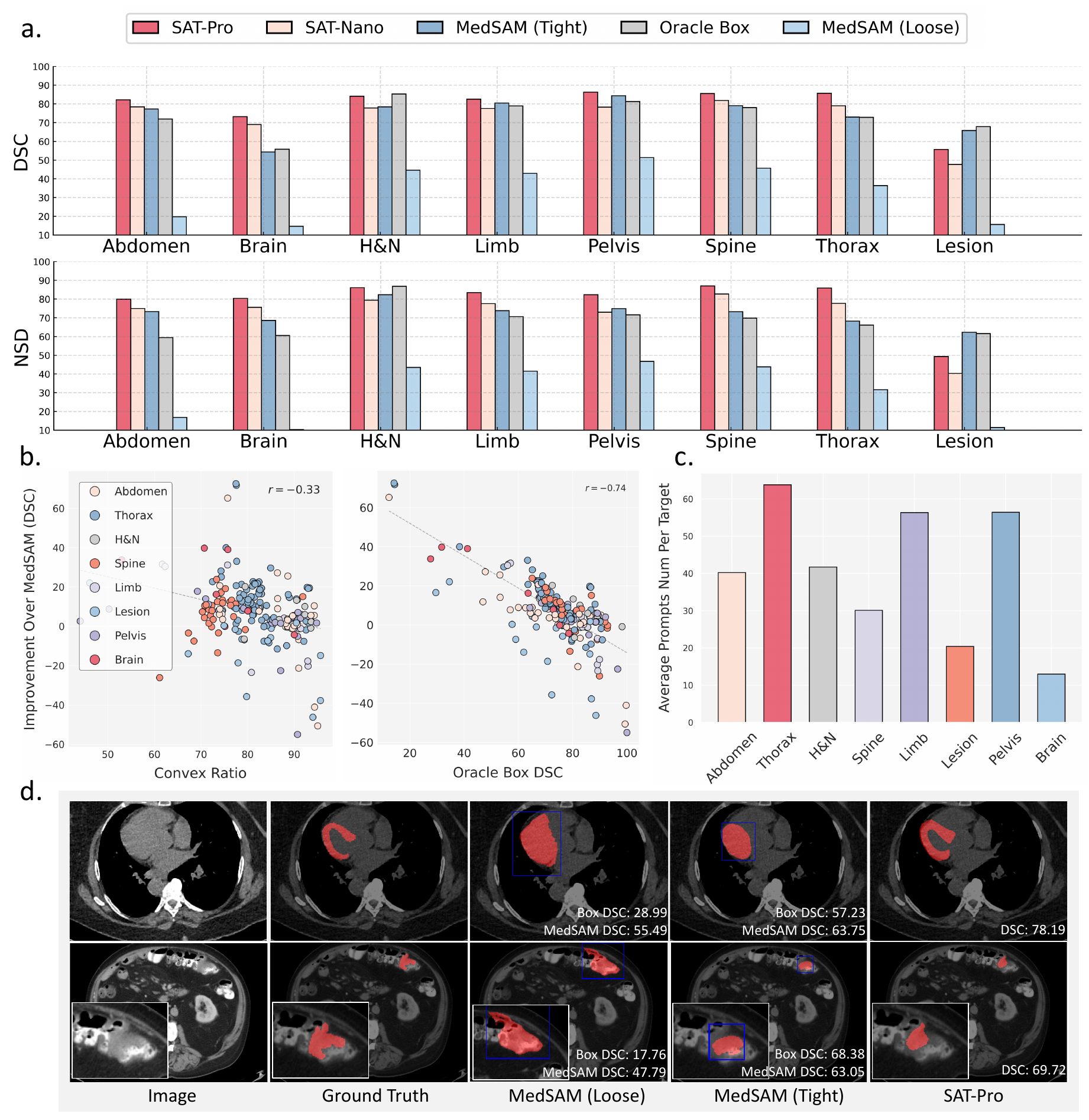}
    \vspace{-0.1cm}
    \caption{\textcolor{black}{\textbf{Internal evaluation between SAT-Pro, SAT-Nano, and MedSAMs on 32 datasets from \samdataset{}} Results are merged by different human body regions and lesions. \textbf{a}, Histograms on DSC and NSD results. \textbf{b}, Scatter plots comparing the performance improvement of \model{}-Pro over MedSAM on different segmentation targets (DSC score), with two irregularity metrics: convex ratio and oracle box DSC. Each point represents an anatomical structure or lesion, with a fitted line illustrating the trend. \textbf{c}, Average prompt numbers required by MedSAM to segment a target in 3D radiology scan, averaged over different human body regions. \textbf{d}, Quantitative results of MedSAMs and \model{}-Pro on myocardium (upper row) and colon cancer (lower row). The ground truth and segmentation masks are painted in red, while box prompts of MedSAM are plotted in black. The DSC score is calculated in slice-wise manner.
    H\&N: Head and Neck.}}
    \label{fig:internal_medsam}
    \vspace{-10pt}
\end{figure}

On \textbf{class-wise results}, in Figure 4 (b), 
we present the performance difference between \model{}-Pro and MedSAM on each category, with respect to the spatial irregularity of regions.
Inspired by BiomedParse~\cite{zhao2024foundation}, we define spatial irregularity with two factors: 
the ratio of ground truth to the tightest convex, denoted as `Convex Ratio';
the DSC score between oracle box prompt and ground truth, denoted as `Oracle Box DSC'. We observe that \model{}-Pro achieves greater improvement on targets with more irregular shapes, while MedSAM outperforms on some relatively regular-shaped targets.

We further present \textbf{qualitative results} from two representative examples in Figure 4 (d).
The upper row shows segmentation of myocardium with a relatively irregular shape. MedSAM incorrectly includes the left heart ventricle surrounded by the myocardium. By comparison, \model{}-Pro generates accurate predictions when simply prompted with the word `myocardium'.
The lower row demonstrates colon cancer segmentation on a CT image. 
The tight box prompt to MedSAM can be viewed as an acceptable segmentation, despite its limitation as a rectangle, while MedSAM's prediction is worse.
In addition, in both cases, we observe noticeable performance drops when the box prompt contains certain deviations, {\em i.e.}, MedSAM (Loose).

In Figure 4 (c), we show the average number of prompts required by MedSAM to segment a target in a 3D image scan. As it only allows slice-wise segmentation and the morphology of segmentation targets varies across different body regions, the number ranges from $10+$ to $60+$. By contrast, as a fully automatic segmentation model for 3D radiology images, \model{} requires only a single text prompt to segment the entire 3D scan. This simplicity and scalability advantage become more pronounced for multiple target segmentation.

We present \textbf{dataset-wise results} in Supplementary Tables 5, 6, 7, and 8, and more detailed \textbf{class-wise results} in Supplementary Table 13.

\subsection{Compare with Text-Prompted Segmentation Foundation Model}
\label{sec:biomedparse_internal}

\begin{figure}[!t]
    \centering
    \includegraphics[width = \textwidth]{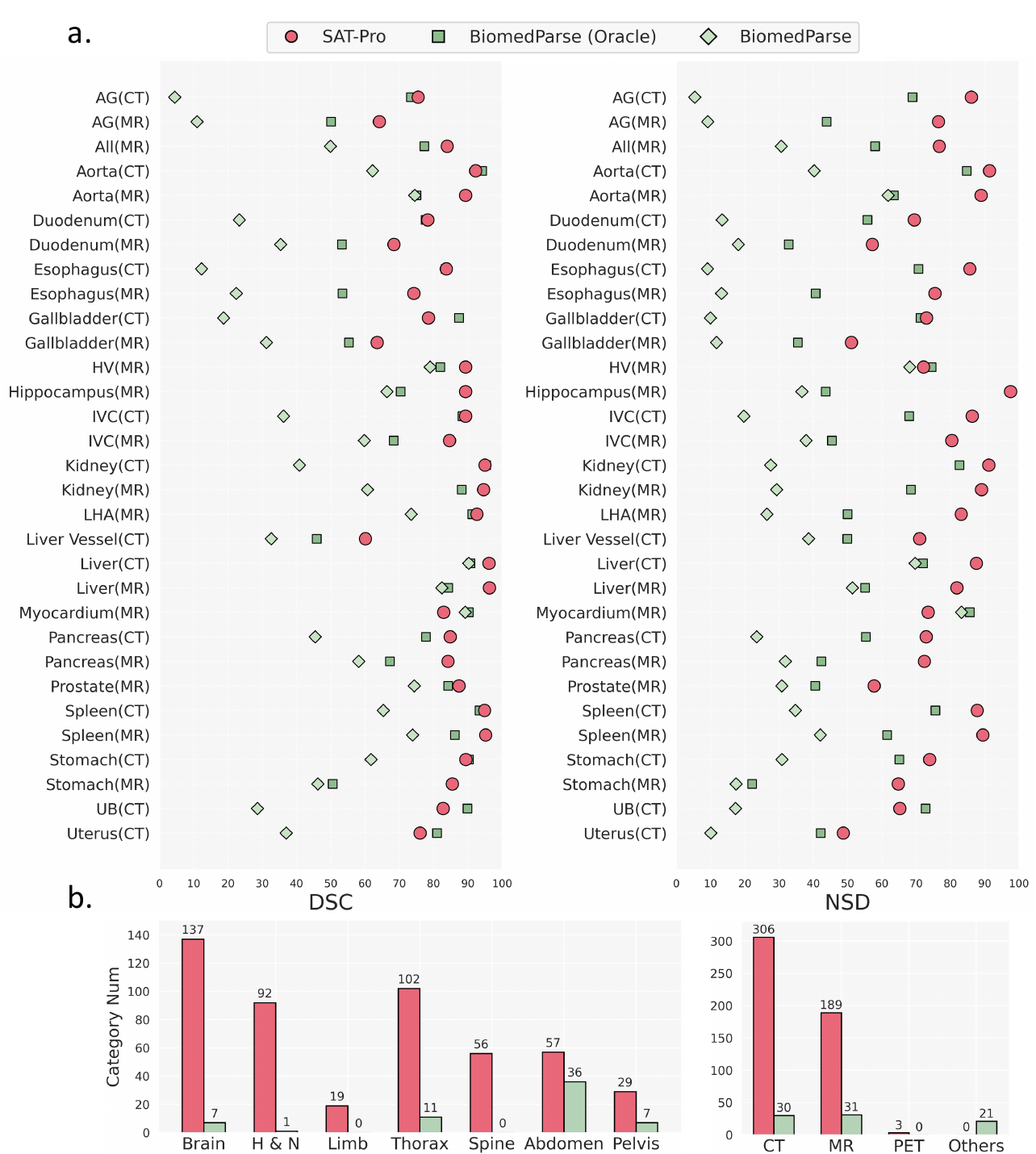}
    \vspace{0.1cm}
    \caption{\textcolor{black}{\textbf{Internal evaluation between \model{}-Pro and BiomedParses on 11 datasets from \samdataset{}}. (a) DSC and NSD scores. Results are merged and presented in class-wise manner. (b) The number of anatomical structures and lesions \model{} and BiomedParse can segment on different human body regions in radiology images, and on different imaging modalities. `Others' denotes non-radiology modalities. AG: adrenal gland; HV: heart ventricle; IVC: inferior vena cava; LHA: left heart atrium; UB: urinary bladder.
    }}
    \label{fig:biomedparse_internal}
    \vspace{-0.2cm}
\end{figure}

In this section, we compare with BiomedParse~\cite{zhao2024foundation}, a concurrent work that proposed a segmentation tool for general 2D biomedical images prompted by text. 
\textcolor{black}{
Due to inconsistent training data, we focus the internal evaluation on all the 11 datasets (out of 72) that were involved in training BiomedParse for fair comparison.
}
We report two results for BiomedParse: 
(i) Based on the ground truth, we only prompt targets present in the current slice, which follows its official evaluation setting. Similar to MedSAM, this approach avoids potential false positives on unannotated slices and thus represents performance under ideal conditions. 
We denote these results as BiomedParse (Oracle); 
(ii) Consistent with \model{}, we prompt all targets available in the dataset and filter out potential false positive predictions by p-values, as suggested by the official implementation.

Figure 5 (a) and Supplementary Table 14 present the \textbf{class-wise} performance of \model{} and BiomedParse. 
Across all categories, BiomedParse (Oracle) consistently achieves higher DSC and NSD scores compared to BiomedParse. This highlights that BiomedParse is prone to generating false positive predictions when prompted with non-existing targets, likely because BiomedParse is a 2D slice segmentation model that overlooks critical information from adjacent slices. 
\textcolor{black}{
\textbf{\model{}-Pro} consistently outperforms BiomedParse in all 30 categories except myocardium. 
Even compared to BiomedParse (Oracle), \model{}-Pro demonstrates superior performance on 23 out of 30 categories and notably excels in overall performance.
On average across all categories, both \textbf{\model{}-Pro} and \textbf{\model{}-Nano} significantly outperforms BiomedParse (Oracle) (paired $t$-test  $p < 7 \times 10^{-3}$ for DSC and $p < 2 \times 10^{-6}$ for NSD).
}

Furthermore, as illustrated in Figure 5 (b), BiomedParse is primarily designed as a segmentation tool for 2D biomedical images. In contrast, \model{}, developed as a large-vocabulary segmentation model specifically for 3D radiology images, demonstrates significantly broader applicability and superior performance on 3D radiology images.

\subsection{Evaluation on External Datasets.}
\label{sec:external}

\begin{figure}[t]
    \centering
    \includegraphics[width = \textwidth]{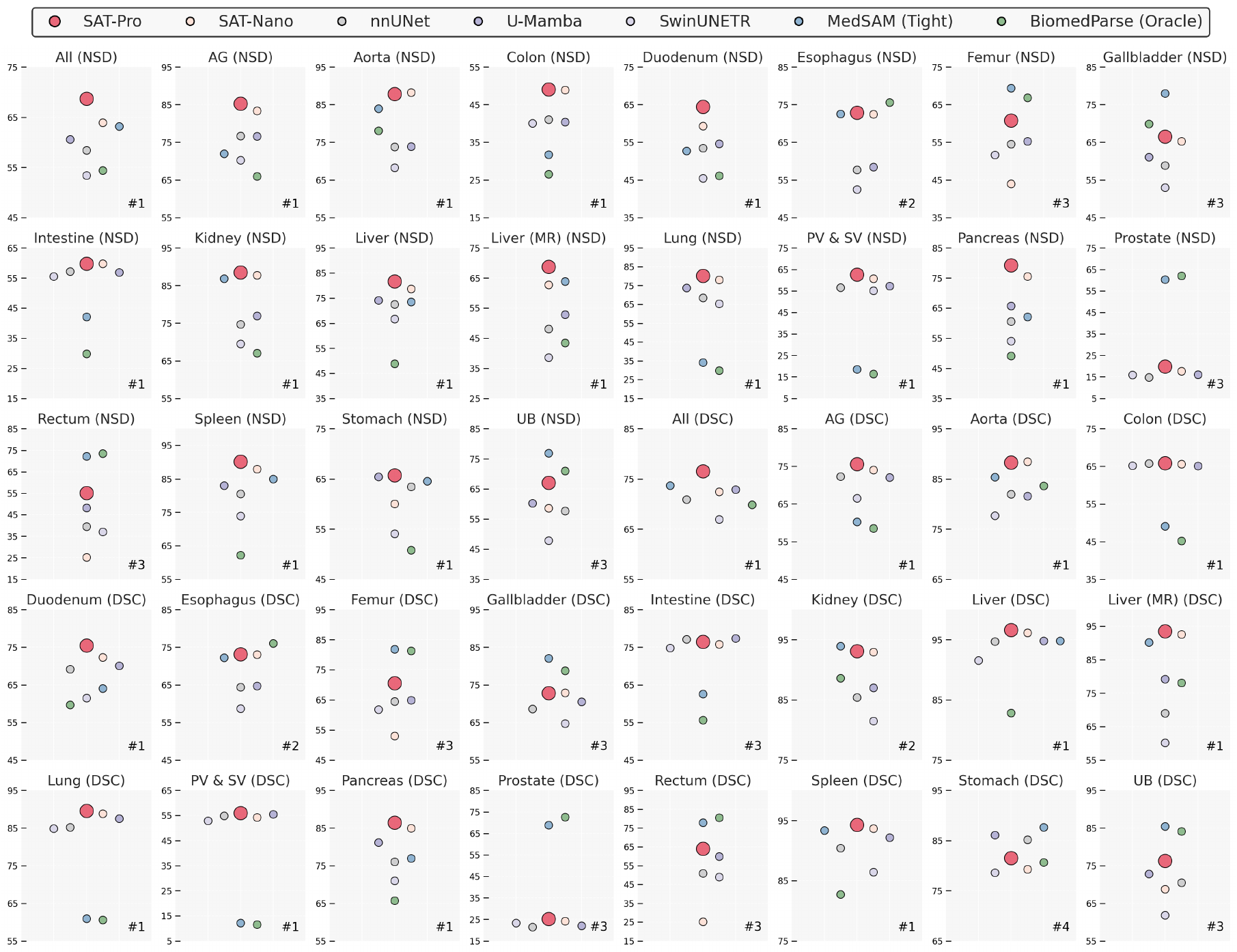}
    \vspace{0.1cm}
    \caption{\textcolor{black}{\textbf{External experiments between \model{}, specialist models, MedSAM and BiomedParse on AbdomenAtlas and LiQA.} Both DSC and NSD results are presented for each class in each dataset. We enlarge the size of \model{}-Pro in each sub-figure for distinction, and annotated the ranking of \model{} on each class. AG: adrenal gland; PV \& SV: portal vein and splenic vein; UB: urinary bladder.}
    }
    \label{fig:external}
\end{figure}

Here, we aim to evaluate the generalization performance of segmentation models on images from different medical centers. 
As generalist models, \model{}, MedSAM, and BiomedParse are directly evaluated on two unseen datasets. For specialist models, considering their customized configurations on each dataset, we systematically evaluate 21 out of 72 specialist models on target datasets for shared categories. For example, to evaluate the generalization performance on `lung' in AbdomenAtlas, we use specialist models trained on CT-ORG and LUNA16, as they all involve this class, and then average the results. 
The details of the overlapped label spaces are shown in Supplementary Figure 5. To maintain performance for specialist models, the pre-processing of target datasets is kept the same as the source dataset in evaluation.

We report DSC and NSD results in Figure 6 and Supplementary Table 15, with the following observations:
(i) For specialist models, U-Mamba achieves more competitive results than nnU-Nets on both DSC and NSD scores, while SwinUNETR remains the worst; 
(ii) For generalist models for 2D images, MedSAM (Tight) consistently outperforms BiomedParse (Oracle) on all categories, implying that accurate box prompts provide strong priors when extending to out-of-domain images;
\textcolor{black}{
(iii) \model{}-Pro achieves the best performance on average over all categories, exceeding the second-best candidate MedSAM by 2.9 on DSC (paired $t$-test  $p < 7 \times 10^{-4}$) and 5.52 on NSD ($p < 9 \times 10^{-6}$). 
Meanwhile, \model{}-Pro consistently outperforms the specialist models on all categories in terms of NSD score and on 17 out of 19 categories in terms of DSC score.
}

\subsection{Ablation Study on Text Encoder}
\label{sec:ab_study_text}

As will be illustrated in Section~\ref{sec:knowledge_encoding}, 
to build a large-vocabulary segmentation model driven by text prompts,
we inject domain knowledge into the text encoder to provide precise prompts for the target of interest, {\em i.e.}, the encoding of terminology. 
In this section, we conduct experiments and discuss the effect of domain knowledge. 
To save computational cost, the experiment have been conducted on \textbf{\samdataset{}-Nano} dataset.

\begin{figure}[t]
    \centering
    \includegraphics[width = .9\textwidth]{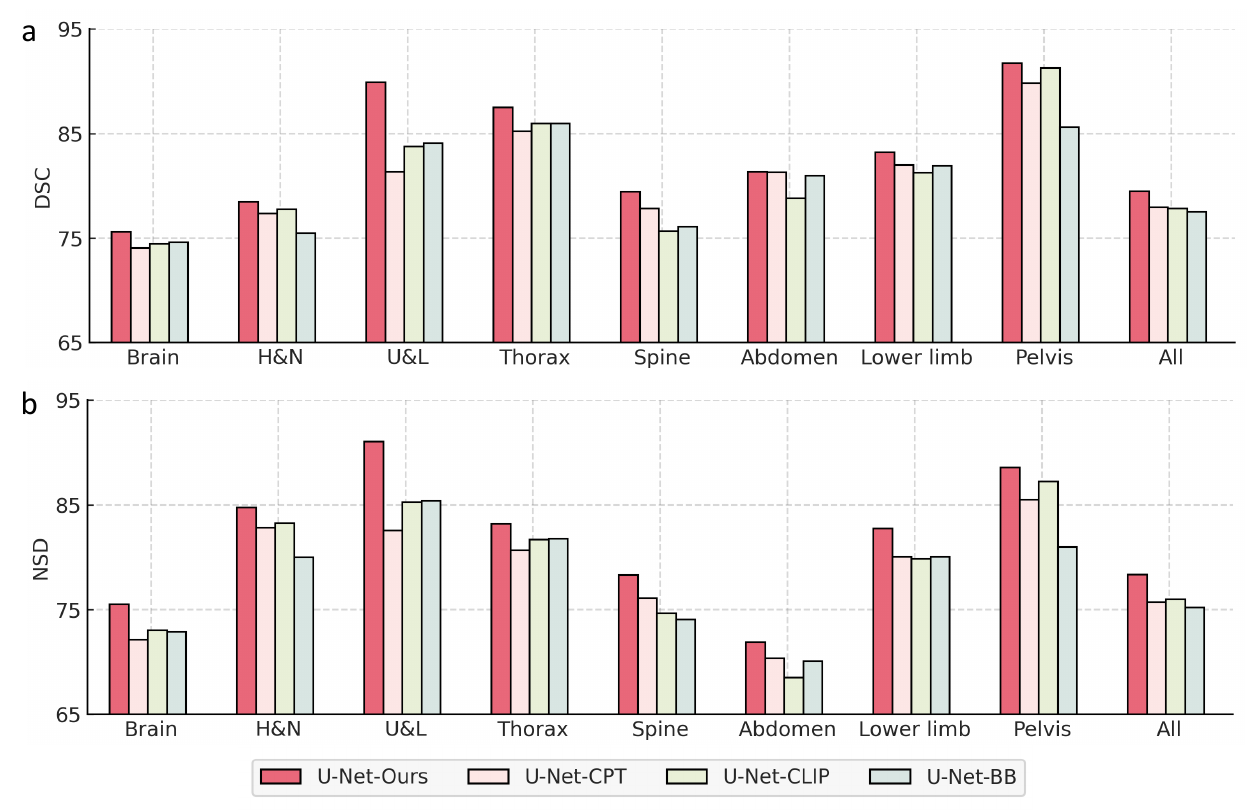}
    \vspace{0.1cm}
    \caption{\textbf{Evaluations on \samdataset{}-Nano variants with different text encoders.} `All' denote the average scores over all the classes(n=429), including lesion classes. \textbf{a}, DSC comparison; \textbf{b}, NSD comparison.}
    \label{fig:text_ab_hist}
\end{figure}

Specifically, we train four \textbf{\model{}-Nano} variants with different text encoders: 
(i) BERT-Base, a prevalent text encoder in natural language processing, but not specifically fine-tuned on medical corpora; 
(ii) the text encoder of CLIP, a state-of-the-art model pretrained on 400M image-text pairs and widely used in vision-language tasks; 
(iii) the state-of-the-art text encoder for medical retrieval tasks, {\em e.g.}, MedCPT; 
(iv) the text encoder pre-trained on our multimodal medical knowledge graph, as illustrated in Section~\ref{sec:knowledge_encoding}. For all variants, we use U-Net as the visual backbone and denote them as \textbf{U-Net-BB}, \textbf{U-Net-CLIP}, \textbf{U-Net-CPT}, and \textbf{U-Net-Ours}.

As shown in Figure 7 and Supplementary Table 17, the performance of U-Net-BB, U-Net-CLIP, and U-Net-CPT is close. Overall, U-Net-BB performs the worst, while U-Net-CPT slightly exceeds others on DSC (+0.1) and U-Net-CLIP slightly exceeds others on NSD (+0.29) scores averaged over all classes.
By contrast, U-Net-Ours surpasses all other variants consistently across all regions, 
with notable margins on both DSC (+1.54) and NSD (+2.36) scores on average over all classes. This demonstrates the effectiveness of our proposed multimodal knowledge injection.

We further investigate the effect on different classes.
As illustrated in Figure 8 (a) and (b), the 429 classes in \textbf{\samdataset{}-Nano} typically follow a long-tail distribution.
The 10 `head' classes account for 12.75\% of the annotations in \samdataset{}-Nano. In contrast, the 150 classes with minimum annotations account for only 3.25\%, even though they comprise 34.97\% of the 429 classes. 
We compare U-Net-Ours, U-Net-CPT, U-Net-CLIP, and U-Net-BB on the `head' classes, `tail' classes, and the rest (denoted as `middle' classes). 
In Figure 8 (c), the performance of the model variants drops from head to tail classes, showing that the long-tailed distribution poses a significant challenge for medical segmentation. 
Using our proposed knowledge-enhanced text encoder, 
U-Net-Ours achieves the best performance across all three scenarios. 
On `head' classes, it outperforms the second-best variant by 0.71 on DSC and 2.44 on NSD. On `tail' classes, the improvement is even more pronounced.
For more detailed results on each class and dataset, we refer the reader to Supplementary Tables 22, 23, 24, and 25. 

In addition to segmentation performance, we evaluate the text encoders on `concept-to-definition' retrieval using human anatomy knowledge. 
In total, we collect 6,502 anatomy concept-definition pairs. We find that the Recall@1~(R1) for BERT-Base is only 0.08\%, suggesting it can hardly understand these anatomy concepts and possesses almost no domain knowledge. 
The R1 is 4.13\% for CLIP and 11.19\% for MedCPT. Though this is a significant improvement over BERT-Base, they still struggle to distinguish these concepts. By contrast, our proposed text encoder achieves 99.18\% R1, indicating that the knowledge is successfully injected into the text embedding for each anatomy concept.

\begin{figure}[!t]
    \centering
    \includegraphics[width = \textwidth]{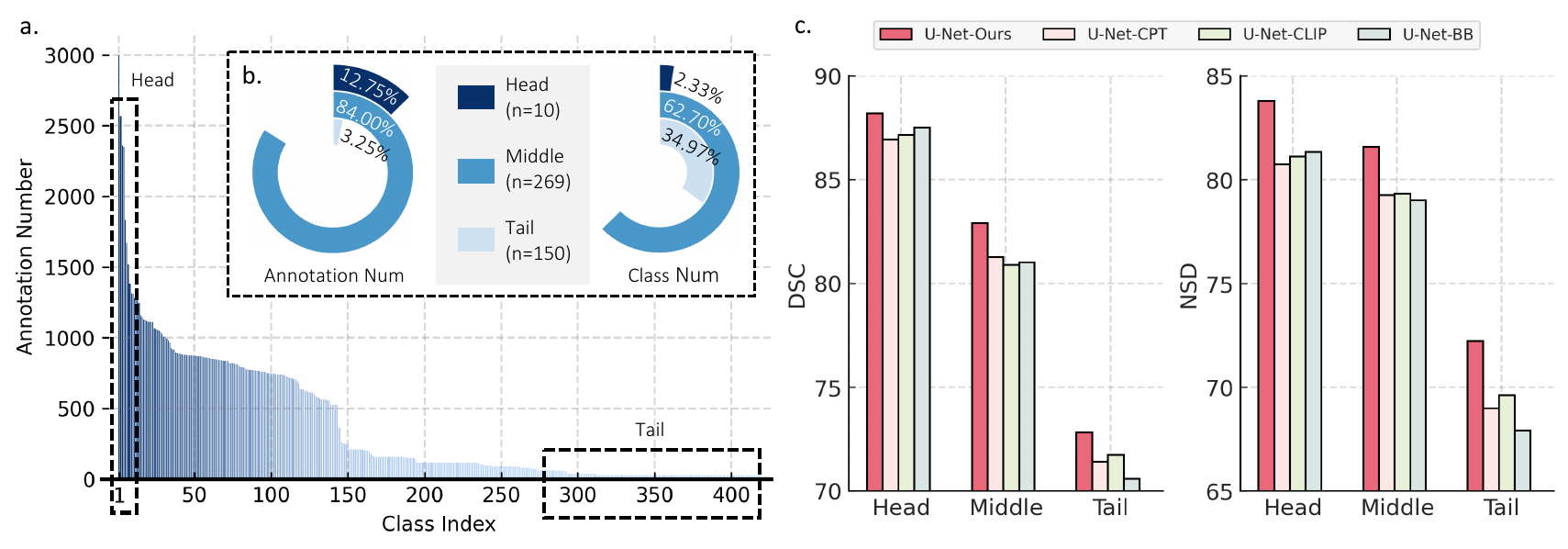}
    \vspace{0.1cm}
    \caption{\textbf{The impact of domain knowledge on a long-tail distribution.} \textbf{a}, The annotation number of all the 429 classes in \samdataset{}-Nano, sorted from highest to lowest. \textbf{b}, The proportion of `head', `middle' and `tail' in class number and annotation number. The DSC and NSD comparison on `head', `middle' and `tail' classes.}
    \label{fig:long_tail}
\end{figure}

\subsection{Qualitative Results -- \model{} as an Interface Between Language and Segmentation}
\label{sec:clinic_demo}

Thanks to the text-driven features of \model{}, it can be seamlessly applied as an interface between natural language and segmentation, 
{\em i.e.}, acting as a high-performance and efficient agent for language models. 
Here, we demonstrate three potential applications in Figure 9:
(i) We demonstrate a scenario where GPT-4~\cite{GPT4} analyzes and extracts the anatomical targets of interest from a real clinical report and prompts \model{} to segment them on the clinical image. As can be seen in the upper row, the targets in reports can be well detected by the language model~(GPT-4) and commendably segmented by \model{}-Pro, which provides visual cues for the clinical report and enhances its interpretability; 
(ii) We show that \model{} can help LLMs handle segmentation requests in free-form conversations with any users. The LLM can easily recognize these requests and leverage \model{} to deliver precise segmentation results, which greatly extends the conversational interface.
(iii) We explore more complicated situations, where \model{} can ground the lesions based on comprehensive analysis of radiology images as well as contextual EHR data such as patient complaints, establishing a complete automated pipeline from diagnosis to segmentation.

\begin{figure}[!t]
    \centering
    \includegraphics[width = \textwidth]{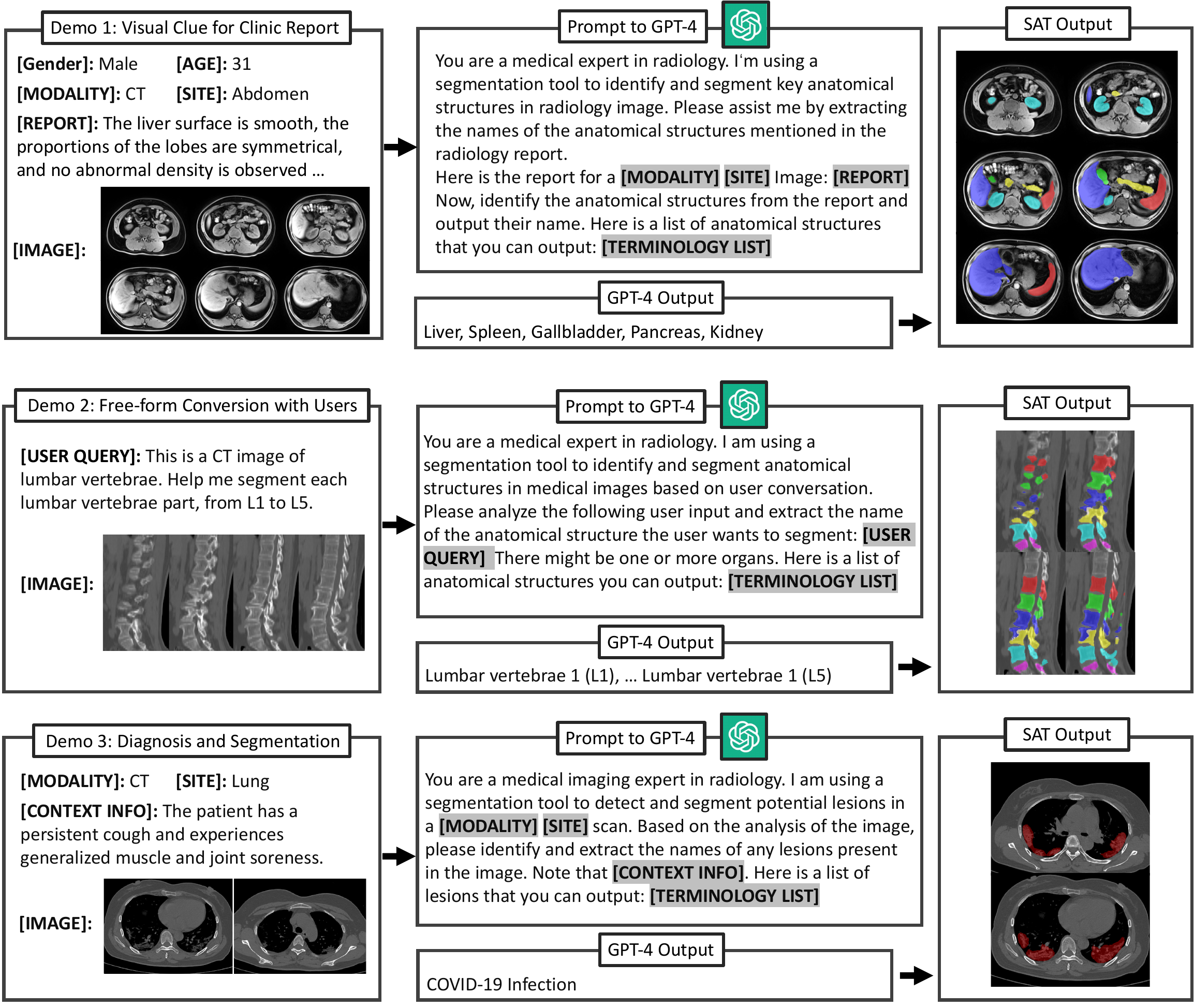}
    \vspace{0.1cm}
    \caption{
    \textcolor{black}{
    \textbf{\model{} as agent for large language models.} Combining \model{}-Pro and GPT4, we demonstrate three potential applications: providing visual clues for clinic report, handling segmentation request in free-form conversation, and an automated pipeline from diagnosis to segmentation. 
    For each application, the specific prompt template in use is shown.
    The \textbf{[TERMINOLOGY LIST]} contains anatomical structures that SAT can segment, which can be customized based on different clinicians' requirements ({\em e.g.}, in demo 3, 
    we only provide lesion categories).
    While the other bolded components in the templates 
    ({\em e.g.}, \textbf{[MODALITY]}) are variable placeholders that need to be filled with case-specific information.
    We show one example case for each application on the leftmost column.
    Target names are extracted from GPT's text output by string parsing, and serve as the exact text prompts for SAT.
    We extract representative slices from the image volume for demonstration.
    }
    }
    \label{fig:clinic_demo}
\end{figure}
\section{Discussion}
\label{sec:discussion}

Developing specialist models on individual tasks has been the dominant solution for medical image segmentation for years~\cite{nnUNET,dou20173d,li2018h,yu2018recurrent,zhou2019unet++,schlemper2019attention,zhang2022decoupled,UNetr,cao2022swin,SwinUNetr,xie2021cotr,zhang2022mmformer,chen20233d,zhou2023nnformer}. In this paper, we aim to build a large-vocabulary, effective, and flexible medical segmentation foundation model by training on an unprecedented dataset collection and driven by knowledge-enhanced text prompts. 
The significance of the proposed \model{} is demonstrated through multiple dimensions.


First, our work represents an important step towards a universal segmentation model in medical scenarios.
Despite the diverse images and segmentation targets from different clinical scenarios, \model{} can flexibly handle them within a single generalist model, effectively replacing the need for dozens of specialist models.
\textcolor{black}{
Through comprehensive internal evaluation, \model{}-Pro has demonstrated competitive results against the ensemble of 72 specialist models, achieving comparable performance to nnU-Net and U-Mamba, and superior performance to SwinUNETR.
Remarkably, \model{}-Pro achieves this with a model size reduced to 20\% or less of the ensemble, greatly improving efficiency.
}
When evaluating on external multi-center datasets, \model{}-Pro exhibits advanced generalization ability compared to all specialist models, highlighting its excellent cross-center transferability.
With dataset-specific fine-tuning, \model{}-Ft can further improve the performance, thus balancing the clinical requirements between generalist solutions and specialist models.

Second, as an automatic method prompted by text, \model{} offers an alternative approach to recent works, such as interactive segmentation foundation models~\cite{MedSAM,wang2023sammed3d}.
Through both qualitative and quantitative comparisons, \model{} demonstrates enhanced segmentation accuracy and robustness, particularly on targets with irregular shapes.
Unlike interactive methods that rely on spatial prompts and may suffer from inaccurate prompts, leading to performance fluctuations, \model{} can effectively automate segmentation on 3D images with text prompts, significantly reducing user inference time and associated costs.
In addition, compared to our concurrent work on text-prompted 2D segmentation foundation model, namely BiomedParse~\cite{zhao2024foundation},
\model{} demonstrates significantly broader applicability in 3D radiology images and consistently outperforms it in both in-domain and out-of-domain scenarios.

Third, our work implies that scaling laws---observed in large language models---also apply to large-vocabulary medical segmentation.
In this work, we build \model{}-Nano (110M) and \model{}-Pro (447M).
In both region-wise and class-wise evaluations, \model{}-Pro shows a clear performance boost over \model{}-Nano, outperforming the latter on most regions and classes. 
These findings indicate a promising way to continuously improve the performance of segmentation foundation models.

Fourth, we construct the first multimodal knowledge graph on human anatomy and demonstrate that knowledge injection can significantly improve segmentation performance, particularly for `tail' classes. 
As the scope of medical segmentation expands to include an increasing number of targets, the long-tail problem will become more pronounced, underscoring the critical importance of knowledge enhancement in addressing this challenge.

Lastly, \model{} can be used as an agent to bridge language and segmentation.
In Section~\ref{sec:clinic_demo}, we show that \model{} can be applied to segment targets based on the output from language models and support visual grounding across various clinical scenarios. This highlights the potential of \model{} as a high-performance, efficient, and out-of-the-box tool agent, seamlessly collaborating with ever-evolving large language models.
In addition, \model{} has recently been applied to provide comprehensive grounding annotations for medical visual-language datasets in a scalable manner~\cite{zhang2024radgenome, xie2024medtrinity}.

\textcolor{black}{
As one of the first exploratory work in this field, 
several limitations remain to be addressed in our work, and we propose the following future works:
(i) The performance of \model{}-Pro still lags behind some specialist models, {\em e.g.}, nnU-Nets, in some region. Further scaling up the model can be a promising direction;
(ii) Although \model{} is capable of segmenting 497 types of targets on medical images, its generalization ability to unseen categories (including unseen lesions/pathologies) remains limited. Inspired by recent advances in language-grounded segmentation for natural images and videos~\cite{xu2023learning,wang2024ov,lai2024lisa,yan2024visa}, exploring open vocabulary segmentation in medical imaging represents a promising direction for future work;
(iii) For practical deployment, while our current inference speed is suitable for clinical use (as shown in Supplementary Tables 1 and 2), further optimization for standard clinical hardware remains important;
We will explore approaches for more efficient deployment, such as our subsequent work on knowledge distillation~\cite{li2024lorkd};
(iv) Although SAT-DS includes datasets from multiple countries/regions (United States, Europe, China, Africa, etc.), distribution biases still persist. 
Many regions remain uncovered, and the dataset is heavily skewed toward adult populations with limited pediatric/fetal data (e.g., FETA2022). 
These demographic imbalances may affect model generalization across different populations and age groups, necessitating bias mitigation strategies in future work;
}
\clearpage
\section{Method}

In this section, we first describe the two types of data collected to build \model{}: multimodal domain knowledge (Section~\ref{sec:knowledge_source}), and medical segmentation data (Section~\ref{sec:dataset4segmentation}).
Based on them, we detail the development of \model{}, starting with the task formulation (Section~\ref{sec:problem_formulation}), then the multimodal knowledge injection (Section~\ref{sec:knowledge_encoding}) and segmentation training (Section~\ref{sec:segmentation}).
Finally, we present the evaluation settings, including the datasets (Section~\ref{sec:evaluation_datasets}), baselines (Section~\ref{sec:baselines}), protocols (Section~\ref{sec:experiment_settings}) and metrics (Section~\ref{sec:metrics}).



\subsection{Domain Knowledge}
\label{sec:knowledge_source}

To acquire textual knowledge, we mainly exploit two sources of domain knowledge:
the Unified Medical Language System (UMLS)~\cite{UMLS},
a comprehensive medical knowledge graph consisting of concept definitions and their interrelations; 
search engines, which are prompted to organize knowledge into a graph of the same format, specifically refined for the human anatomy corpus.
Regarding visual knowledge, we have compiled 72 medical segmentation datasets, creating an atlas that covers over 497 anatomical structures of the human body. 
Examples from these sources are illustrated in Figure 10 (a) and (b).
In the following, we detail each knowledge source in sequence.

Unified Medical Language System (UMLS)~\cite{UMLS} is a knowledge source of biomedical vocabulary developed by the US National Library of Medicine~\cite{nlm_nih}. 
It integrates a wide range of concepts from more than 60 families of biomedical vocabularies, each equipped with a Concept Unique Identifier (CUI) and definition. It also contains the relations among these concepts. Following~\cite{zhang2023knowledge}, we extract 229,435 biomedical terminologies and definitions, as well as 1,048,575 relationship triplets, composing a knowledge graph of these terminologies.

Although UMLS is widely acknowledged and adopted as a general medical knowledge corpus~\cite{wu2023medklip, zhang2023knowledge, lei2023unibrain, zheng2023large}, it lacks a fine-grained description of anatomy concepts critical for segmentation tasks. 
For example, for `liver', the definition is `A large lobed glandular organ in the abdomen of vertebrates that is responsible for detoxification, metabolism, synthesis, and storage of various substances.', which erases the morphological features and focuses on functionality. Meanwhile, more comprehensive knowledge on human anatomy may be scattered across various authoritative websites online, {\em e.g.}, Wikipedia, ScienceDirect, {\em etc.}. 
To harvest such knowledge, we select 6,502 anatomy concepts, 
and prompt a search engine to retrieve and summarize definitions for them. We use the following prompt template:
\vspace{8pt}
\begin{mdframed}[backgroundcolor=gray!10, linewidth=0pt]
Definition of xxx. Include the location, shape, appearance, structure, and spatial relations to other anatomical structures. No need to include functionality. End with `END'.
\end{mdframed}
\vspace{8pt}
For illustration, the search engine referred to authority websites including Columbiasurgery, Hopkins Medicine and summarized the definition for `liver' as: `A large organ found in the upper right quadrant of the abdomen, it stands as the largest gland within the human body, with a weight of about 1.5 kilograms. This structure exhibits a reddish-brown hue and is cone or wedge-shaped ... ...'. 
While constructing the knowledge graph, we also adopt GPT4~\cite{GPT4} to extract 38,344 relations between anatomical structures in the generated information-dense definitions with the following prompt: 
\vspace{8pt}
\begin{mdframed}[backgroundcolor=gray!10, linewidth=0pt]
This is the description of xxx. Please help me find its relations with other anatomical structures in radiological images. Summarize them with the template: Relation: xxx (relational preposition), Anatomical structure: xxx (name of another anatomical structure).\\
For example, ``Relation: situated below, Anatomical structure: xxx'', ``Relation: connected to (via xxx), Anatomical structure: xxx'' ... ...
\end{mdframed}
\vspace{8pt}

Segmentation datasets naturally provide visual features for anatomy concepts corresponding to or complementary to the textual description, such as the texture, spatial location, shape, and so on. 
Details on our collected segmentation datasets are described in Section~\ref{sec:dataset4segmentation}. 
Here, we use them as a large-scale and diverse visual atlas library, 
and link the visual regions to corresponding concepts in the textual knowledge graph, bridging the knowledge between visual and language modality.

In summary, by mixing these data, we construct a multimodal medical knowledge tree. As demonstrated in Figure 10 (c), the concepts (including both anatomical structures and lesions) are linked via the relations and further extended with their definitions, containing their characteristics. Additionally, some are further mapped to the visual atlas, demonstrating their visual features that may hardly be described purely by text. 
More examples on the curated knowledge dataset are shown in Supplementary Table 34, 35, and 36.

\begin{figure}[t]
    \centering
    \includegraphics[width = \textwidth]{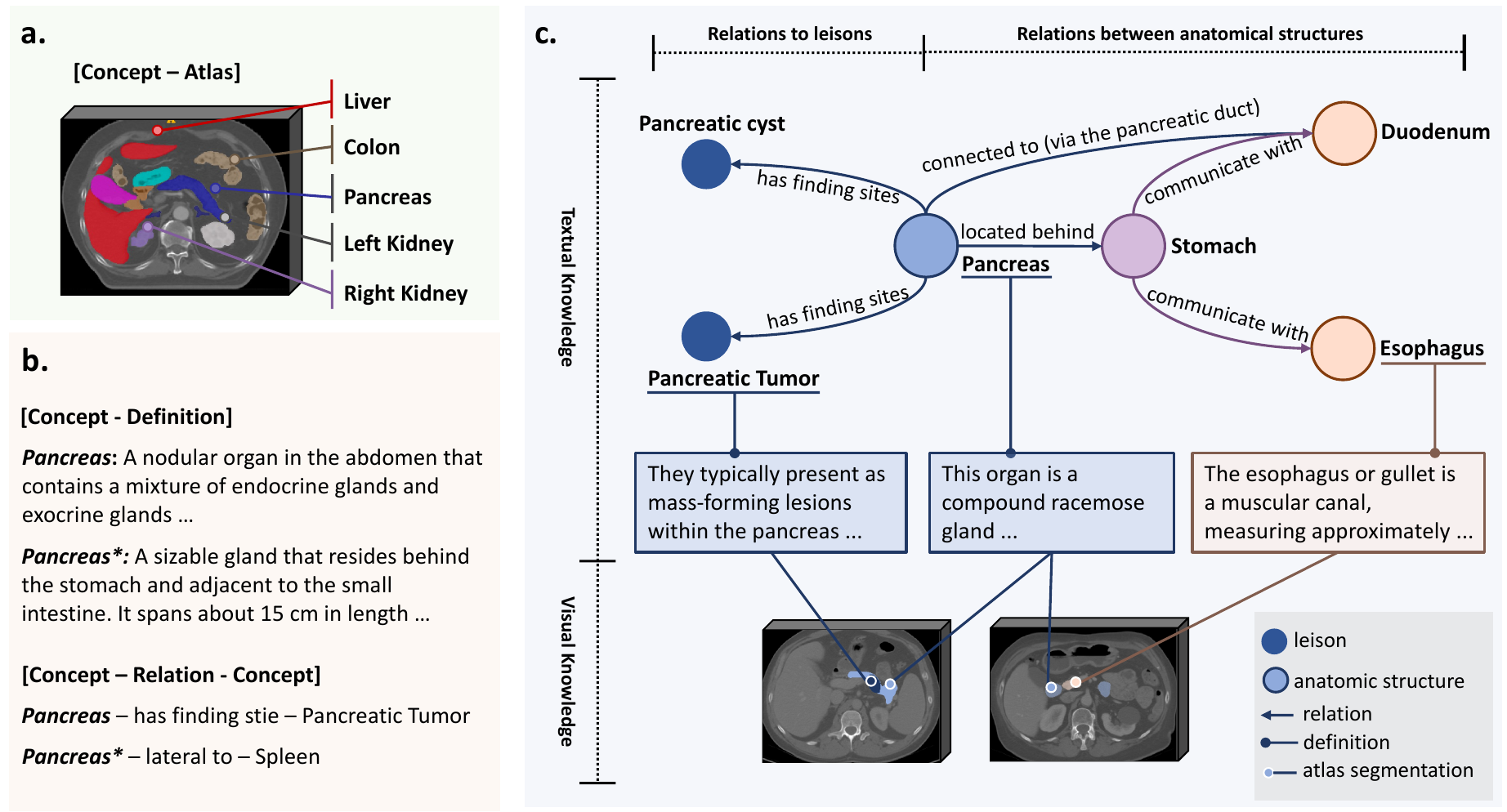}
    \vspace{-10pt}
    \caption{\textbf{The medical knowledge used in the visual-language pretraining.} (a) Segmentation datasets provide an atlas for extensive anatomy concepts. In this example, atlas segmentation is marked with different colors. (b) Knowledge generated from UMLS and search engines encompasses a broad array of concept-definition pairs and extensive relationships. (c) By integrating all collected knowledge sources, a medical knowledge tree is constructed. All definitions are partially displayed for conciseness. Definition and relation denoted with * are derived from the search engine, otherwise from UMLS.
    }
    \label{fig:knowledge_source_examples}
    \vspace{-10pt}
\end{figure}

\subsection{Segmentation Dataset}
\label{sec:dataset4segmentation}

To train our segmentation model with the ability to handle segmentation tasks of different targets, across various modalities and anatomical regions, 
we collect and integrate 72 diverse publicly available medical segmentation datasets, totaling 22,186 scans including CT, MRI, and PET, and 302,033 segmentation annotations spanning 8 different regions of the human body: Brain, Head and Neck, Upper Limb, Thorax, Abdomen, Pelvis, and Lower Limb. 
The dataset is termed as \textbf{\samdataset{}}.
More details are present in Supplementary Table 26 and 27. 
\textbf{Note that}, some public datasets are not mutually exclusive, {\em e.g.}, KiTS23 and KiTS21~\cite{KiTS23}, 
we thus only collect the latest version, to avoid redundancy and potential data leakage in train-test split.

Before mixing these datasets for training, two challenges remain: 
(i) the anatomical targets from each dataset must be integrated into a unified annotation system. The clinic demands beneath each dataset collection might be different, resulting in different annotation standards and granularity. Meanwhile, since most datasets are annotated for training specialist models like nnU-Net~\cite{nnUNET}, precise and consistent terminology or expression for anatomical targets is often ignored. 
Therefore, a unified label system is demanded to avoid potential contradictions when training on mixed datasets.
(ii) some critical image statistics, such as intensity distribution and voxel spacing vary from dataset to dataset, hindering the model from learning consistent image representations across datasets. 
In the following, we present details for dataset integration and pre-processing, and how we address the abovementioned challenges.

\begin{figure}[t]
    \centering
    \includegraphics[width = \textwidth]{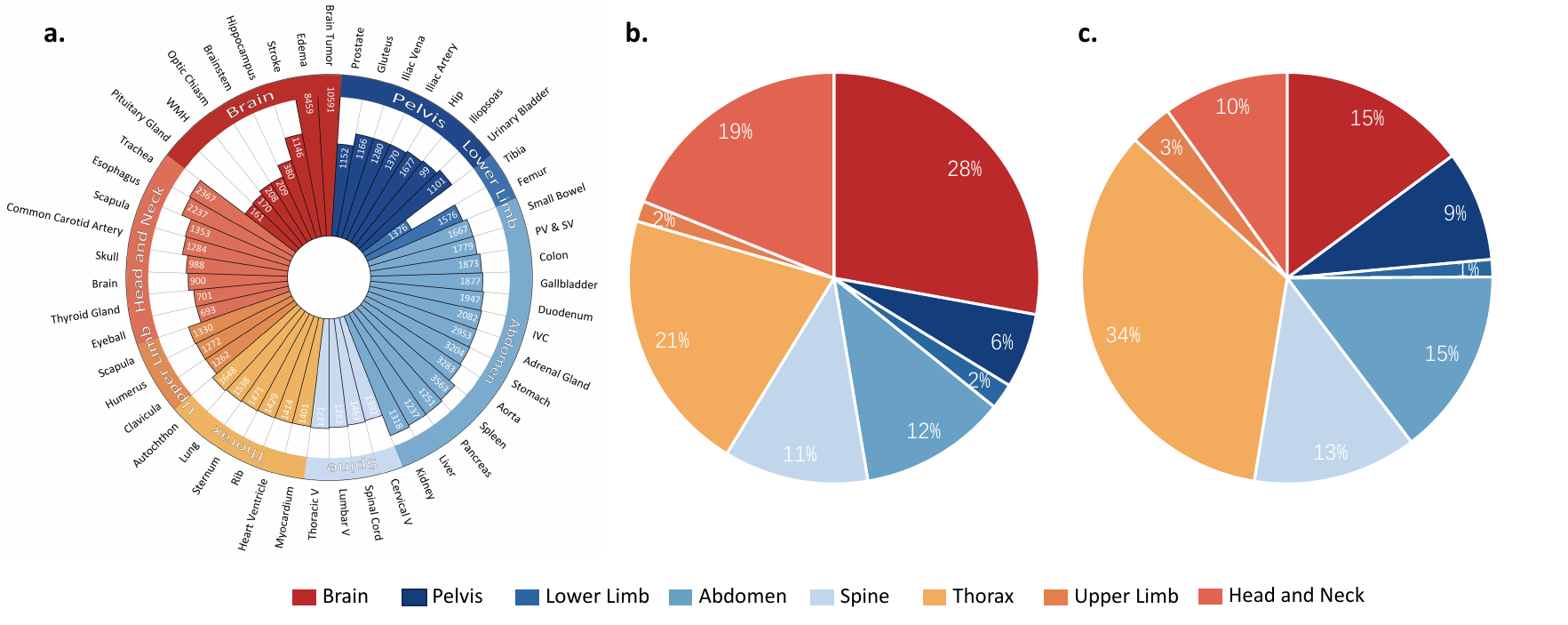}
    \caption{\textbf{Statistics of ~\samdataset{} across different anatomical regions.} (a) Annotation num of some representative classes in each anatomical region; (b) Number of classes in each anatomical region; (c) Number of annotations in each anatomical region. LC/HC: laryngeal/hypopharyngeal cancer, WHM: white matter hyperintensities, PV\&SV: portal vein and splenic vein, IVC: inferior vena cava, Thoracic V: thoracic vertebrae, Cervical V: cervical vertebrae, Lumbar V: lumbar vertebrae.
    }
    \label{fig:region_statistic}
\end{figure}

To ensure a unified annotation system, we take three procedures while integrating different datasets:
(i) we manually check each anatomical target in each dataset and assign a medical term to it, which is guaranteed to be precise and unambiguous across datasets. For instance, the targets that require distinction between orientations, such as the left lung and right lung, are always identified according to the left and right of the human body. And the same anatomical targets from different datasets are named consistently. 
For example, the $i$-th lumbar vertebrae in both TotalSegmentator~\cite{Totalsegmentator} and MRSpineSeg~\cite{MRSpineSeg} are named with the format ``lumbar vertebrae $i$ (L$i$)''; 
(ii) we adjust the annotations to minimize contradictions between overlapped classes. For example, considering that many organ segmentation datasets do not exclude lesions within organs, {\em e.g.}, AbdomenCT-1K and CT-ORG, we merged the lesion annotations with the corresponding infected organ annotations in other datasets to maintain consistency.
(iii) the same anatomy may have been annotated with different hierarchies in different datasets. In such cases, we manually merge the fine-grained classes to generate additional classes as a complement to close the gap between datasets. For example, sub-regions of the liver in Couinaud Liver~\cite{Couinaud} are merged and added as a new class ``liver''. As we will keep collecting datasets to scale up \samdataset{}, such a label system will be maintained and updated continuously.

As properties of each dataset may greatly impact the training of the segmentation network~\cite{nnUNET}, such as intensity distribution and voxel spacing, we deliberately apply some normalization procedures to all the datasets to ensure uniformity and compatibility between them. 
{\em Firstly}, all the images are reoriented to specific axcodes, respaced to a voxel size of $1 \times 1 \times 3~mm^2$ and cropped to the non-zero region. {\em Secondly}, we apply different intensity normalization strategies to CT, MRI and PET images. Specifically, for CT images, intensity values are truncated to $[-500, 1000]$ and applied z-score normalization. 
For MRI and PET images, intensity values are clipped by $0.5\%$ and $99.5\%$ of the image, and then z-score normalized. During training, we randomly crop the image patch with a fixed size of $288 \times 288 \times 96$. Random zoom-in, zoom-out, and intensity scaling are applied for data augmentation.

After integrating datasets, we derive a segmentation data collection that covers 497 segmentation classes, far outpacing each single dataset in both diversity and scale. 
Specifically, the data collection is more than \textbf{fourth times} the size of the largest dataset (BraTS2023-GLI) in terms of volume number, and nearly \textbf{triple} the most comprehensive dataset (DAP Atlas) in terms of the class number. 
We divide the human body into eight regions and classify each class into them manually. Figure 11 (b) and (c) show the distribution of classes and annotations across different human body regions. 
We further show the distribution of some example classes in each region in Figure 11 (a). 
The extensive range of categories and regions lays the foundation for the \model{}'s wide application scenarios.

In the process of building \samdataset{}, 
we merge a wide range of segmentation tasks, 
and establish a unified label system by using natural language/text. Generally speaking, 
there are three advantages to doing this: 
(i) natural language is powerful and discriminative, 
which enables better differentiation of the medical terminologies in the language embedding space;  
(ii) as shown in previous work~\cite{wu2023medklip, zhang2023knowledge, lei2023unibrain, zheng2023large}, knowledge-enhanced representation learning for the text encoder demonstrates promising performance, allowing to learn the implicit or explicit relationships between these segmentation targets. For example, segmenting a specific lobe of the liver requires the exact segmentation of the liver as an organ in the abdominal cavity, and shall be facilitated by referring to other parts of the liver. Therefore, establishing such connections via systematic medical knowledge shall be beneficial. 
(iii) text prompts can be given automatically without any human intervention, for instance, from large language models. This would pave the way for building a segmentation model that can be flexibly integrated into foundation models for generalist medical artificial intelligence, as a powerful grounding tool.



\subsection{Large-Vocabulary Segmentation Prompted by Text}
\label{sec:problem_formulation}

Assuming we have a segmentation dataset collection, 
{\em i.e.}, $\mathcal{D} = \{(x_1, y_1; T_1), ..., (x_K, y_K; T_K)\}$, 
where $x_i \in \mathbb{R}^{H\times W\times D \times C}$ denotes the image scan, 
$y_i \in \mathbb{R}^{H\times W\times D\times M}$ is the binary segmentation annotations of the anatomical targets in the image 
and $T_i = \{t_1, t_2, ..., t_M\}$ denotes the corresponding medical terminology set, 
the segmentation task can be formulated as:
\begin{equation}
    \hat{y}_i = \Phi_{\text{SAT}}(\Phi_{\text{visual}}(x_i), \Phi_{\text{text}}(T_i)),
\end{equation}
where $\Phi_{\text{visual}}$ is a visual encoder,  $\Phi_{\text{text}}$ is a text encoder, 
$\Phi_{\text{SAT}}$ is a large-vocabulary segmentation foundation model. 
Ideally $x_i$ can be an image scan from any modality and anatomical region, and $T_i$ can contain an arbitrary number of text-based medical terminologies of interest.

To build such a model, we consider two main stages, namely, multimodal knowledge injection and segmentation training.
In the following, we firstly show how to structure multimodal medical knowledge and inject it into a text encoder~(Section~\ref{sec:knowledge_encoding}). 
Then, we employ the text encoder to guide our segmentation model training on \samdataset{} dataset~(Section~\ref{sec:segmentation}).
\textcolor{black}{
In addition, we provide more details about the model architecture and training strategies in the ``Technical Details'' Section in Supplementary.
}

\subsection{Multimodal Knowledge Injection}
\label{sec:knowledge_encoding}
Here, we aim to inject rich multimodal medical knowledge into the visual and text encoders. The section starts from the procedure for structuring the multimodal medical knowledge data and further presents details to use them for visual-language pre-training.

As shown in Figure 12 (a), 
the data from UMLS, search engine, 
and segmentation datasets can be aggregated into two formats:
\vspace{-0.2cm}
\begin{itemize}
\setlength\itemsep{0.15cm}
    \item \textbf{Textual Medical Concept Pair.} 
    For text-only knowledge, 
    each concept $t_i$ is associated with a definition $p_i$, 
    constructing pairs of text $(t_i; p_i)$. We also derive a knowledge graph that connects the medical concepts through abundant triplet relationships $(t_i, r_{ij}, t_j)$. This graph can be alternatively seen as a specialized text pair, $(t_i+r_{ij}; t_j)$ or $(t_i; r_{ij}+t_j)$, 
    where `$+$' refers to string concatenation. In this way, we can thus unify the two kinds of textual knowledge.
    
    \item \textbf{Visual Medical Concept Pair.} To align with the segmentation task, we gather pairs consisting of a concept (can be either an anatomical structure or lesion) and its image atlas. Note that, multiple pairs could be extracted from a single image. These pairs share a similar format to the segmentation data, denoted as $(x_i, y_i; t_i)$, where $x_i$ and $t_i$ are consistent with their definition in Section~\ref{sec:problem_formulation} and $y_i \in \mathbb{R}^{H\times W\times D\times 1}$ is a binary segmentation mask for $t_i$.
\end{itemize}

In summary, all the knowledge can either be represented as pure text description, {\em e.g.}, $t_i$, $p_i$, $t_i+r_{ij}$, $r_{ij}+t_j$,
or atlas segmentation $(x_i, y_i)$, and paired further. 

\begin{figure}[t]
    \centering
    \includegraphics[width = \textwidth]{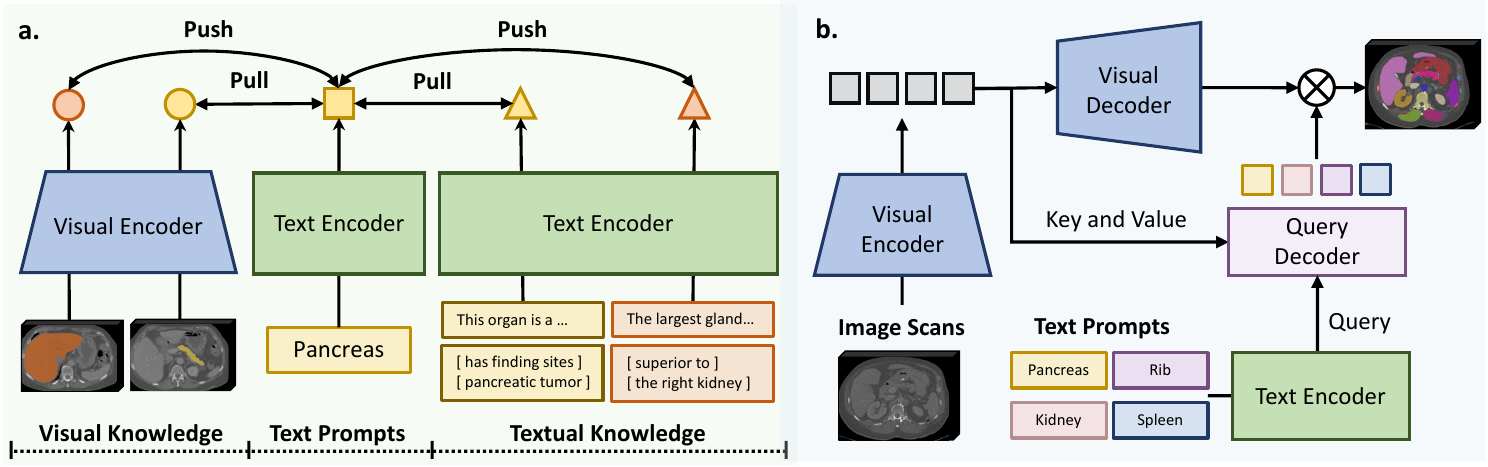}
    \vspace{1pt}
    \caption{\textbf{Overview of \model{}.} (a) We inject multimodal medical knowledge into knowledge encoders via contrastive learning. The knowledge is in different formats: atlas segmentation, concepts(terminologies), definitions, and relationships between concepts. We devise visual and text encoder to embed them; (b) We train a segmentation network based on the text prompts from the pre-trained text encoder. It is capable of segmenting a wide range of targets for image scans from different modalities and anatomical regions.
    }
    \label{fig:framework}
\end{figure}

As shown in Figure 12 (a), for pure text description, 
we encode them with a BERT~\cite{BERT} pre-trained on PubMed abstracts~\cite{PubmedBERT}:
\begin{equation}
    z = \Phi_{\text{text}}(\textbf{t}),\text{ } \textbf{t}\in[t_i, p_i, t_i+r_{ij}, r_{ij}+t_j], \text{ } z\in\mathbb{R}^{d},
\label{equ:text_enc}
\end{equation}
where $d$ refers to the feature dimension.
For visual concepts, we adopt the visual encoder $\Phi_{\text{visual}}$. Given the excellent robustness and performance of U-Net~\cite{nnUNET,UNet}, we apply a standard 3D U-Net encoder to extract multi-scale image embeddings:
\begin{equation}
    \label{equ:vis_enc}
    V_i = \{v_{i1}, v_{i2}, ..., v_{iS}\} = \Phi_{\text{visual}}(x_i), \text{ } v_{is} \in \mathbb{R}^{H_s \times W_s \times D_s \times d_s},
\end{equation}
where $V_i$ is the multi-scale feature maps from U-Net encoder layers, 
and $H_s, W_s, D_s, d_s$ are the spatial resolutions and channel width at different layers. 
We further average ROI pooling on these feature maps respectively, based on the down-sampled segmentation mask fitting resolutions at different layers. The concatenated pooled features can thus be treated as a representation of the anatomical target on this image, containing multi-scale visual clues for it.
\begin{equation}
    \label{equ:pool}
    z = \mathcal{F}_{\text{pooling}}(\Phi_{\text{visual}}(x_i); y_i), 
    \text{ } z\in\mathbb{R}^{d}.
\end{equation}

We train the visual encoder and text encoder by maximizing the similarities between all positive knowledge pairs, linked by text prompts~(medical terminology names), as shown in Figure 12 (a). Specifically, given $(x_i, y_i; t_i), (t_i; p_i), (t_i+r_{ij}; t_j), (t_i; r_{ij}+t_j)$, for simplicity, we denote all the encoded features as $z$, regardless of their modality format. 
For a batch of $N$ pairs $\{(z_1, z_1'), ... (z_{N}, z_{N}')\}$, we have:
\begin{equation}
    \mathcal{L}_{\text{knowledge}} = -\frac{1}{N} \sum^{N}_{i=1} (\log \frac{\exp(z_i \cdot z_i'/\tau)}{\sum^{N}_{k=1} \mathbbm{1}_{i \neq k} \exp(z_i \cdot z_k'/\tau)} + \log \frac{\exp(z_i \cdot z_i'/\tau)}{\sum^{N}_{k=1} \mathbbm{1}_{i \neq k} \exp(z_k \cdot z_i'/\tau)}),
\end{equation}
with $\tau = 0.07$ as temperature.

On formulation, this procedure resembles a typical contrastive learning pipeline~\cite{radford2021learning,lin2023pmc,zhang2023biomedclip}. 
However, different from previous work that directly contrasts the paired visual-language data, we aim for knowledge-enhanced representation learning. 
By maximizing the similarities between the constructed positive textual and visual feature pairs, we force the text encoders to construct neural representations for medical concepts based on domain knowledge from two aspects: 
(i) through the well-established knowledge graph in text form, the text encoder enables encoding relationships between concepts in the latent space; (ii) the model captures the characteristics of anatomical structures and lesions via both visual atlas segmentations and detailed definitions. Therefore, in contrast to the one-hot labeling that treats each anatomical target as being independent, such continuous neural representation shall provide more helpful guidance for the segmentation task.

\subsection{Segmentation Training}

\label{sec:segmentation}

With the pre-trained visual and text encoder, 
we now continue the procedure for building the segmentation model with text as prompts. Figure 12 (b) demonstrates the overall framework. 
Specifically, apart from the pre-trained visual and text encoders, the segmentation model consists of three more components: a visual decoder $\Phi_{\text{dec}}$, a query decoder $\Phi_{\text{query}}$, and a mask generator.
Although a sample in the segmentation dataset collection $(x_i, y_i; T_i)$ may contain multiple annotations, {\em i.e.}, $T_i = \{t_1, t_2, ..., t_M\}$, for simplicity, we first describe the segmentation procedure for one target~$t_i$ in the following.

Given an anatomical terminology $t_i$, we employ the pre-trained text encoder to generate its neural embedding, which serves as the text prompt for segmentation:
\begin{equation}
    z_i = \Phi_{\text{text}}(t_i), \text{ } z_i\in\mathbb{R}^{d}.
    \label{eq:terminology_encoding}
\end{equation}
Note that, after pre-training the text encoder with domain knowledge injection, $z_i$ should contain both the textual background information and visual information from atlas samples.

For image scan $y_i$, we first adopt the pre-trained visual encoder to derive the multi-scale image embeddings $V_i$, as explained in Equa.~\ref{equ:vis_enc}, and continue training it.
Then, in the visual decoder, the feature maps from the encoder are gradually upsampled with skip connections, effectively following the U-Net architecture~\cite{nnUNET, UNet}, ending up with per-pixel dense features:
\begin{equation}
    u_i = \Phi_{\text{dec}}(V_i), \text{ } u_i \in \mathbb{R}^{H \times W \times D \times d'},
\end{equation}
where $d'$ is the dimension for the per-pixel dense feature after recovering to the original resolution.

Although a general representation of the anatomical target is derived from the pre-trained text encoder with a text prompt, 
visual variations may still exist from patient to patient, 
we thus insert a transformer-based query decoder to further enhance the text prompts with visual clues. 
In practice, it consists of 6 standard transformer decoders~\cite{Attention}, that treat text embedding as query, 
and the pooled multi-scale visual features from the U-Net encoder as key, values, formulated as:
\begin{equation}
    q_i = \Phi_{\text{query}}(V_i, z_i), \text{ } q_i\in\mathbb{R}^{d}.
    \label{eq:query_decoding}
\end{equation}
Where $z_i$ is consistent with $z$ in Equa.~\ref{equ:pool}. Therefore $q_i$ can be seen as an adapted representation of the anatomical target in a specific image scan $x_i$. 

Finally, by conducting a pixel-wise dot product between the representation of the anatomical target and the fine-grained per-pixel embedding, we can acquire a per-pixel prediction:
\begin{equation}
    \hat{y}_i = \sigma(g(q_i) \cdot u_i), \text{ }  \hat{y}_i \in \mathbb{R}^{H \times W \times D},
    \label{eq:mask_generating}
\end{equation}
where $g(\cdot)$ is a feed-forward layer projecting $q_i$ to a consistent dimension with the dense feature map $u_i$, and $\sigma(\cdot)$ denotes the sigmoid function.
Note that, the whole forward procedure does not involve any operation between different text prompts. Therefore, for input with multiple text prompts or segmentation targets, {\em i.e.}, $T_i = \{t_1, t_2, ..., t_M\}$, the processes described in Equation~\ref{eq:terminology_encoding}, \ref{eq:query_decoding} and \ref{eq:mask_generating} will be executed for each target in parallel, and we could derive $\hat{y}_i \in \mathbb{R}^{H \times W \times D \times M}$.

Following~\cite{nnUNET}, we adopt a loss function as the sum of binary cross-entropy loss and dice loss. For a sample with $M$ classes and $C$ voxels, we denote $p_{\text{c},\text{m}}$ and $s_{\text{c,m}}$ as the prediction and ground truth for $c$-th pixel respectively on class $m$, the loss is: 
\begin{equation}
\mathcal{L} = \underbrace{-\frac{1}{M} \sum_{\text{m=1}}^{M} \frac{1}{C}\sum_{\text{c=1}}^{C} p_{\text{c,m}} \cdot \log s_{\text{c,m}}}_{\text{Binary Cross Entropy Loss}} + (\underbrace{1-\frac{2 \sum_{\text{i=1}}^{M} \sum_{\text{c=1}}^{C} p_{\text{c,m}} \cdot s_{\text{c,m}}}{\sum_{\text{m=1}}^{M} \sum_{\text{c=1}}^{C} p_{\text{c,m}}^{2}+ \sum_{\text{m=1}}^{M} \sum_{\text{c=1}}^{C} s_{\text{c,m}}^{2}}}_{\text{Dice Loss}})
\end{equation}

\subsection{Evaluation Datasets}
\label{sec:evaluation_datasets}
To strike a balance between extensive experiments and computational costs, we utilize two collections of datasets in evaluation:
\vspace{-0.4em}
\begin{itemize}
\setlength\itemsep{0.4em}
    \item \textbf{\samdataset{}}. As describe in Section~\ref{sec:dataset4segmentation}, this contains all the 72 datasets, 497 classes from all human body regions, 22,186 image scans and 302,033 segmentation annotations. 
    \item \textbf{\samdataset{}-Nano}. A subset of \samdataset{}, including only 49 datasets, 13,303 images and 151,461 annotations. Note that \samdataset{}-Nano also covers 429 classes from all human body regions, adequate to evaluate the large-vocabulary segmentation task. 
\end{itemize}
The detailed composition of \samdataset{} and \samdataset{}-Nano can be found in Supplementary Table 32 and 33.
As there is no existing benchmark for evaluating the large-vocabulary segmentation foundation model, we randomly split each dataset into 80\% for training and 20\% for testing:
(i) datasets may share the same images but with different classes. 
For example, Couinaud Liver provides fine-grained liver segmentation on a subset of MSD Hepatic Vessel. We carefully split the Couinaud Liver to make sure the test set will not be leaked in the train set of MSD Hepatic Vessel; 
(ii) scans of the same patient but different modalities are treated as different samples during training and evaluation. For example, MSD Prostate contains T2 and ADC scans of each patient. However, they share the same structure on the image. To avoid potential data leaking, we carefully split such datasets by patient id. Note that when involving segmentation datasets in the visual-language pretraining, we only use the training data to avoid potential data leaking. For datasets involved in \samdataset{}-Nano, we keep their splits the same as in \samdataset{}.
The download link for each dataset can be found in Section~\ref{sec:data_availability}, and we have released our dataset processing code and train-test splits to the research community for reproduction and benchmarking.

\subsection{Baselines}
\label{sec:baselines}
We take nnU-Net~\cite{nnUNET}, U-Mamba~\cite{ma2024u} and SwinUNETR~\cite{SwinUNetr} as representative types of \textbf{specialist model} and strong baselines for comparison.
For a comprehensive evaluation, we train one specialist model on each of the datasets. 
Note that, following~\cite{Totalsegmentator}, 
we split Totalsegmentator into 6 subsets and treat them as different datasets. 
Similarly, datasets such as CHAOS with both MRI and CT images are treated as two different datasets. 
When training specialist models on each dataset, we adopt a multi-class segmentation setting and deliver the masks of all categories in this dataset at once. 
We derive the optimal network architecture and pre-processing pipeline with the default setting of each specialist model. 
We present the detailed network design of nnU-Nets in Supplementary Table 26 and Table 27 for a straightforward comparison.
In summary, we train an ensemble of 72 models for each type of specialist model, 
that are customized on each dataset. 
We adopt the latest official implementation of nnU-Net v2 and U-Mamba in practice. 
The SwinUNETR is adopted to the same auto-configuration framework as U-Mamba.

We take MedSAM~\cite{MedSAM} as a representative \textbf{interactive segmentation model} and competitive baseline. 
MedSAM finetunes SAM~\cite{SAM} on 86 segmentation datasets, and supports 2D medical image segmentation with box prompts. 
We follow the official implementation to process and infer image slice by slice, and calculate the metrics on the finally stacked 3D prediction. For each single target on a slice, to simulate box prompts towards it, we both take the minimum rectangle containing the ground truth segmentation (Tight), and follow the official data augmentation procedure, randomly shift each corner up to 8\% of the whole image resolution (Loose). In addition, we consider directly using the tight box prompts as predictions (Oracle Box).

We take BiomedParse~\cite{zhao2024foundation}, a concurrent \textbf{text-prompted segmentation model} for 2D biomedical images, as a baseline.
We follow the official implementation for data processing, inference, and post-filtering. 
Similar to MedSAM, we process and infer image slice by slice, and calculate the metrics on the finally stacked 3D prediction. 
As BiomedParse may fail to detect the target on the slice, we evaluate it under two settings: only prompt target present in the current slice (Oracle) and prompt all the targets available in the dataset and post-filter out potential false positive predictions by p-values.

\subsection{Evaluation Protocols}
\label{sec:experiment_settings}
Given our goal is to develop a large-vocabulary medical segmentation foundation model, 
this provides opportunities to evaluate novel perspectives in addition to the traditional evaluation per dataset. Specifically, we conduct the internal evaluations from three dimensions:
\begin{itemize}
\setlength\itemsep{0.15cm}    
    \item \textbf{Class-wise Evaluation.} 
    As \model{} is capable of segmenting a wide range of anatomical targets across the human body, we merge the results from the same classes across datasets to indicate the performance on each anatomical target. Specifically, we follow macro-average method: for a class annotated in multiple datasets, we first calculate its average scores within each dataset, and then average them over all datasets. Note that, the same anatomical structures or lesions from different modalities are treated as different classes in this work, {\em e.g.,} liver in both CT and MRI images.
    
    \item \textbf{Region-wise Evaluation.} 
    In general, anatomical structures from the same human body region are closely connected and more likely to be involved in diagnosis within the same hospital department. Here, we consider the region-wise evaluation: based on class-wise evaluation, we merge results from all classes in the same body region, 
    as to indicate the general performance in this region. For classes existing in multiple regions, we classify them into `Whole Body' category.
    In addition, we report results for lesions classes independently as a category `lesion', instead of merging them into specific regions.

    \item \textbf{Dataset-wise Evaluation.} 
    Results of the classes within the same dataset are averaged to indicate the performance on this dataset. 
    This is the same as the conventional evaluation protocol of specialist segmentation models trained on a single dataset.
\end{itemize}

\subsection{Evaluation Metrics}
\label{sec:metrics}
We quantitatively evaluate the segmentation performance from the perspective of region and boundary metrics~\cite{maier2022metrics}, \emph{e.g.}, Dice Similarity Coefficient~(DSC) and Normalized Surface Distance~(NSD) respectively.

Dice Similarity Coefficient~(DSC) is a standard region-based metric for medical image segmentation evaluation. It measures the overlap between the model's prediction $P$ and ground truth $G$, formally defined as:
\begin{equation}
    DSC(P,G) = \frac{2|P \bigcap G|}{|P|+|G|}.
\end{equation}

\vspace{-6pt}
Normalized Surface Distance~(NSD)~\cite{nikolov2021clinically} is a boundary-based metric that measures the consistency at the boundary area of the model's prediction $P$ and ground truth $G$, which is defined as:
\begin{equation}
    NSD(P,G) = \frac{|\partial P \bigcap B_{\partial G}|+|\partial G \bigcap B_{\partial P}|}{|\partial P|+|\partial G|},
\end{equation}
where $B_{\partial P}=\{x\in\mathbf{R}^3|\exists \hat{x}\in \partial P, ||x-\hat{x}|| \leq \tau\}$ and $B_{\partial G}=\{x\in\mathbf{R}^3|\exists \hat{x}\in \partial G, ||x-\hat{x}|| \leq \tau\}$ are the boundary areas of the model's prediction and ground truth at a tolerance $\tau$, respectively. We set $\tau$ as 1 in the experiments.

\section{Code Availability}
The code is available at \href{https://github.com/zhaoziheng/SAT}{https://github.com/zhaoziheng/SAT}.

\section{Data Availability of \samdataset{}}
\label{sec:data_availability}
The access to each dataset can be found in Table~\ref{tab:dataset_links} and Table~\ref{tab:dataset_links2}. The data process code to build \samdataset{} and our train-test splits for reproducibility and benchmarking are available at \href{https://github.com/zhaoziheng/SAT-DS}{https://github.com/zhaoziheng/SAT-DS}. 
\begin{table*}[htpb]
\center
\caption{Download links of the 72 datasets in \samdataset{}.}
\begin{tabular}{ll}
\toprule
\rowcolor{lightgray} Dataset & Download Link \\
\midrule
AbdomenCT1K~\cite{AbdomenCT1K} & \href{https://github.com/JunMa11/AbdomenCT-1K}{https://github.com/JunMa11/AbdomenCT-1K} \\\hline
ACDC~\cite{ACDC} & \href{https://humanheart-project.creatis.insa-lyon.fr/database/}{https://humanheart-project.creatis.insa-lyon.fr/database/} \\\hline
AMOS CT~\cite{AMOS22} & \href{https://zenodo.org/records/7262581}{https://zenodo.org/records/7262581} \\\hline
AMOS MRI~\cite{AMOS22} & \href{https://zenodo.org/records/7262581}{https://zenodo.org/records/7262581} \\\hline
ATLASR2~\cite{ATLASR2} & \href{http://fcon_1000.projects.nitrc.org/indi/retro/atlas.html}{http://fcon\_1000.projects.nitrc.org/indi/retro/atlas.html} \\\hline
ATLAS~\cite{ATLAS} & \href{https://atlas-challenge.u-bourgogne.fr}{https://atlas-challenge.u-bourgogne.fr} \\\hline
autoPET~\cite{autoPET} & \href{https://wiki.cancerimagingarchive.net/pages/viewpage.action?pageId=93258287}{https://wiki.cancerimagingarchive.net/pages/viewpage.action?pageId=93258287} \\\hline
Brain Atlas~\cite{Brain_Atlas} & \href{http://brain-development.org/}{http://brain-development.org/} \\\hline
BrainPTM~\cite{BrainPTM} & \href{https://brainptm-2021.grand-challenge.org/}{https://brainptm-2021.grand-challenge.org/} \\\hline
BraTS2023 GLI~\cite{BraTS2023GLI} & \href{https://www.synapse.org/#!Synapse:syn51514105}{https://www.synapse.org/\#!Synapse:syn51514105} \\\hline
BraTS2023 MEN~\cite{BraTS2023MEN} & \href{https://www.synapse.org/#!Synapse:syn51514106}{https://www.synapse.org/\#!Synapse:syn51514106} \\\hline
BraTS2023 MET~\cite{BraTS2023MET} & \href{https://www.synapse.org/#!Synapse:syn51514107}{https://www.synapse.org/\#!Synapse:syn51514107} \\\hline
BraTS2023 PED~\cite{BraTS2023PED} & \href{https://www.synapse.org/#!Synapse:syn51514108}{https://www.synapse.org/\#!Synapse:syn51514108} \\\hline
BraTS2023 SSA~\cite{BraTS2023SSA} & \href{https://www.synapse.org/#!Synapse:syn51514109}{https://www.synapse.org/\#!Synapse:syn51514109} \\\hline
BTCV Abdomen~\cite{BTCV} & \href{https://www.synapse.org/#!Synapse:syn3193805/wiki/217789}{https://www.synapse.org/\#!Synapse:syn3193805/wiki/217789} \\\hline
BTCV Cervix~\cite{BTCV} & \href{https://www.synapse.org/#!Synapse:syn3193805/wiki/217790}{https://www.synapse.org/\#!Synapse:syn3193805/wiki/217790} \\\hline
CHAOS CT~\cite{CHAOS} & \href{https://chaos.grand-challenge.org/}{https://chaos.grand-challenge.org/} \\\hline
CHAOS MRI~\cite{CHAOS} & \href{https://chaos.grand-challenge.org/}{https://chaos.grand-challenge.org/} \\\hline
CMRxMotion~\cite{CMRxMotion} & \href{https://www.synapse.org/#!Synapse:syn28503327/files/}{https://www.synapse.org/\#!Synapse:syn28503327/files/} \\\hline
Couinaud~\cite{Couinaud} & \href{https://github.com/GLCUnet/dataset}{https://github.com/GLCUnet/dataset} \\\hline
COVID-19 CT Seg~\cite{COVID19} & \href{https://github.com/JunMa11/COVID-19-CT-Seg-Benchmark}{https://github.com/JunMa11/COVID-19-CT-Seg-Benchmark} \\\hline
CrossMoDA2021~\cite{CrossMoDA2021} & \href{https://crossmoda.grand-challenge.org/Data/}{https://crossmoda.grand-challenge.org/Data/} \\\hline
CT-ORG~\cite{CTORG} & \href{https://wiki.cancerimagingarchive.net/pages/viewpage.action?pageId=61080890}{https://wiki.cancerimagingarchive.net/pages/viewpage.action?pageId=61080890} \\\hline
CTPelvic1K~\cite{CTPelvic1K} & \href{https://zenodo.org/record/4588403#.YEyLq_0zaCo}{https://zenodo.org/record/4588403\#\.YEyLq\_0zaCo} \\\hline
DAP Atlas~\cite{DAPAtlas} & \href{https://github.com/alexanderjaus/AtlasDataset}{https://github.com/alexanderjaus/AtlasDataset} \\\hline
FeTA2022~\cite{FeTA2022} & \href{https://feta.grand-challenge.org/data-download/}{https://feta.grand-challenge.org/data-download/} \\\hline
FLARE22~\cite{FLARE22} & \href{https://flare22.grand-challenge.org/}{https://flare22.grand-challenge.org/} \\\hline
FUMPE~\cite{FUMPE} & \href{https://www.kaggle.com/datasets/andrewmvd/pulmonary-embolism-in-ct-images}{https://www.kaggle.com/datasets/andrewmvd/pulmonary-embolism-in-ct-images} \\\hline
HAN Seg~\cite{HANSeg} & \href{https://zenodo.org/record/}{https://zenodo.org/record/} \\\hline
HECKTOR2022~\cite{HECTOR2022} & \href{https://hecktor.grand-challenge.org/Data/}{https://hecktor.grand-challenge.org/Data/} \\\hline
INSTANCE~\cite{INSTANCE} & \href{https://instance.grand-challenge.org/Dataset/}{https://instance.grand-challenge.org/Dataset/} \\\hline
ISLES2022~\cite{ISLES2022} & \href{http://www.isles-challenge.org/}{http://www.isles-challenge.org/} \\\hline
KiPA22~\cite{KiPA22} & \href{https://kipa22.grand-challenge.org/dataset/}{https://kipa22.grand-challenge.org/dataset/} \\\hline
KiTS23~\cite{KiTS23} & \href{https://github.com/neheller/kits23}{https://github.com/neheller/kits23} \\\hline
LAScarQS2022 Task 1~\cite{LAScarQS2022} & \href{https://zmiclab.github.io/projects/lascarqs22/data.html}{https://zmiclab.github.io/projects/lascarqs22/data.html} \\\hline
LAScarQS2022 Task 2~\cite{LAScarQS2022} & \href{https://zmiclab.github.io/projects/lascarqs22/data.html}{https://zmiclab.github.io/projects/lascarqs22/data.html} \\\hline
LNDb~\cite{LNDb} & \href{https://zenodo.org/record/7153205#.Yz_oVHbMJPZ}{https://zenodo.org/record/7153205\#\.Yz\_oVHbMJPZ} \\\hline
LUNA16~\cite{LUNA16} & \href{https://luna16.grand-challenge.org/}{https://luna16.grand-challenge.org/} \\\hline
MM-WHS CT~\cite{MMWHS} & \href{https://mega.nz/folder/UNMF2YYI#1cqJVzo4p_wESv9P_pc8uA}{https://mega.nz/folder/UNMF2YYI\#1cqJVzo4p\_wESv9P\_pc8uA} \\\hline
MM-WHS MR~\cite{MMWHS} & \href{https://mega.nz/folder/UNMF2YYI#1cqJVzo4p_wESv9P_pc8uA}{https://mega.nz/folder/UNMF2YYI\#1cqJVzo4p\_wESv9P\_pc8uA} \\\hline
MRSpineSeg~\cite{MRSpineSeg} & \href{https://www.cg.informatik.uni-siegen.de/en/spine-segmentation-and-analysis}{https://www.cg.informatik.uni-siegen.de/en/spine-segmentation-and-analysis} \\\hline
MSD Cardiac~\cite{MSD} & \href{http://medicaldecathlon.com/}{http://medicaldecathlon.com/} \\\hline
MSD Colon~\cite{MSD} & \href{http://medicaldecathlon.com/}{http://medicaldecathlon.com/} \\\hline
MSD HepaticVessel~\cite{MSD} & \href{http://medicaldecathlon.com/}{http://medicaldecathlon.com/} \\\hline
MSD Hippocampus~\cite{MSD} & \href{http://medicaldecathlon.com/}{http://medicaldecathlon.com/} \\\bottomrule
\end{tabular}
\label{tab:dataset_links}
\end{table*}

\begin{table*}[t]
\center
\small
\caption{(Continued) Download links of the 72 datasets in \samdataset{}.}
\begin{tabular}{ll}
\toprule
\rowcolor{lightgray} Dataset & Download Link \\
\midrule
MSD Liver~\cite{MSD} & \href{http://medicaldecathlon.com/}{http://medicaldecathlon.com/} \\\hline
MSD Lung~\cite{MSD} & \href{http://medicaldecathlon.com/}{http://medicaldecathlon.com/} \\\hline
MSD Pancreas~\cite{MSD} & \href{http://medicaldecathlon.com/}{http://medicaldecathlon.com/} \\\hline
MSD Prostate~\cite{MSD} & \href{http://medicaldecathlon.com/}{http://medicaldecathlon.com/} \\\hline
MSD Spleen~\cite{MSD} & \href{http://medicaldecathlon.com/}{http://medicaldecathlon.com/} \\\hline
MyoPS2020~\cite{MyoPS2020} & \href{https://mega.nz/folder/BRdnDISQ#FnCg9ykPlTWYe5hrRZxi-w}{https://mega.nz/folder/BRdnDISQ\#FnCg9ykPlTWYe5hrRZxi-w} \\\hline
NSCLC~\cite{NSCLC} & \href{https://wiki.cancerimagingarchive.net/pages/viewpage.action?pageId=68551327}{https://wiki.cancerimagingarchive.net/pages/viewpage.action?pageId=68551327} \\\hline
Pancreas CT~\cite{PancreasCT} & \href{https://wiki.cancerimagingarchive.net/display/public/pancreas-ct}{https://wiki.cancerimagingarchive.net/display/public/pancreas-ct} \\\hline
Parse2022~\cite{PARSE2022} & \href{https://parse2022.grand-challenge.org/Dataset/}{https://parse2022.grand-challenge.org/Dataset/} \\\hline
PDDCA~\cite{PDDCA} & \href{https://www.imagenglab.com/newsite/pddca/}{https://www.imagenglab.com/newsite/pddca/} \\\hline
PROMISE12~\cite{PROMISE12} & \href{https://promise12.grand-challenge.org/Details/}{https://promise12.grand-challenge.org/Details/} \\\hline
SEGA~\cite{SEGA} & \href{https://multicenteraorta.grand-challenge.org/data/}{https://multicenteraorta.grand-challenge.org/data/} \\\hline
SegRap2023 Task1~\cite{SegRap2023} & \href{https://segrap2023.grand-challenge.org/}{https://segrap2023.grand-challenge.org/} \\\hline
SegRap2023 Task2~\cite{SegRap2023} & \href{https://segrap2023.grand-challenge.org/}{https://segrap2023.grand-challenge.org/} \\\hline
SegTHOR~\cite{SegTHOR} & \href{https://competitions.codalab.org/competitions/21145#learn_the_details}{https://competitions.codalab.org/competitions/21145\#learn\_the\_details} \\\hline
SKI10~\cite{SKI10} & \href{https://ambellan.de/sharing/QjrntLwah}{https://ambellan.de/sharing/QjrntLwah} \\\hline
SLIVER07~\cite{SLIVER07} & \href{https://sliver07.grand-challenge.org/}{https://sliver07.grand-challenge.org/} \\\hline
ToothFairy~\cite{ToothFairy} & \href{https://ditto.ing.unimore.it/toothfairy/}{https://ditto.ing.unimore.it/toothfairy/} \\\hline
TotalSegmentator Cardiac~\cite{Totalsegmentator} & \href{https://zenodo.org/record/6802614}{https://zenodo.org/record/6802614} \\\hline
TotalSegmentator Muscles~\cite{Totalsegmentator} & \href{https://zenodo.org/record/6802614}{https://zenodo.org/record/6802614} \\\hline
TotalSegmentator Organs~\cite{Totalsegmentator} & \href{https://zenodo.org/record/6802614}{https://zenodo.org/record/6802614} \\\hline
TotalSegmentator Ribs~\cite{Totalsegmentator} & \href{https://zenodo.org/record/6802614}{https://zenodo.org/record/6802614} \\\hline
TotalSegmentator Vertebrae~\cite{Totalsegmentator} & \href{https://zenodo.org/record/6802614}{https://zenodo.org/record/6802614} \\\hline
TotalSegmentator V2~\cite{Totalsegmentator} & \href{https://zenodo.org/record/6802614}{https://zenodo.org/record/6802614} \\\hline
VerSe~\cite{VerSe} & \href{https://github.com/anjany/verse}{https://github.com/anjany/verse} \\\hline
WMH~\cite{WMH} & \href{https://wmh.isi.uu.nl/}{https://wmh.isi.uu.nl/} \\\hline
WORD~\cite{WORD} & \href{https://github.com/HiLab-git/WORD}{https://github.com/HiLab-git/WORD} \\
\bottomrule
\end{tabular}
\label{tab:dataset_links2}
\end{table*}

\section{Acknowledgments}

This work is supported by Science and Technology Commission of Shanghai Municipality (No. 22511106101, No. 18DZ2270700, No. 21DZ1100100), 111 plan (No. BP0719010), State Key Laboratory of UHD Video and Audio Production and Presentation, National Key R\&D Program of China (No. 2022ZD0160702).

\section{Author Contributions}

All authors make contributions to the conception or design of the work. 
Specifically, Z.Z. contributed to the technical implementation.
Z.Z. and Y.Z. (Yao) contributed to data collection and processing.
Z.Z., Y.Z. (Yao) and X.Z. (Xiao) contributed to the baseline implementation. 
All authors contributed to the drafting and revising of the manuscript.

\section{Competing Interests}

The authors declare no competing interests.

\clearpage

\bibliographystyle{sn-mathphys} 
\bibliography{references} 
\clearpage

\appendix
\setcounter{figure}{0}
\setcounter{table}{0}

\section{Technique Details}
\label{sec:technique_details}

\subsection{Model Architecture}

{
\color{black}

We provided the detailed architecture of the vision encoder, 
vision decoder, text encoder, and query decoder in Supplementary Figure~\ref{fig:Detailed_architecture}.
Specifically, the vision encoder and decoder follow a 6-layer U-Net architecture; the text encoder is a 12-layer BERT model; and the query decoder consists of 6 transformer decoder layers.

\begin{figure}[htbp]
    \centering
    \includegraphics[width = 0.9\textwidth]{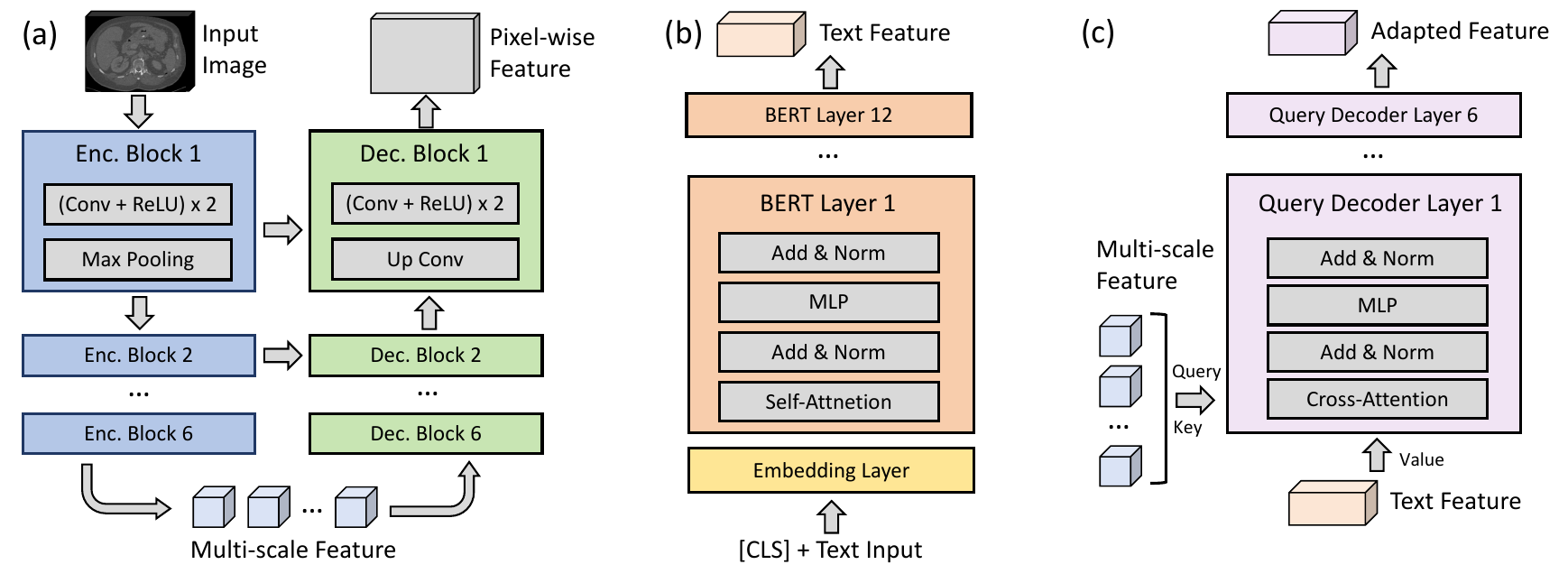}
    \vspace{10pt}
    \caption{
    \textcolor{black}{
    \textbf{Architecture details of SAT.} (a) Vision encoder and decoder follow a 6-layer U-Net architecture; (b) Text encoder is a 12-layer BERT model; (c) Query decoder consists of 6 transformer decoder layers.
    }
    }
    \label{fig:Detailed_architecture}
\end{figure}

Based on Supplementary Figure~\ref{fig:arch_w_formula}, we present a more detailed illustration of how \model{} generates segmentation predictions based on text prompts. The whole procedure can be divided into three parts:
\vspace{-0.4em}
\begin{itemize}
    \setlength\itemsep{0.4em}
    \item \textbf{On the visual backbone side}, given a 3D volume input, 
    we adopt the vision encoder to derive the multi-scale features $V_i = \{v_{i1},v_{i2},...,v_{iS}\}$, where $v_{is}\in \mathbb{R}^{H_s\times W_s\times D_s\times d}$ is from the $s$-th encoder layer. 
    We then derive the pixel-wise dense feature $u_i\in \mathbb{R}^{H\times W\times D\times d'}$ from the vision decoder. This corresponds to the upper pathway in Supplementary Figure~\ref{fig:arch_w_formula}. 
    
    \item \textbf{On the text prompt side}, given an arbitrary number of medical terminologies, $T_i = \{t_1, t_2, ..., t_M\}$, as text prompts, we \textbf{first} derive text embeddings $z_m$ for each term from the knowledge-enhanced text encoder. 
    \begin{equation}
        z_m = \Phi_{\text{text}}(t_m), \text{ } z_m\in\mathbb{R}^{d}.
    \end{equation}
    \textbf{Then}, the query decoder enables the text embedding to iteratively attend to the image and update its embeddings, {\em i.e.}, $q_m$. 
    This corresponds to the lower pathway in Supplementary Figure~\ref{fig:arch_w_formula}. 
    \begin{equation}
        q_m = \Phi_{\text{query}}(V_i, z_m), \text{ } q_m\in\mathbb{R}^{d}.
    \end{equation}
    
    \item \textbf{To generate the final prediction}, we \textbf{first} use a feed-forward layer $g(\cdot)$ to project each text embedding $q_m$ to dimension $d'$, aligned with the pixel-wise dense feature. \textbf{Then}, we compute the dot product between each projected text embedding and the pixel-wise image feature to get the predicted one-channel heatmap, {\em i.e.}, the score that each pixel belongs to this anatomical structure:
    \begin{equation}
    \hat{y}_i = \sigma(g(q_i) \cdot u_i), \text{ }  \hat{y}_i \in \mathbb{R}^{H \times W \times D},
    \end{equation}
    where $\sigma(\cdot)$ denotes the sigmoid function.
\end{itemize}

\begin{figure}[htbp]
    \centering
    \includegraphics[width = 0.9\textwidth]{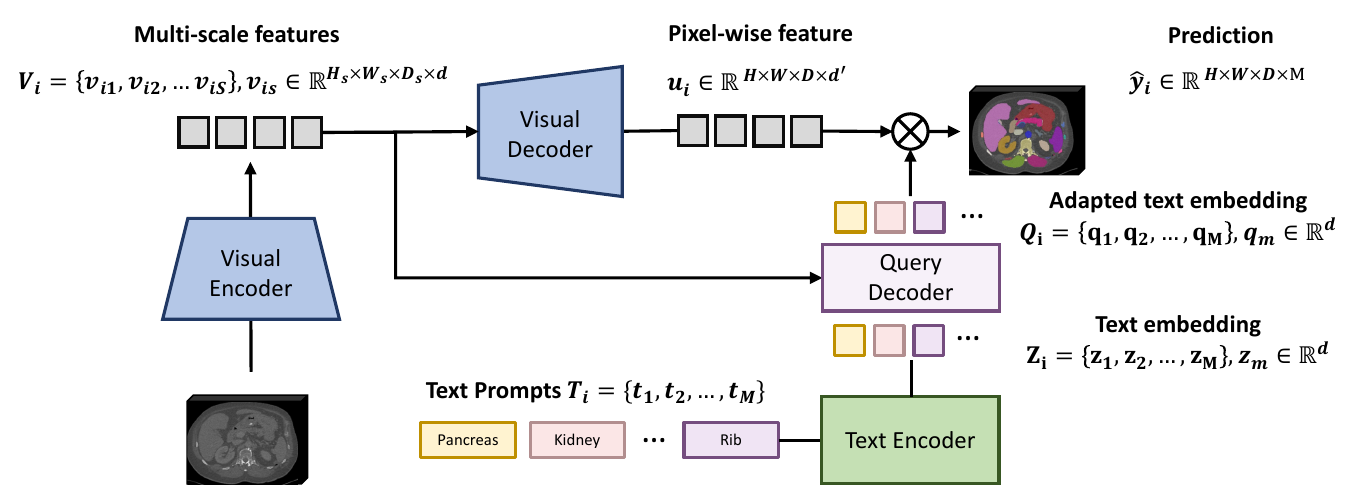}
    \vspace{10pt}
    \caption{
    \textcolor{black}{
    \textbf{Workflow details of SAT.} \model{} take 3D radiology images as input and can be prompted by an arbitrary number of terminologies in text form. Binary segmentation prediction is generated for each prompt. Key variables and their dimensional information are annotated on the figure.}
    }
    \label{fig:arch_w_formula}
\end{figure}
}

\subsection{Training Strategies}
\label{sec:training strategies}

We encounter several challenges in training on the combined large number of heterogeneous medical datasets in 3D format. In this section, we provide details on the adopted training strategies.

\noindent \textbf{Progressive Knowledge Injection.} 
For the stability of training, we implement the multimodal knowledge injection procedure introduced in Section 4.4 progressively. 
At the beginning, the text encoder is pretrained on the text knowledge via contrastive learning, {\em i.e.,} the textual medical concept pairs. The maximal sequence length after tokenization is 256, for even longer text input, we apply random truncation to fully exploit the knowledge in the long text. After convergence, to align text representations and visual features, we apply contrastive learning between the finetuned text encoder and the visual encoder on the visual medical concept pair.

\noindent \textbf{Pre-processing Segmentation Dataset.} 
We take voxel spacing as $1 \times 1 \times 3\ mm^{2}$ and patch size $288 \times 288 \times 96$, based on two empirical considerations: (i) when mixing the various datasets, scans with a wide range of voxel spacings ought to be normalized before processing in the same convolutional network. While resampling to larger voxel spacing may lose information, smaller voxel spacing will generate artifacts; (ii) a larger receptive field generally ensures better segmentation performance for most targets; however, increasing patch size will result in higher computational cost. We strike a balance between them based on the computational resources in use. 

\noindent \textbf{Balancing Segmentation Datasets.} 
To balance between all datasets, 
we set the sampling strategy for each scan based on two intuitions respectively: 
(i) training case number varies significantly from dataset to dataset, which should be alleviated. We follow ~\cite{ulrich2023multitalent} and set the sampling weight of all scans in a dataset as the inverse proportion to $\sqrt{N}$, $N$ is the number of training cases in the dataset; 
(ii) scans with larger annotation areas or more annotated classes should be sampled more as they are often harder to learn. 
Thus, we repeat such scans for $R$ times in the sampling pool, where $R=\frac{S_{roi}}{288 \times 288 \times 96} \times \frac{M}{32}$. $S_{roi}$ is the size of the annotated area in an image scan, namely, its foreground area; $288 \times 288 \times 96$ is the patch size to crop; $c$ is the number of annotated classes on it; and 32 is the maximal number of text prompts in a batch.

\noindent \textbf{Balancing Segmentation Classes.} 
While cropping the image scan, it's a common practice~\cite{nnUNET} to over-sample foreground crops, {\em i.e.}, crops containing at least one segmentation target. However, weighting these crops evenly may ignore the unbalanced spatial distribution of segmentation targets. For example, in large scans with numerous annotations, tiny targets are harder to sample and thus may be ignored by the model. Thus, in foreground oversampling, we give more weight to regions with more segmentation targets.

\vspace{3pt}\noindent \textbf{Difference between \model{}-Pro and \model{}-Nano.} 
We devise two variants of \model{} with different model sizes in the visual backbone. For \model{}-Nano, we adopt a U-Net with 6 blocks in depth, each block has 2 convolutional layers and $3 \times 3$ size each kernel. The channel widths for each stage are $[64, 64, 128, 256, 512, 768]$; \model{}-Pro shares the same architecture with \model{}-Nano, 
except that each block consists of 3 convolutional layers, and the channel widths are doubled to $[128, 128, 256, 512, 1024, 1536]$.

\vspace{3pt}\noindent \textbf{Hyperparameters.} 
The query decoder is a 6-layer standard transformer decoder with 8 heads in each attention module. Feature dimensions of a text prompt $d = 768$ and per-pixel embedding $d' = 64$. 
To unify the features from visual backbone variants with varying channel widths, they are projected to $d = 768$ with different feed-forward layers, and input as key and value to the query decoder. 
For images with multiple segmentation targets, we set the maximal text prompts sampled in a batch of up to 32. 
A combination of cross-entropy loss and dice loss is applied as supervision at training time. 
We use AdamW~\cite{AdamW} as the optimizer with cosine annealing schedule, maximal $lr = 1 \times 10^{-4}$, and 10000 steps for warm-up. 
\textcolor{black}{
The \model{}-Nano is trained on 8 NVIDIA A100 GPUs with 80GB memory for 14 days (approximately 2688 GPU hours), using maximal batch size 2; 
while the \model{}-Pro is trained on 16 NVIDIA A100 GPUs with 80GB memory for 14 days (approximately 5376 GPU hours), using maximal batch size 1.
}

\clearpage

\section{Inference Speed Test}
{
\color{black}
The inference time depends heavily on the size of the scan and the number of text prompts (categories). 
We demonstrate the inference speed of SAT-Pro and SAT-Nano on each dataset in Supplementary Table~\ref{tab:inference_speed1} and \ref{tab:inference_speed2}.
}
\begin{table}[htbp]
\centering
\footnotesize
\begin{tabular}{lcccccc}
\toprule
Dataset & Region & Size & \#Categories & SAT-Pro (s) & SAT-Nano (s) \\
\midrule

\rowcolor{lightgray!50} \textbf{MRI Data} & & & & & \\
AMOS22 MRI                 & Abdomen       & 385×275×96  &  16 &  2.48 & 0.99 \\
ATLAS                      & Abdomen       & 417×336×73  &   2 &  2.14 & 0.66 \\
ATLASR2                    & Brain         & 197×233×64  &   1 &  0.49 & 0.19 \\
BraTS2023 GLI              & Brain         & 137×171×47  &   4 &  0.49 & 0.18 \\
BraTS2023 MEN              & Brain         & 133×166×46  &   4 &  0.49 & 0.19 \\
BraTS2023 MET              & Brain         & 134×172×46  &   4 &  0.48 & 0.18 \\
BraTS2023 PED              & Brain         & 137×166×46  &   4 &  0.49 & 0.18 \\
BraTS2023 SSA              & Brain         & 136×174×46  &   4 &  0.80 & 0.54 \\
Brain Atlas                & Brain         & 168×186×50  & 108 &  0.49 & 0.22 \\
BrainPTM                   & Brain         & 130×171×45  &   7 &  0.49 & 0.18 \\
CHAOS MRI                  & Abdomen       & 431×326×80  &   5 &  2.53 & 0.78 \\
CMRxMotion                 & Thorax        & 302×335×33  &   4 &  1.35 & 0.41 \\
CrossMoDA2021              & Head and Neck & 210×211×59  &   2 &  0.48 & 0.18 \\
FeTA2022                   & Brain         & 91×107×32   &   7 &  0.48 & 0.18 \\
ISLES2022                  & Brain         & 139×166×44  &   1 &  0.48 & 0.18 \\
LAScarQS2022 Task 1        & Thorax        & 380×380×37  &   2 &  1.96 & 0.60 \\
LAScarQS2022 Task 2        & Thorax        & 448×448×35  &   1 &  1.93 & 0.60 \\
MM-WHS MRI                 & Thorax        & 208×258×110 &   9 &  1.31 & 0.42 \\
MRSpineSeg                 & Spine         & 161×304×19  &  23 &  0.96 & 0.32 \\
MSD Cardiac                & Thorax        & 298×400×53  &   1 &  1.45 & 0.44 \\
MSD Hippocampus            & Brain         & 35×50×13    &   3 &  0.48 & 0.18 \\
MSD Prostate               & Pelvis        & 197×197×23  &   3 &  0.49 & 0.19 \\
MyoPS2020                  & Thorax        & 345×342×17  &   6 &  1.94 & 0.60 \\
PROMISE12                  & Pelvis        & 200×200×29  &   1 &  0.49 & 0.18 \\
SKI10                      & Upper Limb    & 113×138×37  &   4 &  0.49 & 0.19 \\
WMH                        & Brain         & 190×240×63  &   1 &  0.49 & 0.18 \\

\rowcolor{lightgray!50} \textbf{PET Data} & & & & & \\
HECKTOR2022                & Head and Neck & 515×515×183 &   2 & 11.29 & 3.30 \\
autoPET                    & Whole Body    & 708×708×172 &   1 & 21.58 & 6.46 \\

\bottomrule
\end{tabular}
\vspace{.3cm}
\caption{
\textcolor{black}{
Inference speed comparison between SAT-Pro and SAT-Nano on each dataset. The size is averaged over all the volumes in the dataset. All the inferences are conducted on one A100 GPU. The speed is measured in seconds (s) and averaged over all the volumes in the dataset.}
}
\label{tab:inference_speed1}
\end{table}

\begin{table}[htbp]
\centering
\footnotesize
\begin{tabular}{lcccccc}
\toprule
Dataset & Region & Avg. Size & \#Categories & SAT-Pro (s) & SAT-Nano (s) \\
\midrule

\rowcolor{lightgray!50} \textbf{CT Data} & & & & & \\
AbdomenCT1K                & Abdomen       & 387×324×111 &   4 &  3.47 & 1.11 \\
ACDC                       & Thorax        & 332×357×29  &   4 &  1.91 & 0.59 \\
AMOS22 CT                  & Abdomen       & 368×295×164 &  16 &  5.43 & 1.60 \\
BTCV Abdomen               & Abdomen       & 400×332×155 &  15 &  6.13 & 1.89 \\
BTCV Cervix                & Abdomen       & 466×350×148 &   4 &  6.94 & 2.33 \\
CHAOS CT                   & Abdomen       & 359×303×60  &   1 &  1.47 & 0.45 \\
COVID-19 CT Seg            & Thorax        & 350×304×96  &   4 &  2.92 & 0.88 \\
CT-ORG                     & Whole Body    & 393×323×189 &   6 &  6.20 & 1.82 \\
CTPelvic1K                 & Lower Limb    & 414×309×91  &   5 &  1.35 & 0.42 \\
Couinaud                   & Abdomen       & 381×331×80  &  11 &  2.34 & 0.72 \\
DAP Atlas                  & Whole Body    & 436×425×311 & 191 & 14.88 & 6.58 \\
FLARE22                    & Abdomen       & 376×337×96  &  15 &  1.95 & 0.61 \\
FUMPE                      & Thorax        & 330×294×76  &   1 &  1.51 & 0.42 \\
HAN Seg                    & Head and Neck & 530×388×133 &  41 &  5.74 & 1.93 \\
INSTANCE                   & Brain         & 199×202×49  &   1 &  0.48 & 0.18 \\
KiPA22                     & Abdomen       & 98×98×43    &   4 &  0.48 & 0.18 \\
KiTS23                     & Abdomen       & 400×338×136 &   3 &  4.70 & 1.43 \\
LNDb                       & Thorax        & 356×302×106 &   1 &  2.32 & 0.69 \\
LUNA16                     & Thorax        & 353×302×106 &   4 &  3.87 & 1.08 \\
MM-WHS CT                  & Thorax        & 211×209×54  &   9 &  0.48 & 0.19 \\
MSD Colon                  & Abdomen       & 394×323×148 &   1 &  7.54 & 2.21 \\
MSD HepaticVessel          & Abdomen       & 366×316×78  &   2 &  2.52 & 0.79 \\
MSD Liver                  & Abdomen       & 403×335×168 &   2 &  7.52 & 2.06 \\
MSD Lung                   & Thorax        & 425×353×108 &   1 &  4.25 & 1.27 \\
MSD Pancreas               & Abdomen       & 382×321×89  &   2 &  1.94 & 0.60 \\
MSD Spleen                 & Abdomen       & 391×337×117 &   1 &  3.08 & 0.94 \\
NSCLC                      & Thorax        & 444×379×125 &   2 &  5.59 & 1.69 \\
PDDCA                      & Head and Neck & 536×385×135 &  12 &  6.01 & 1.76 \\
Pancreas CT                & Abdomen       & 410×332×73  &   1 &  2.34 & 0.71 \\
Parse2022                  & Thorax        & 334×286×105 &   1 &  2.53 & 0.76 \\
SEGA                       & Thorax        & 402×326×215 &   1 &  5.92 & 1.78 \\
SLIVER07                   & Abdomen       & 376×320×113 &   1 &  3.38 & 1.03 \\
SegRap2023 Task 1          & Head and Neck & 506×348×127 &  61 &  5.31 & 1.74 \\
SegRap2023 Task 2          & Head and Neck & 506×348×127 &   2 &  5.17 & 1.57 \\
SegTHOR                    & Thorax        & 467×376×134 &   4 &  5.40 & 1.60 \\
ToothFairy                 & Head and Neck & 377×346×57  &   1 &  1.92 & 0.73 \\
TotalSegmentor Cardiac     & Whole Body    & 351×297×133 &  17 &  2.20 & 0.71 \\
TotalSegmentor Muscles     & Whole Body    & 351×297×133 &  31 &  2.23 & 0.72 \\
TotalSegmentor Organs      & Whole Body    & 351×297×133 &  24 &  2.59 & 0.70 \\
TotalSegmentor Ribs        & Whole Body    & 351×297×133 &  39 &  2.24 & 0.75 \\
TotalSegmentor Vertebrae   & Whole Body    & 351×297×133 &  29 &  2.22 & 0.73 \\
TotalSegmentor V2          & Whole Body    & 351×297×133 &  24 &  2.20 & 0.69 \\
VerSe                      & Spine         & 255×458×95  &  30 &  1.34 & 0.44 \\
WORD                       & Abdomen       & 463×356×197 &  18 & 10.40 & 3.06 \\

\bottomrule
\end{tabular}
\vspace{.3cm}
\caption{
\textcolor{black}{
(Continued) Inference speed comparison between SAT-Pro and SAT-Nano on each dataset. The size is averaged over all the volumes in the dataset. All the inferences are conducted on one A100 GPU. The speed is measured in seconds (s) and averaged over all the volumes in the dataset.}
}
\label{tab:inference_speed2}
\end{table}
\newpage

\section{Calibration Analysis}
{
\color{black}
We conduct detailed calibration analysis in both internal and external validation. 
We first calculate the Expected Calibration Error (ECE) for each category at the pixel level with 20 bins.
As we formulate a binary segmentation task for each text prompt~(category), we define the pixel-wise confidence as:
\begin{equation}
\text{conf} = 
\begin{cases}
\sigma, & \text{if } \sigma \geq 0.5 \\
1 - \sigma, & \text{if } \sigma < 0.5 \\
\end{cases}
\end{equation}
Where $\sigma\in[0, 1]$ is the pixel logit. And the pixel-wise accuracy is:
\begin{equation}
\text{acc} = \mathbf{1}[(\sigma \geq 0.5) = (y = 1)]
\end{equation}
Where $y \in \{0, 1\}$ is the ground truth label for the pixel (0 for background, 1 for foreground).
On 72 internal datasets, SAT-Pro achieves an average ECE score of 4.2\% across all 497 categories, demonstrating well-calibrated predictions. 
On external datasets, SAT-Pro achieves ECE scores of 1.7\% on LiQA and 6.07\% on AbdomenAtlas, both outperforming the strongest baseline MedSAM (Oracle) with ECE scores of 3.2\% and 8.94\%, respectively.
We further provide reliability diagrams for 2 datasets in the external validation, shown in Supplementary Figure~\ref{fig:reliability}.
It further illustrates that
even though SAT exhibits slight overconfidence under distribution shifts, especially on AbdomenAtlas, it still demonstrates significantly more trustworthy predictions than MedSAM (Oracle).
\begin{figure}[htbp]
    \centering
    \includegraphics[width = 0.9\textwidth]{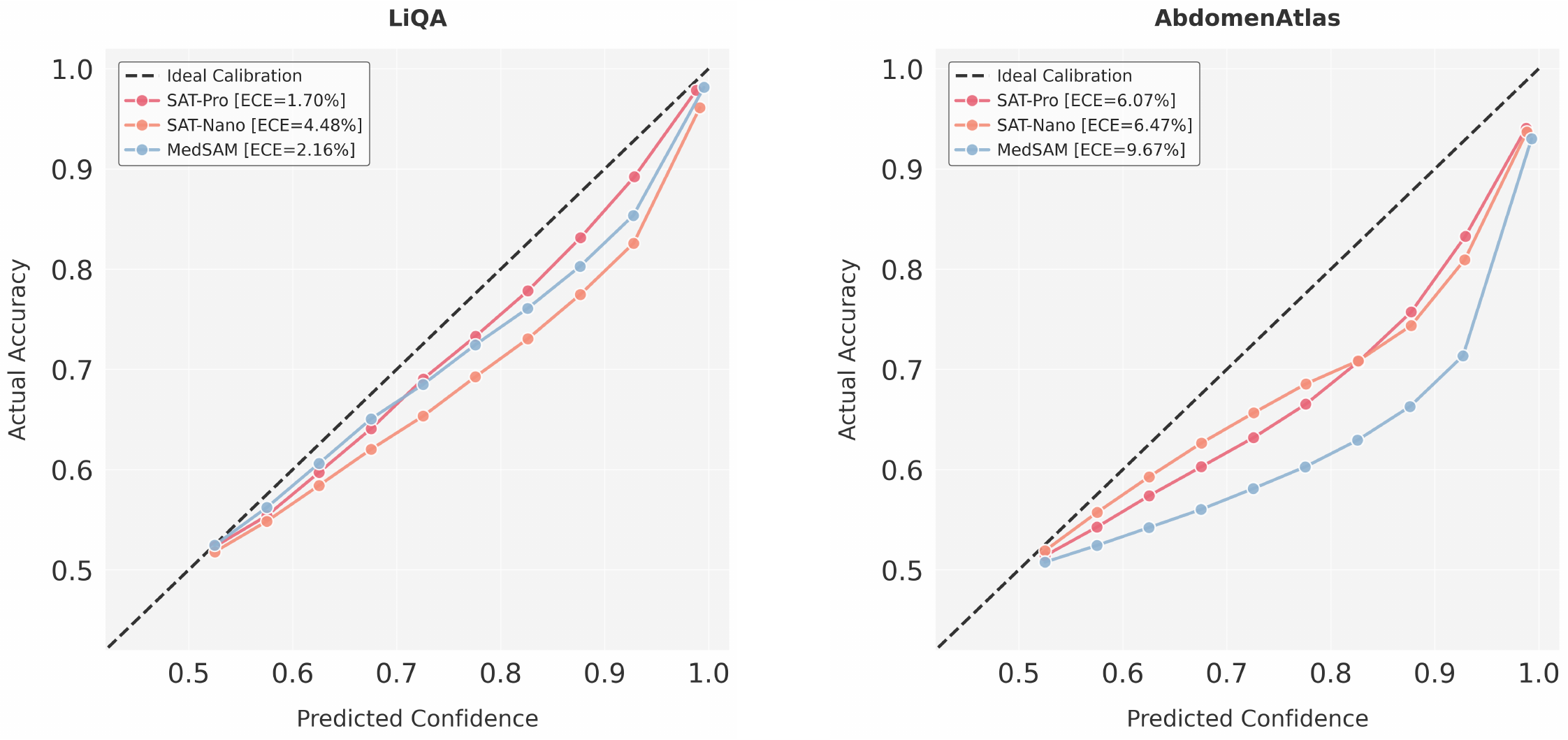}
    \vspace{.3cm}
    \caption{\textcolor{black}{
    \textbf{Reliability diagram on two external datastes.} SAT-Pro, SAT-Nano and the strongest baseline MedSAM (Oracle) are compared.
    The confidence and accuracy are averaged across all categories.
    }
    }
    \label{fig:reliability}
\end{figure}
}
\newpage

\section{Detailed Internal Evaluation Results}
\label{sec:supple_internal_results}

\begin{table}[htpb]
\center
\caption{Region-wise results of SAT-Nano, SAT-Pro, SAT-Ft, nnU-Nets, U-Mamba, and SwinUNETR on different human body regions and lesions. 
H\&N: head and neck, LL: lower limb, UL: upper limb, 
WB: whole body, All: average over all the 497 classes. 
The best results are \textbf{bolded}.}
\vspace{0.3cm}
\resizebox{\textwidth}{!}{
\begin{tabular}{clcccccccccc|c}
\toprule
\rowcolor{lightgray} \textbf{Metric} & \textbf{Method} & \textbf{Brain} & \textbf{H\&N} & \textbf{UL} & \textbf{Thorax} & \textbf{Spine} & \textbf{Abdomen} & \textbf{LL} & \textbf{Pelvis} & \textbf{WB} & \textbf{Lesion} & \textbf{All} \\ 
\midrule

\multirow{6}{*}{DSC$\uparrow$} & \model{}-Nano & 73.90 & 74.08 & 89.58 & 82.09 & 75.55 & 77.12 & 81.79 & 79.50 & 77.80 & 46.04 & 75.51 \\ 

& \model{}-Pro & 77.70 & 77.29 & 91.34 & 86.38 & 74.73 & 81.47 & 84.01 & 83.01 & 82.19 & 51.39 & 78.81 \\

& \model{}-Ft & 78.80 & \textbf{78.36} & \textbf{93.78} & \textbf{89.43} & \textbf{82.48} & 84.11 & \textbf{87.12} & \textbf{86.84} & 84.24 & 53.71 & \textbf{81.41} \\

& nnU-Nets & \textbf{81.93} & 72.45 & 89.20 & 85.60 & 81.62 & 86.96 & 82.89 & 84.29 & 84.20 & 57.52 & 80.54 \\

& U-Mamba & 81.73 & 70.44 & 87.54 & 84.46 & 78.82 & \textbf{87.65} & 86.43 & 85.45 & \textbf{84.74} & \textbf{58.64} & 79.80 \\

& SwinUNETR & 80.76 & 56.46 & 86.60 & 80.34 & 76.74 & 84.47 & 82.56 & 78.01 & 83.21 & 55.65 & 75.01 \\

\midrule

\multirow{6}{*}{NSD$\uparrow$} & \model{}-Nano & 72.95 & 79.15 & 90.94 & 79.68 & 73.68 & 67.83 & 78.42 & 74.89 & 75.36 & 38.17 & 73.76 \\ 

& \model{}-Pro & 77.77 & 82.45 & 93.56 & 85.15 & 72.87 & 72.94 & 82.50 & 78.66 & 79.61 & 44.87 & 77.82 \\

& \model{}-Ft & 80.43 & \textbf{84.01} & \textbf{95.64} & \textbf{86.38} & \textbf{82.37} & 78.42 & \textbf{87.25} & \textbf{82.92} & \textbf{82.48} & 47.74 & \textbf{81.60} \\

& nnU-Nets & 81.96 & 74.51 & 86.04 & 82.66 & 77.47 & 79.54 & 80.20 & 76.41 & 79.30 & 49.89 & 78.15 \\

& U-Mamba & \textbf{82.06} & 72.19 & 86.40 & 81.94 & 74.45 & \textbf{80.63} & 82.98 & 78.02 & 79.99 & \textbf{50.88} & 77.59 \\

& SwinUNETR & 80.66 & 56.04 & 82.11 & 75.72 & 71.37 & 75.03 & 77.94 & 66.20 & 77.65 & 47.05 & 71.15 \\

\bottomrule

\end{tabular}
}
\label{tab:region_results}
\end{table}
\begin{table}[htbp]
\center
\footnotesize
\caption{Region-wise results of SAT-Pro, SAT-Nano and MedSAMs on different human body regions and lesion.
H\&N: head and neck, All: average over all the classes. The best results are bolded. \textbf{Note that \model{}-Pro and \model{}-Nano are fully automatic methods, while MedSAM is interactive.}}
\vspace{0.3cm}
\resizebox{\textwidth}{!}{
\begin{tabular}{clccccccccc}
\toprule
\rowcolor{lightgray} \textbf{Metric} & \textbf{Method} & \textbf{Brain} & \textbf{H\&N} & \textbf{Thorax} & \textbf{Spine} & \textbf{Abdomen} & \textbf{Limb} & \textbf{Pelvis} & \textbf{Lesion} & \textbf{All} \\
\midrule
\multirow{5}{*}{DSC$\uparrow$}& SAT-Pro & \textbf{73.23} & \textbf{84.16} & \textbf{85.72} & \textbf{85.52} & \textbf{82.23} & \textbf{82.56} & \textbf{86.35} & 55.69 & \textbf{82.47} \\
& SAT-Nano & 69.05 & 77.93 & 79.06 & 81.88 & 78.43 & 77.6 & 78.31 & 47.7 & 76.66 \\
\cmidrule[\lightrulewidth]{2-11}

& Oracle Box & 55.84 & 85.32 & 72.85 & 78.11 & 71.92 & 78.96 & 81.3 & \textbf{67.94} & 63.57 \\
& MedSAM (Tight) & 54.35 & 78.48 & 73.01 & 79.09 & 77.35 & 80.53 & 84.42 & 65.85 & 75.39 \\
& MedSAM (Loose) & 14.7 & 44.68 & 36.49 & 45.72 & 19.79 & 42.97 & 51.42 & 15.71 & 35.18 \\
\midrule

\multirow{5}{*}{NSD$\uparrow$}& SAT-Pro & \textbf{80.46} & \textbf{86.12} & \textbf{85.87} & \textbf{87.01} & \textbf{79.98} & \textbf{83.53} & \textbf{82.39} & 49.35 & \textbf{81.79} \\
& SAT-Nano & 75.62 & 79.44 & 77.74 & 82.7 & 74.96 & 77.58 & 72.98 & 40.38 & 74.82 \\
\cmidrule[\lightrulewidth]{2-11}
& Oracle Box & 60.55 & 86.83 & 66.16 & 69.86 & 59.51 & 70.67 & 71.59 & 61.59 & 48.93 \\

& MedSAM (Tight) & 68.62 & 82.37 & 68.26 & 73.29 & 73.39 & 73.84 & 74.9 & \textbf{62.28} & 70.99 \\

& MedSAM (Loose) & 10.29 & 43.57 & 31.61 & 43.84 & 16.81 & 41.54 & 46.74 & 11.46 & 31.50 \\
\bottomrule

\end{tabular}
}
\label{tab:region_results_medsam}
\end{table}
\begin{table*}[!tbh]
\renewcommand{\arraystretch}{1.3} 
\center
\caption{Dataset-wise \textbf{DSC} scores of \model{}-Ft, \model{}-Pro, \model{}-Nano, nnU-Nets, U-Mamba, SwinUNETR and MedSAMs on 72 datasets in \samdataset{}. Datasets uninvolved in traning MedSAM are excluded when evaluating MedSAM and marked as /. `SwinU' stands for SwinUNETR, `MS-T' stands for MedSAM Tight (with Oracle Box as prompt) while `MS-L' stands for MedSAM Loose.}
\vspace{0.3cm}
\label{tab:dataset_dsc_results}
\resizebox{1.0\textwidth}{!}{
\begin{tabular}{lcccccccc}
\toprule
\rowcolor{lightgray} \textbf{Dataset} & \textbf{ SAT-Ft } & \textbf{SAT-Pro} & \textbf{SAT-Nano} & \textbf{nnU-Net} & \textbf{U-Mamba} & \textbf{ SwinU } & \textbf{MS-T } & \textbf{ MS-L} \\
\midrule
AbdomenCT1K~\cite{AbdomenCT1K}                               & 94.9 & 94.28 & 93.32 & 95.09 & 95.35 & 93.73 & 86.03 & 10.97 \\\hline
ACDC~\cite{ACDC}                                             & 89.64 & 87.74 & 85.5 & 90.76 & 90.83 & 86.49 & / & / \\\hline
AMOS22 CT~\cite{AMOS22}                                      & 88.75 & 86.37 & 84.93 & 89.77 & 90.57 & 87.32 & 81.21 & 5.27 \\\hline
AMOS22 MRI~\cite{AMOS22}                                     & 84.82 & 78.9 & 78.76 & 86.43 & 87.06 & 84.08 & 79.17 & 10.97 \\\hline
ATLAS~\cite{ATLAS}                                           & 76.26 & 76.02 & 68.32 & 78.83 & 78.33 & 70.88 & / & / \\\hline
ATLASR2~\cite{ATLASR2}                                       & 61.77 & 61.44 & 53.95 & 53.69 & 68.94 & 65.0 & 61.19 & 3.96 \\\hline
autoPET~\cite{autoPET}                                       & 71.66 & 68.56 & 62.27 & 74.98 & 74.48 & 71.46 & / & / \\\hline
Brain Atlas~\cite{Brain_Atlas}                               & 79.71 & 78.57 & 74.89 & 83.78 & 83.81 & 83.56 & / & / \\\hline
BrainPTM~\cite{BrainPTM}                                     & 65.33 & 66.8 & 64.42 & 68.37 & 67.74 & 66.31 & / & / \\\hline
BraTS2023 GLI~\cite{BraTS2023GLI}                            & 68.18 & 67.92 & 65.05 & 73.22 & 73.87 & 71.67 & 64.49 & 14.28 \\\hline
BraTS2023 MEN~\cite{BraTS2023MEN}                            & 63.89 & 58.04 & 52.75 & 64.4 & 70.47 & 66.41 & 77.03 & 35.96 \\\hline
BraTS2023 MET~\cite{BraTS2023MET}                            & 44.22 & 43.76 & 40.6 & 47.54 & 51.95 & 48.28 & 49.98 & 15.5 \\\hline
BraTS2023 PED~\cite{BraTS2023PED}                            & 59.74 & 51.83 & 49.02 & 57.48 & 61.56 & 55.55 & 72.04 & 24.83 \\\hline
BraTS2023 SSA~\cite{BraTS2023SSA}                            & 55.68 & 55.73 & 53.19 & 59.02 & 57.15 & 52.6 & 65.61 & 20.12 \\\hline
BTCV~\cite{BTCV}                                      & 81.6 & 80.71 & 79.6 & 77.66 & 74.88 & 71.82 & / & / \\\hline
BTCV Cervix~\cite{BTCV}                                & 74.86 & 73.73 & 72.99 & 71.57 & 78.7 & 75.0 & / & / \\\hline
CHAOS CT~\cite{CHAOS}                                        & 97.24 & 97.02 & 96.54 & 97.08 & 97.34 & 96.88 & / & / \\\hline
CHAOS MRI~\cite{CHAOS}                                       & 87.99 & 87.28 & 82.07 & 88.8 & 87.27 & 80.84 & / & / \\\hline
CMRxMotion~\cite{CMRxMotion}                                 & 90.28 & 88.19 & 87.45 & 91.14 & 91.94 & 88.06 & / & / \\\hline
Couinaud Liver~\cite{Couinaud}                               & 85.54 & 81.23 & 72.01 & 87.86 & 87.8 & 86.16 & / & / \\\hline
COVID19~\cite{COVID19}                                       & 87.09 & 83.18 & 71.51 & 91.53 & 91.78 & 88.58 & 61.45 & 58.22 \\\hline
CrossMoDA2021~\cite{CrossMoDA2021}                           & 79.15 & 78.34 & 73.16 & 81.77 & 83.94 & 82.45 & 90.41 & 0.87 \\\hline
CT ORG~\cite{CTORG}                                          & 92.21 & 90.12 & 85.44 & 75.27 & 76.05 & 71.43 & 74.24 & 31.87 \\\hline
CTPelvic1K~\cite{CTPelvic1K}                                 & 96.58 & 95.86 & 95.14 & 76.43 & 77.26 & 77.81 & / & / \\\hline
DAP Atlas~\cite{DAPAtlas}                                    & 85.79 & 85.79 & 84.39 & 87.73 & 88.33 & 80.81 & / & / \\\hline
FeTA2022~\cite{FeTA2022}                                     & 76.24 & 73.23 & 69.05 & 75.83 & 75.39 & 75.05 & 54.35 & 14.7 \\\hline
FLARE22~\cite{FLARE22}                                       & 91.78 & 91.12 & 88.79 & 93.36 & 93.46 & 90.91 & / & / \\\hline
FUMPE~\cite{FUMPE}                                           & 22.94 & 22.04 & 36.13 & 47.65 & 34.19 & 35.74 & / & / \\\hline
HAN Seg~\cite{HANSeg}                                        & 73.15 & 72.11 & 69.73 & 62.18 & 61.23 & 54.59 & / & / \\\hline
Hecktor2022~\cite{HECTOR2022}                                & 61.99 & 58.3 & 55.75 & 64.52 & 65.68 & 62.29 & / & / \\\hline
Instance22~\cite{INSTANCE}                                   & 67.84 & 70.18 & 55.7 & 81.53 & 80.43 & 71.9 & 71.22 & 3.85 \\\hline
ISLES2022~\cite{ISLES2022}                                   & 53.7 & 55.12 & 43.64 & 64.95 & 64.59 & 63.77 & 54.74 & 9.61 \\\hline
KiPA22~\cite{KiPA22}                                         & 76.59 & 74.87 & 64.39 & 90.34 & 90.5 & 89.61 & 60.9 & 20.0 \\\hline
KiTS23~\cite{KiTS23}                                         & 71.53 & 67.98 & 55.96 & 74.69 & 74.33 & 68.86 & 67.93 & 18.54 \\\hline
LAScarQS22 Task1~\cite{LAScarQS2022}                         & 66.83 & 68.97 & 66.45 & 71.47 & 70.25 & 69.09 & / & / \\\hline
LAScarQS22 Task2~\cite{LAScarQS2022}                         & 92.36 & 92.03 & 90.0 & 85.28 & 92.73 & 91.59 & / & / \\\hline
LNDb~\cite{LNDb}                                             & 36.2 & 37.08 & 28.0 & 24.45 & / & 23.91 & / & / \\\hline
LUNA16~\cite{LUNA16}                                         & 97.16 & 96.32 & 95.97 & 96.64 & 96.88 & 95.94 & / & / \\\hline
MM WHS CT~\cite{MMWHS}                                       & 91.14 & 89.97 & 88.23 & 88.64 & 91.56 & 91.25 & / & / \\\hline
MM WHS MRI~\cite{MMWHS}                                      & 87.73 & 86.7 & 84.37 & 30.88 & 21.20 & 20.87 & / & / \\\hline
MRSpineSeg~\cite{MRSpineSeg}                                 & 79.78 & 71.97 & 74.06 & 68.97 & 67.96 & 66.3 & / & / \\
\bottomrule
\end{tabular}
}
\end{table*}

\begin{table*}[!tbh]
\renewcommand{\arraystretch}{1.3} 
\center
\caption{(Continued) Dataset-wise \textbf{DSC} scores of \model{}-Ft, \model{}-Pro, \model{}-Nano, nnU-Nets, U-Mamba, SwinUNETR and MedSAMs on 72 datasets in \samdataset{}. Datasets uninvolved in traning MedSAM are excluded when evaluating MedSAM and marked as /. `SwinU' stands for SwinUNETR, `MS-T' stands for MedSAM Tight (with Oracle Box as prompt) while `MS-L' stands for MedSAM Loose. `TS' stands for TotalSegmentator.}
\label{tab:dataset_dsc_results2}
\vspace{0.3cm}
\resizebox{1.0\textwidth}{!}{
\begin{tabular}{lcccccccc}
\toprule
\rowcolor{lightgray} \textbf{Dataset} & \textbf{SAT-Ft} & \textbf{SAT-Pro} & \textbf{SAT-Nano} & \textbf{nnU-Net} & \textbf{U-Mamba} & \textbf{SwinUNETR} & \textbf{MS-T} & \textbf{MS-L} \\
\midrule
MSD Cardiac~\cite{MSD}                                       & 93.38 & 92.61 & 90.28 & 94.28 & 93.8 & 93.46 & 83.6 & 3.11 \\\hline
MSD Colon~\cite{MSD}                                         & 38.45 & 35.29 & 23.43 & 54.39 & 54.25 & 41.4 & 71.02 & 2.59 \\\hline
MSD HepaticVessel~\cite{MSD}                                 & 63.43 & 63.14 & 53.56 & 67.74 & 68.73 & 66.07 & / & / \\\hline
MSD Hippocampus~\cite{MSD}                                   & 87.62 & 87.62 & 86.25 & 89.18 & 89.03 & 89.03 & / & / \\\hline
MSD Liver~\cite{MSD}                                         & 78.86 & 76.63 & 68.16 & 77.92 & 77.99 & 75.46 & / & / \\\hline
MSD Lung~\cite{MSD}                                          & 61.28 & 62.65 & 51.01 & 71.74 & 66.07 & 64.59 & 68.43 & 1.23 \\\hline
MSD Pancreas~\cite{MSD}                                      & 59.23 & 59.32 & 58.2 & 68.64 & 69.7 & 57.87 & 69.96 & 2.84 \\\hline
MSD Prostate~\cite{MSD}                                      & 77.98 & 78.33 & 73.38 & 71.32 & 77.72 & 67.73 & 74.09 & 69.18 \\\hline
MSD Spleen~\cite{MSD}                                        & 94.97 & 94.12 & 93.5 & 92.95 & 86.11 & 84.83 & 93.33 & 81.14 \\\hline
MyoPS2020~\cite{MyoPS2020}                                   & 61.06 & 58.69 & 59.77 & 14.85 & 12.41 & 11.94 & / & / \\\hline
NSCLC~\cite{NSCLC}                                           & 77.97 & 77.51 & 75.48 & 78.58 & 78.83 & 78.56 & 77.51 & 49.94 \\\hline
Pancreas CT~\cite{PancreasCT}                                & 84.69 & 85.57 & 84.35 & 87.52 & 87.6 & 86.89 & / & / \\\hline
PARSE2022~\cite{PARSE2022}                                   & 79.41 & 74.9 & 71.04 & 85.85 & 85.77 & 85.03 & / & / \\\hline
PDDCA~\cite{PDDCA}                                           & 73.75 & 76.68 & 72.83 & 57.45 & 53.07 & 51.65 & / & / \\\hline
PROMISE12~\cite{PROMISE12}                                   & 87.28 & 86.51 & 84.55 & 88.86 & 89.53 & 87.46 & 85.51 & 8.67 \\\hline
SEGA~\cite{SEGA}                                             & 89.59 & 83.9 & 81.48 & 89.43 & 89.95 & 87.23 & / & / \\\hline
SegRap2023 Task1~\cite{SegRap2023}                           & 86.46 & 84.86 & 82.8 & 79.98 & 76.78 & 57.32 & / & / \\\hline
SegRap2023 Task2~\cite{SegRap2023}                           & 72.01 & 70.9 & 65.98 & 74.48 & 74.69 & 71.09 & / & / \\\hline
SegTHOR~\cite{SegTHOR}                                       & 88.98 & 86.69 & 82.6 & 91.32 & 91.37 & 89.92 & 74.9 & 5.37 \\\hline
SKI10~\cite{SKI10}                                           & 84.7 & 83.36 & 80.51 & 88.15 & 88.27 & 87.23 & / & / \\\hline
SLIVER07~\cite{SLIVER07}                                     & 97.63 & 97.43 & 97.03 & 97.3 & 96.77 & 93.56 & / & / \\\hline
ToothFairy~\cite{ToothFairy}                                 & 78.17 & 77.95 & 63.65 & 83.08 & 83.28 & 79.85 & / & / \\\hline
TS Cardiac~\cite{Totalsegmentator}             & 92.52 & 88.96 & 76.77 & 93.3 & 93.73 & 91.23 & 81.26 & 36.35 \\\hline
TS Muscles~\cite{Totalsegmentator}             & 93.33 & 88.04 & 82.17 & 91.6 & 92.0 & 90.21 & 82.23 & 43.74 \\\hline
TS Organs~\cite{Totalsegmentator}              & 90.42 & 87.53 & 83.4 & 93.22 & 93.23 & 90.41 & 82.71 & 35.52 \\\hline
TS Ribs~\cite{Totalsegmentator}                & 91.53 & 83.73 & 75.78 & 92.1 & 90.85 & 88.51 & 68.85 & 30.54 \\\hline
TS v2~\cite{Totalsegmentator}                  & 86.71 & 78.46 & 72.48 & 92.39 & 91.53 & 88.85 & 80.11 & 65.89 \\\hline
TS Vertebrae~\cite{Totalsegmentator}           & 90.42 & 85.21 & 81.92 & 95.37 & 95.68 & 94.08 & 79.13 & 44.83 \\\hline
VerSe~\cite{VerSe}                                           & 81.01 & 61.55 & 61.18 & 81.82 & 69.13 & 70.17 & / & / \\\hline
WMH ~\cite{WMH}                        & 69.22 & 69.05 & 62.55 & 77.02 & 77.77 & 75.22 & / & / \\\hline
WORD~\cite{WORD}                                             & 87.92 & 86.77 & 86.57 & 85.49 & 87.75 & 85.27 & / & / \\
\bottomrule
\end{tabular}
}
\end{table*}

%
%

\begin{table*}[!tbh]
\renewcommand{\arraystretch}{1.3} 
\center
\caption{Dataset-wise \textbf{NSD} scores of \model{}-Ft, \model{}-Pro, \model{}-Nano, nnU-Nets, U-Mamba, SwinUNETR and MedSAMs on 72 datasets in \samdataset{}. Datasets uninvolved in traning MedSAM are excluded when evaluating MedSAM and marked as /. `SwinU' stands for SwinUNETR, `MS-T' stands for MedSAM Tight (with Oracle Box as prompt) while `MS-L' stands for MedSAM Loose.}
\vspace{0.3cm}
\label{tab:dataset_nsd_results}
\resizebox{1.0\textwidth}{!}{
\begin{tabular}{lcccccccc}
\toprule
\rowcolor{lightgray} \textbf{Dataset} & \textbf{ SAT-Ft } & \textbf{SAT-Pro} & \textbf{SAT-Nano} & \textbf{nnU-Net} & \textbf{U-Mamba} & \textbf{ SwinU } & \textbf{MS-T } & \textbf{ MS-L} \\
\midrule
AbdomenCT1K~\cite{AbdomenCT1K}                               & 89.53 & 87.3 & 84.41 & 88.24 & 88.7 & 85.38 & 72.85 & 2.37 \\\hline
ACDC~\cite{ACDC}                                             & 78.79 & 72.47 & 66.77 & 97.62 & 80.89 & 73.92 & / & / \\\hline
AMOS22 CT~\cite{AMOS22}                                      & 87.66 & 82.98 & 80.21 & 87.11 & 88.44 & 82.52 & 75.72 & 2.25 \\\hline
AMOS22 MRI~\cite{AMOS22}                                     & 83.39 & 75.22 & 74.01 & 79.56 & 80.59 & 75.07 & 74.32 & 7.71 \\\hline
ATLAS~\cite{ATLAS}                                           & 48.93 & 46.99 & 40.21 & 45.19 & 45.99 & 35.88 & / & / \\\hline
ATLASR2~\cite{ATLASR2}                                       & 62.14 & 61.24 & 51.57 & 48.93 & 69.06 & 63.26 & 66.74 & 1.11 \\\hline
autoPET~\cite{autoPET}                                       & 60.47 & 56.3 & 48.17 & 57.94 & 58.58 & 55.37 & / & / \\\hline
Brain Atlas~\cite{Brain_Atlas}                               & 81.28 & 78.48 & 73.34 & 85.35 & 85.66 & 85.11 & / & / \\\hline
BrainPTM~\cite{BrainPTM}                                     & 54.62 & 54.02 & 50.68 & 33.9 & 34.08 & 32.5 & / & / \\\hline
BraTS2023 GLI~\cite{BraTS2023GLI}                            & 63.79 & 62.27 & 58.26 & 68.25 & 69.3 & 66.55 & 62.66 & 4.83 \\\hline
BraTS2023 MEN~\cite{BraTS2023MEN}                            & 60.84 & 54.55 & 47.95 & 61.57 & 67.09 & 61.53 & 76.31 & 31.25 \\\hline
BraTS2023 MET~\cite{BraTS2023MET}                            & 42.18 & 41.62 & 37.22 & 45.95 & 50.21 & 45.63 & 57.26 & 11.78 \\\hline
BraTS2023 PED~\cite{BraTS2023PED}                            & 54.09 & 45.94 & 42.89 & 51.38 & 55.86 & 48.81 & 70.69 & 19.75 \\\hline
BraTS2023 SSA~\cite{BraTS2023SSA}                            & 45.96 & 46.35 & 42.94 & 46.73 & 45.09 & 41.16 & 59.84 & 9.26 \\\hline
BTCV~\cite{BTCV}                                      & 79.52 & 77.19 & 76.22 & 74.85 & 54.5 & 50.12 & / & / \\\hline
BTCV Cervix~\cite{BTCV}                                & 51.85 & 50.93 & 49.21 & 48.31 & 76.26 & 69.28 & / & / \\\hline
CHAOS CT~\cite{CHAOS}                                        & 85.88 & 84.63 & 81.12 & 81.04 & 81.33 & 79.71 & / & / \\\hline
CHAOS MRI~\cite{CHAOS}                                       & 59.78 & 53.4 & 47.39 & 64.35 & 64.02 & 52.99 & / & / \\\hline
CMRxMotion~\cite{CMRxMotion}                                 & 80.54 & 73.09 & 70.39 & 86.47 & 87.88 & 80.66 & / & / \\\hline
Couinaud Liver~\cite{Couinaud}                               & 64.95 & 53.05 & 42.72 & 70.44 & 70.58 & 66.72 & / & / \\\hline
COVID19~\cite{COVID19}                                       & 78.44 & 72.83 & 48.68 & 77.02 & 80.23 & 73.61 & 27.03 & 19.29 \\\hline
CrossMoDA2021~\cite{CrossMoDA2021}                           & 96.24 & 95.69 & 93.06 & 93.16 & 96.57 & 95.24 & 96.49 & 1.38 \\\hline
CT ORG~\cite{CTORG}                                          & 82.5 & 77.57 & 72.0 & 66.15 & 67.32 & 61.97 & 57.13 & 20.85 \\\hline
CTPelvic1K~\cite{CTPelvic1K}                                 & 98.48 & 97.57 & 96.28 & 73.45 & 79.43 & 73.42 & / & / \\\hline
DAP Atlas~\cite{DAPAtlas}                                    & 86.51 & 86.51 & 85.49 & 87.35 & 88.3 & 76.86 & / & / \\\hline
FeTA2022~\cite{FeTA2022}                                     & 84.31 & 80.46 & 75.62 & 81.34 & 81.01 & 80.19 & 68.62 & 10.29 \\\hline
FLARE22~\cite{FLARE22}                                       & 90.88 & 89.47 & 85.76 & 91.15 & 91.44 & 87.0 & / & / \\\hline
FUMPE~\cite{FUMPE}                                           & 21.7 & 18.07 & 32.15 & 42.02 & 28.31 & 29.32 & / & / \\\hline
HAN Seg~\cite{HANSeg}                                        & 79.93 & 77.76 & 74.29 & 63.52 & 62.04 & 53.43 & / & / \\\hline
Hecktor2022~\cite{HECTOR2022}                                & 49.97 & 45.49 & 43.21 & 43.84 & 45.59 & 40.88 & / & / \\\hline
Instance22~\cite{INSTANCE}                                   & 64.65 & 66.74 & 45.25 & 79.73 & 78.76 & 66.88 & 70.73 & 1.95 \\\hline
ISLES2022~\cite{ISLES2022}                                   & 53.53 & 54.67 & 42.15 & 60.1 & 59.88 & 58.85 & 58.6 & 5.66 \\\hline
KiPA22~\cite{KiPA22}                                         & 77.58 & 74.83 & 61.68 & 89.54 & 90.29 & 87.73 & 61.4 & 10.16 \\\hline
KiTS23~\cite{KiTS23}                                         & 65.81 & 59.8 & 47.06 & 70.14 & 70.04 & 62.64 & 51.16 & 15.49 \\\hline
LAScarQS22 Task1~\cite{LAScarQS2022}                         & 78.3 & 80.45 & 75.39 & 77.06 & 76.95 & 73.2 & / & / \\\hline
LAScarQS22 Task2~\cite{LAScarQS2022}                         & 81.08 & 80.23 & 72.58 & 64.02 & 78.81 & 74.24 & / & / \\\hline
LNDb~\cite{LNDb}                                             & 43.21 & 45.52 & 31.24 & 29.38 & / & 27.36 & / & / \\\hline
LUNA16~\cite{LUNA16}                                         & 96.79 & 94.12 & 92.42 & 93.85 & 94.44 & 90.03 & / & / \\\hline
MM WHS CT~\cite{MMWHS}                                       & 79.62 & 75.61 & 70.1 & 64.45 & 73.79 & 72.95 & / & / \\\hline
MM WHS MRI~\cite{MMWHS}                                      & 72.42 & 69.32 & 64.67 & 23.92 & 7.74 & 7.43 & / & / \\\hline
MRSpineSeg~\cite{MRSpineSeg}                                 & 78.25 & 67.01 & 68.62 & 58.82 & 57.59 & 56.02 & / & / \\
\bottomrule
\end{tabular}
}
\end{table*}

\begin{table*}[!tbh]
\renewcommand{\arraystretch}{1.3} 
\center
\caption{(Continued) Dataset-wise \textbf{NSD} scores of \model{}-Ft, \model{}-Pro, \model{}-Nano, nnU-Nets, U-Mamba, SwinUNETR and MedSAMs on 72 datasets in \samdataset{}. Datasets uninvolved in traning MedSAM are excluded when evaluating MedSAM and marked as /. `SwinU' stands for SwinUNETR, `MS-T' stands for MedSAM Tight (with Oracle Box as prompt) while `MS-L' stands for MedSAM Loose. `TS' stands for TotalSegmentator.}
\label{tab:dataset_nsd_results2}
\vspace{0.3cm}
\resizebox{1.0\textwidth}{!}{
\begin{tabular}{lcccccccc}
\toprule
\rowcolor{lightgray} \textbf{Dataset} & \textbf{SAT-Ft} & \textbf{SAT-Pro} & \textbf{SAT-Nano} & \textbf{nnU-Net} & \textbf{U-Mamba} & \textbf{SwinUNETR} & \textbf{MS-T} & \textbf{MS-L} \\
\midrule
MSD Cardiac~\cite{MSD}                                       & 85.93 & 83.14 & 77.29 & 64.21 & 62.63 & 61.68 & 60.84 & 1.49 \\\hline
MSD Colon~\cite{MSD}                                         & 34.21 & 29.79 & 15.7 & 51.07 & 50.42 & 36.31 & 63.65 & 1.52 \\\hline
MSD HepaticVessel~\cite{MSD}                                 & 56.89 & 56.07 & 46.49 & 61.85 & 63.15 & 59.18 & / & / \\\hline
MSD Hippocampus~\cite{MSD}                                   & 96.46 & 96.46 & 95.64 & 97.92 & 97.7 & 97.7 & / & / \\\hline
MSD Liver~\cite{MSD}                                         & 67.19 & 62.89 & 52.66 & 63.78 & 63.27 & 57.17 & / & / \\\hline
MSD Lung~\cite{MSD}                                          & 53.59 & 52.42 & 38.62 & 58.32 & 54.8 & 50.9 & 59.58 & 0.73 \\\hline
MSD Pancreas~\cite{MSD}                                      & 46.57 & 46.84 & 45.25 & 53.31 & 54.77 & 42.31 & 52.36 & 0.96 \\
\bottomrule
\end{tabular}
}
\end{table*}

\begin{table*}[!tbh]
\renewcommand{\arraystretch}{1.3} 
\center
\footnotesize
\caption{Class-wise \textbf{DSC} results of \model{}-Ft, \model{}-Pro, \model{}-Nano, nnU-Nets, U-Mamba and SwinUNETR on common anatomical structures in abdomen, brain, head and neck, and spine.}
\vspace{0.3cm}
\resizebox{1.0\textwidth}{!}{\begin{tabular}{lclcccccc}
\toprule
\rowcolor{lightgray} \textbf{Region} & \textbf{Modality} & \textbf{Anatomical Target} & \textbf{\model{}-Ft} & \textbf{\model{}-Pro} & \textbf{\model{}-Nano} & \textbf{nnU-Nets} & \textbf{U-Mamba} & \textbf{SwinUNETR} \\\midrule

\multirow{17}{*}{Abdomen} & CT & Adrenal gland        & 82.21 & 81.54 & 79.24 & 84.43 & 84.76 & 80.15 \\
& CT & Celiac trunk         & 87.56 & 87.56 & 84.55 & 86.86 & 87.57 & 73.19 \\
& CT & Duodenum             & 82.55 & 80.77 & 79.23 & 84.38 & 85.28 & 78.4 \\
& CT & Gallbladder          & 81.14 & 80.37 & 78.96 & 83.98 & 84.96 & 78.74 \\
& CT & Inferior vena cava   & 91.91 & 90.94 & 89.93 & 92.24 & 92.86 & 89.48 \\
& CT & Intestine            & 88.78 & 86.6 & 86.03 & 89.41 & 89.9 & 87.83 \\
& CT & Kidney               & 94.03 & 93.23 & 91.21 & 91.69 & 92.43 & 90.42 \\
\bottomrule

\end{tabular}}
\label{tab:specialist_dsc_class_results}
\end{table*}

\begin{table*}[!htp]
\renewcommand{\arraystretch}{1.3} 
\center
\footnotesize
\caption{(Continued) Class-wise \textbf{DSC} results of \model{}-Ft, \model{}-Pro, \model{}-Nano, nnU-Nets, U-Mamba and SwinUNETR on common anatomical structures in thorax, limbs, pelvis, whole body and common lesions.}
\vspace{0.3cm}
\resizebox{1.0\textwidth}{!}{\begin{tabular}{lclcccccc}
\toprule
\rowcolor{lightgray} \textbf{Region} & \textbf{Modality} & \textbf{Anatomical Target} & \textbf{\model{}-Ft} & \textbf{\model{}-Pro} & \textbf{\model{}-Nano} & \textbf{nnU-Nets} & \textbf{U-Mamba} & \textbf{SwinUNETR} \\\midrule

\multirow{9}{*}{Thorax} & CT & Autochthon           & 96.24 & 95.71 & 93.82 & 94.11 & 93.47 & 91.97 \\
& CT & Breast               & 61.82 & 61.82 & 58.17 & 58.45 & 60.63 & 42.23 \\
& CT & Heart atrium         & 92.91 & 91.17 & 89.33 & 92.56 & 93.7 & 92.32 \\
& MRI & Heart ventricle      & 94.01 & 92.2 & 88.97 & 93.72 & 94.88 & 92.97 \\
& CT & Lung                 & 94.88 & 93.65 & 90.44 & 95.57 & 95.38 & 92.74 \\
& CT & Rib                  & 89.65 & 85.62 & 80.47 & 88.38 & 89.62 & 84.95 \\
& CT & Myocardium           & 92.75 & 88.8 & 83.66 & 92.92 & 94.33 & 91.4 \\
\end{tabular}}
\end{table*}

%
%

\begin{table*}[!tbh]
\renewcommand{\arraystretch}{1.3} 
\center
\footnotesize
\caption{Class-wise \textbf{NSD} results of \model{}-Ft, \model{}-Pro, \model{}-Nano, nnU-Nets, U-Mamba and SwinUNETR on common anatomical structures in abdomen, brain, head and neck, and spine.}
\vspace{0.3cm}
\resizebox{1.0\textwidth}{!}{\begin{tabular}{lclcccccc}
\toprule
\rowcolor{lightgray} \textbf{Region} & \textbf{Modality} & \textbf{Anatomical Target} & \textbf{\model{}-Ft} & \textbf{\model{}-Pro} & \textbf{\model{}-Nano} & \textbf{nnU-Nets} & \textbf{U-Mamba} & \textbf{SwinUNETR} \\\midrule

\multirow{17}{*}{Abdomen} & CT & Adrenal gland        & 90.71 & 90.26 & 87.83 & 89.41 & 89.67 & 84.17 \\
& CT & Celiac trunk         & 94.35 & 94.35 & 92.7 & 91.7 & 92.34 & 77.82 \\
& CT & Duodenum             & 77.94 & 73.62 & 71.01 & 77.1 & 78.91 & 67.05 \\
& CT & Gallbladder          & 79.09 & 77.31 & 74.73 & 78.25 & 81.58 & 70.8 \\
& CT & Inferior vena cava   & 90.49 & 88.77 & 86.6 & 87.22 & 88.24 & 80.82 \\
& CT & Intestine            & 80.73 & 73.51 & 72.01 & 81.87 & 83.34 & 78.82 \\
& CT & Kidney               & 92.21 & 90.28 & 86.64 & 88.67 & 89.56 & 85.55 \\
\bottomrule

\end{tabular}}
\label{tab:specialist_nsd_class_results}
\end{table*}

\begin{table*}[!htp]
\renewcommand{\arraystretch}{1.3} 
\center
\footnotesize
\caption{(Continued) Class-wise \textbf{NSD} results of \model{}-Ft, \model{}-Pro, \model{}-Nano, nnU-Nets, U-Mamba and SwinUNETR on common anatomical structures in thorax, limbs, pelvis, whole body and common lesions.}
\vspace{0.3cm}
\resizebox{1.0\textwidth}{!}{\begin{tabular}{lclcccccc}
\toprule
\rowcolor{lightgray} \textbf{Region} & \textbf{Modality} & \textbf{Anatomical Target} & \textbf{\model{}-Ft} & \textbf{\model{}-Pro} & \textbf{\model{}-Nano} & \textbf{nnU-Nets} & \textbf{U-Mamba} & \textbf{SwinUNETR} \\\midrule

\multirow{9}{*}{Thorax} & CT & Autochthon           & 96.65 & 95.02 & 91.36 & 84.73 & 84.8 & 77.24 \\
& CT & Breast               & 61.33 & 61.33 & 56.36 & 53.85 & 57.3 & 31.32 \\
& CT & Heart atrium         & 84.91 & 82.05 & 78.34 & 74.65 & 78.96 & 72.99 \\
& MRI & Heart ventricle      & 87.07 & 81.78 & 75.05 & 77.99 & 82.09 & 74.79 \\
& CT & Lung                 & 88.74 & 85.66 & 77.8 & 89.71 & 89.69 & 80.8 \\
& CT & Rib                  & 93.0 & 89.86 & 84.98 & 89.48 & 90.41 & 85.12 \\
& CT & Myocardium           & 90.69 & 83.63 & 75.39 & 83.84 & 87.99 & 81.16 \\
\end{tabular}}
\end{table*}
\begin{table*}[!tbh]
\renewcommand{\arraystretch}{1.3} 
\center
\caption{Class-wise results of \model{}-Pro, \model{}-Nano and MedSAMs on common anatomical structures and lesions. `MS-T' stands for MedSAM (Tight) while `MS-L' stands for MedSAM (Loose).}
\vspace{0.3cm}
\resizebox{1.0\textwidth}{!}{\begin{tabular}{lclcccccccc}
\toprule
\rowcolor{lightgray} & & & \multicolumn{4}{c}{\textbf{DSC}$\uparrow$} & \multicolumn{4}{c}{\textbf{NSD}$\uparrow$} \\
\rowcolor{lightgray} \multirow{-2}{*}{\textbf{Region}} & \multirow{-2}{*}{\textbf{Modality}} & \multirow{-2}{*}{\textbf{Anatomical Target}} & \textbf{\model{}-Pro} & \textbf{\model{}-Nano} & \textbf{MS-T} & \textbf{MS-L} & \textbf{\model{}-Pro} & \textbf{\model{}-Nano} & \textbf{MS-T} & \textbf{MS-L} \\ \midrule

\multirow{15}{*}{Abdomen} & CT & Adrenal gland        & 80.98 & 78.37 & 66.8 & 68.99 & 89.82 & 86.78 & 77.2 & 71.56 \\
& CT & Duodenum             & 78.3 & 77.39 & 69.33 & 65.04 & 71.67 & 69.59 & 61.6 & 55.76 \\
& CT & Gallbladder          & 80.89 & 78.69 & 87.66 & 82.76 & 78.07 & 74.24 & 83.82 & 73.54 \\
& CT & Kidney               & 93.6 & 90.17 & 89.1 & 78.46 & 90.57 & 85.68 & 81.01 & 53.16 \\
& CT & Liver                & 94.9 & 93.07 & 95.28 & 76.26 & 82.4 & 78.03 & 77.75 & 38.13 \\
& CT & Pancreas             & 86.03 & 85.06 & 77.1 & 60.02 & 76.96 & 74.58 & 61.52 & 43.42 \\
\bottomrule

\end{tabular}}
\label{tab:class_results_medsam}
\end{table*}
\begin{table*}[!tbh]
\renewcommand{\arraystretch}{1.3} 
\center
\small
\caption{Detailed comparison of \model{}-Pro, \model{}-Nano and BiomedParses on each dataset and each category. `BP' denotes BiomedParse. Subregions of anatomical structures are averaged and presented as a whole, {\em e.g.}, left and right kidney. }
\vspace{0.3cm}
\resizebox{1.0\textwidth}{!}{\begin{tabular}{llcccccccc}
\toprule
\rowcolor{lightgray} & & \multicolumn{4}{c}{\textbf{DSC}$\uparrow$} & \multicolumn{4}{c}{\textbf{NSD}$\uparrow$} \\
\rowcolor{lightgray} \multirow{-2}{*}{\textbf{Dataset}} & \multirow{-2}{*}{\textbf{Anatomical Target}} & \textbf{\model{}-Pro} & \textbf{\model{}-Nano} & \textbf{BP (Oracle)} & \textbf{BP} & \textbf{\model{}-Pro} & \textbf{\model{}-Nano} & \textbf{BP (Oracle)} & \textbf{BP} \\ 
\midrule

\multirow{2}{*}{ACDC (MRI)~\cite{ACDC}}& heart ventricle & 89.34 & 87.36 & 81.97 & 78.98 & 72.13 & 66.63 & 74.5 & 68.11 \\
& myocardium & 82.94 & 79.9 & 90.26 & 89.21 & 73.49 & 67.2 & 85.68 & 83.23 \\\hline

\multirow{12}{*}{AMOS22 CT~\cite{AMOS22}}& adrenal gland & 75.5 & 72.81 & 73.36 & 4.46 & 86.12 & 83.0 & 68.93 & 5.37 \\
& aorta & 92.25 & 92.76 & 94.12 & 62.17 & 91.41 & 92.24 & 84.74 & 40.22 \\
& duodenum & 78.28 & 75.67 & 77.66 & 23.29 & 69.46 & 65.45 & 55.8 & 13.38 \\
& esophagus & 83.73 & 82.25 & 83.47 & 12.26 & 85.67 & 83.8 & 70.7 & 9.08 \\
\bottomrule
\end{tabular}}
\label{tab:biomedparse_internal}
\end{table*}

\clearpage

\section{Detailed External Evaluation Results}

\begin{table*}[htbp]
\renewcommand{\arraystretch}{1.3} 
\center
\caption{\textbf{DSC} results of SAT-Pro, SAT-Nano, nnU-Nets, U-Mamba and SwinUNETR on two held-out dataset AbdomenAtlas 1.1~\cite{abdomenatlas} and LiQA~\cite{LiQA}. 
`PV \& SV' stands for Portal Vein And Splenic Vein.
The best results are bolded. Note that, MedSAM is interactive and semi-automatic method, while the rest are fully automatic.}
\vspace{0.3cm}
\label{tab:zeroshot_results_dsc}
\resizebox{1.0\textwidth}{!}{\begin{tabular}{lccccccc}
\toprule
\rowcolor{lightgray} \textbf{Category} & \textbf{SAT-Pro} & \textbf{SAT-Nano} & \textbf{nnU-Nets} & \textbf{U-Mamba} & \textbf{SwinUNETR} & \textbf{MedSAM} & \textbf{Oracle Box} \\
\midrule
Adrenal Gland       & \textbf{75.63} & 74.08 & 72.35 & 72.08 & 66.5 & 60.24 & 58.55 \\\hline
Aorta               & 88.27 & \textbf{88.46} & 81.93 & 81.59 & 77.65 & 85.37 & 83.61 \\\hline
Colon               & \textbf{65.9} & 65.64 & 65.79 & 65.16 & 65.2 & 49.12 & 45.24 \\\hline
Duodenum            & \textbf{75.48} & 72.33 & 69.15 & 70.04 & 61.46 & 64.01 & 59.71 \\\hline
Esophagus           & 73.16 & 73.04 & 64.35 & 64.68 & 58.65 & 72.26 & \textbf{76.07} \\\hline
Femur               & 70.52 & 53.02 & 64.48 & 64.9 & 61.84 & \textbf{81.85} & 81.33 \\\hline
Gallbladder         & 72.8 & 72.86 & 68.63 & 70.49 & 64.72 & \textbf{82.08} & 78.75 \\\hline
Intestine           & 76.42 & 75.8 & 77.09 & \textbf{77.34} & 74.79 & 62.59 & 55.65 \\\hline
Kidney              & 93.12 & 92.93 & 85.39 & 87.01 & 81.5 & \textbf{93.94} & 88.6 \\\hline
Liver               & \textbf{96.59} & 96.15 & 94.72 & 94.78 & 91.54 & 94.8 & 82.85 \\\hline
Lung                & \textbf{89.59} & 88.82 & 85.17 & 87.53 & 84.85 & 60.9 & 60.61 \\\hline
Pancreas            & \textbf{86.42} & 84.96 & 75.99 & 81.12 & 70.95 & 76.92 & 65.67 \\\hline
PV \& SV            & \textbf{55.94} & 54.32 & 54.84 & 55.52 & 52.96 & 12.14 & 11.52 \\\hline
Prostate            & 25.26 & 24.26 & 21.46 & 22.09 & 23.39 & 68.88 & \textbf{72.66} \\\hline
Rectum              & 63.98 & 25.25 & 50.89 & 59.86 & 48.92 & 77.84 & \textbf{80.48} \\\hline
Spleen              & \textbf{94.31} & 93.64 & 90.4 & 92.2 & 86.43 & 93.34 & 82.74 \\\hline
Stomach             & 81.52 & 79.26 & 85.2 & 86.11 & 78.6 & \textbf{87.63} & 80.66 \\\hline
Urinary Bladder     & 76.31 & 68.75 & 70.47 & 72.77 & 61.82 & \textbf{85.46} & 84.16 \\\hline
Liver (MR)          & \textbf{93.5} & 92.64 & 68.96 & 79.19 & 60.13 & 90.21 & 78.13 \\
\bottomrule
\end{tabular}}
\end{table*}


\begin{table*}[htbp]
\renewcommand{\arraystretch}{1.3} 
\center
\caption{\textbf{NSD} results of SAT-Pro, SAT-Nano, nnU-Nets, U-Mamba and SwinUNETR on two held-out dataset AbdomenAtlas 1.1~\cite{abdomenatlas} and LiQA~\cite{LiQA}. 
`PV \& SV' stands for Portal Vein And Splenic Vein.
The best results are bolded. Note that, MedSAM is interactive and semi-automatic method, while the rest are fully automatic.}
\vspace{0.3cm}
\label{tab:zeroshot_results}
\resizebox{1.0\textwidth}{!}{\begin{tabular}{lccccccc}
\toprule
\rowcolor{lightgray} \textbf{Category} & \textbf{SAT-Pro} & \textbf{SAT-Nano} & \textbf{nnU-Nets} & \textbf{U-Mamba} & \textbf{SwinUNETR} & \textbf{MedSAM} & \textbf{Oracle Box} \\
\midrule
Adrenal Gland       & \textbf{85.25} & 83.42 & 76.68 & 76.61 & 70.26 & 71.95 & 65.94 \\\hline
Aorta               & 87.81 & \textbf{88.29} & 73.8 & 73.85 & 68.25 & 83.95 & 78.12 \\\hline
Colon               & \textbf{49.04} & 48.9 & 41.08 & 40.38 & 40.09 & 31.7 & 26.53 \\\hline
Duodenum            & \textbf{64.43} & 59.31 & 53.48 & 54.65 & 45.49 & 52.73 & 46.12 \\\hline
Esophagus           & 72.86 & 72.45 & 57.73 & 58.42 & 52.45 & 72.52 & \textbf{75.63} \\\hline
Femur               & 60.82 & 43.97 & 54.57 & 55.28 & 51.69 & \textbf{69.39} & 66.84 \\\hline
Gallbladder         & 66.51 & 65.26 & 58.86 & 61.08 & 53.02 & \textbf{78.01} & 69.95 \\\hline
Intestine           & 59.73 & \textbf{59.77} & 57.15 & 56.91 & 55.55 & 42.14 & 29.82 \\\hline
Kidney              & \textbf{88.48} & 87.78 & 74.74 & 76.92 & 69.54 & 86.87 & 67.03 \\\hline
Liver               & \textbf{81.61} & 78.63 & 72.46 & 74.1 & 66.61 & 73.42 & 48.79 \\\hline
Lung                & \textbf{80.14} & 78.0 & 68.52 & 73.74 & 65.37 & 34.17 & 29.83 \\\hline
Pancreas            & \textbf{79.2} & 75.52 & 60.52 & 65.72 & 54.07 & 62.14 & 49.06 \\\hline
PV \& SV            & \textbf{62.56} & 60.71 & 56.58 & 57.32 & 55.16 & 18.51 & 16.44 \\\hline
Prostate            & 19.86 & 17.69 & 14.76 & 16.06 & 15.92 & 60.28 & \textbf{62.09} \\\hline
Rectum              & 55.19 & 25.25 & 39.48 & 48.18 & 37.01 & 72.25 & \textbf{73.48} \\\hline
Spleen              & \textbf{90.17} & 87.96 & 80.48 & 83.06 & 73.91 & 85.0 & 62.17 \\\hline
Stomach             & \textbf{65.72} & 60.03 & 63.46 & 65.39 & 54.09 & 64.53 & 50.81 \\\hline
Urinary Bladder     & 67.11 & 58.57 & 57.66 & 60.3 & 47.86 & \textbf{76.91} & 71.0 \\\hline
Liver (MR)          & \textbf{68.69} & 62.79 & 48.13 & 52.88 & 38.64 & 63.92 & 43.4 \\
\bottomrule
\end{tabular}}
\end{table*}

\begin{figure}[htpb]
    \centering
    \includegraphics[width = 0.9\textwidth]{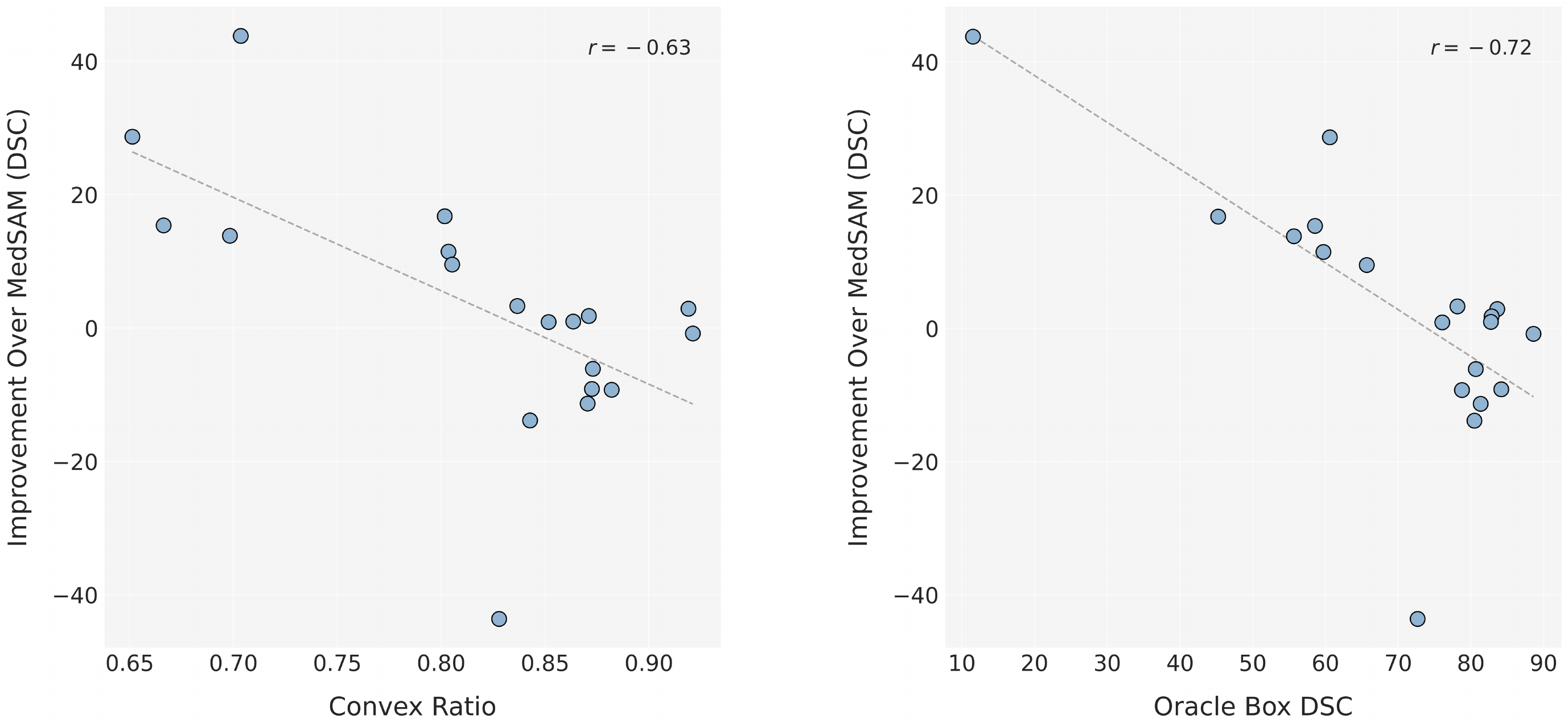}
    \vspace{10pt}
    \caption{Scatter plots comparing the performance improvement of \model{}-Pro over MedSAM on 19 categories in external evaluation (DSC score), with two irregularity metrics: convex ratio and oracle box DSC. Each point represents an anatomical structure or lesion, with a fitted line illustrating the trend.}
    \label{fig:zeroshot_irregularity}
\end{figure}
\clearpage

\begin{figure}[htpb]
    \centering
    \includegraphics[width = 0.9\textwidth]{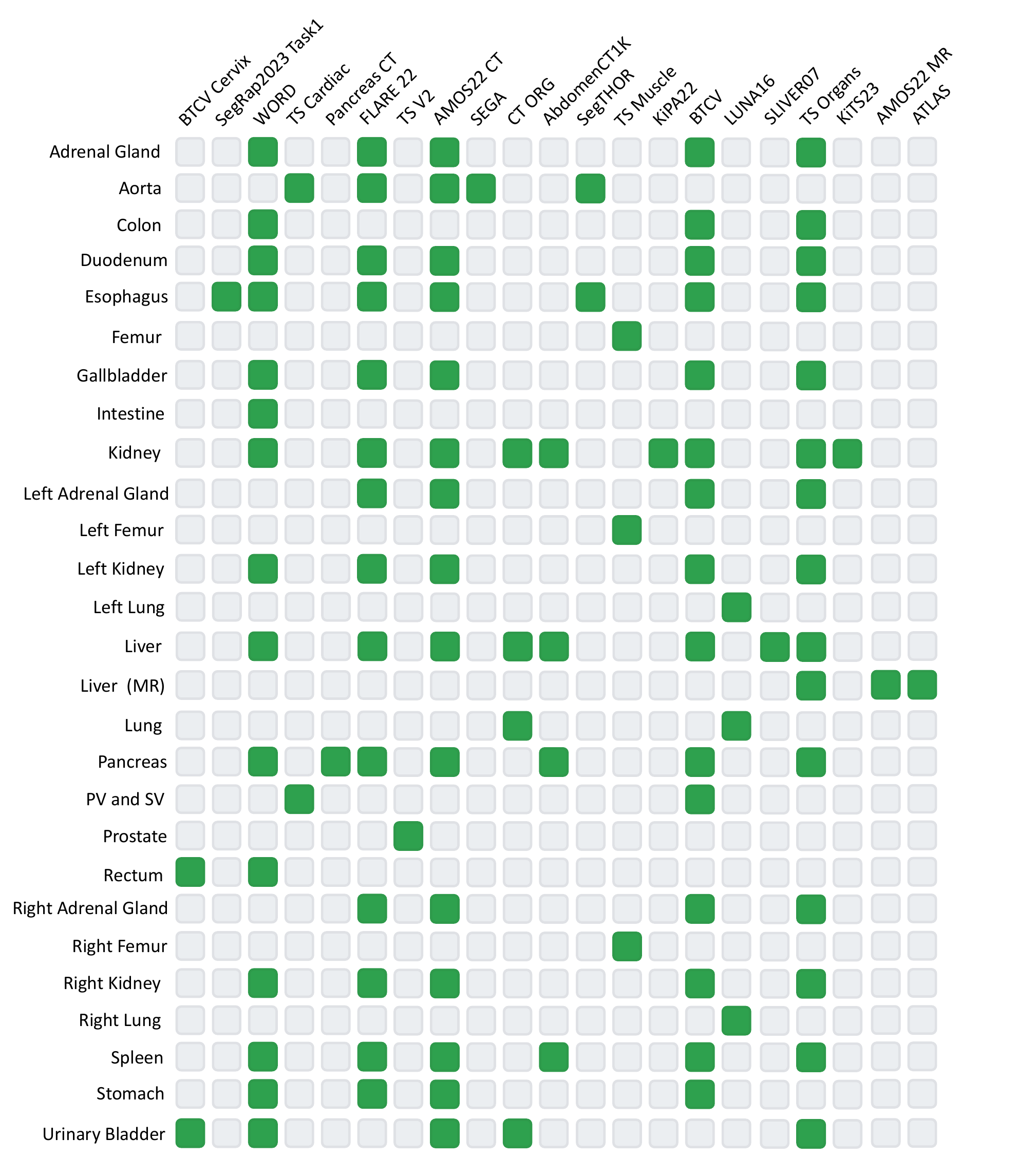}
    \vspace{10pt}
    \caption{The detailed transfer mapping of specialist models from 21 datasets to 2 held-out datasets, in the external evaluation. Liver (MR) is from LiQA, while the other categories are from AbdomenAtlas 1.1.}
    \label{fig:zeroshot_mapping}
\end{figure}
\clearpage

\section{Extended Ablation Studies}

\subsection{Knowledge Source Ablation}
\label{sec:ab_study_knowledge_source}

We have shown the benefits of knowledge enhancement through a comprehensive ablation study in Section 2.5 in the main manuscript.
In this section, we further validate the necessity of both textual and visual knowledge introduced in Section 4.1 in the main manuscript.
Specifically, we train a \model{}-Nano variant with only textual knowledge used in knowledge injection. 
As shown in Supplementary Table~\ref{tab:knowledge_data_ablation}, we have the following observations: 
(i) only utilizing textual knowledge in knowledge injection leads to superior segmentation performance over off-the-shelf language models, for example, MedCPT and BERT-Base. 
While MedCPT, trained on medical corpus, slightly outperforms BERT-Base; 
(ii) combining multimodal knowledge in pre-training further improves the performance and achieves the best results. These findings justify the benefit of both visual and textual domain knowledge in universal segmentation.

\begin{table}[htbp]
\renewcommand{\arraystretch}{1.3} 
\center
\caption{Ablation study on knowledge data in knowledge injection, and comparison with variants with other text encoder. `Multimodal Knowledge' stands for using both visual and textual knowledge data, {\em i.e.}, the whole knowledge tree. While `Text Knowledge Only' stands for using the textual knowledge data only. Results are merged by different regions of the human body and lesions. 
H\&N: head and neck, LL: lower limb, UL: upper limb, All: average over all the 429 classes. 
The best results are \textbf{bolded}.}
\vspace{0.3cm}
\resizebox{\textwidth}{!}{
\begin{tabular}{clccccccccc|c}
\toprule
\rowcolor{lightgray} \textbf{Metric} & \textbf{Method} & \textbf{Brain} & \textbf{H\&N} & \textbf{UL} & \textbf{Thorax} & \textbf{Spine} & \textbf{Abdomen} & \textbf{LL} & \textbf{Pelvis} & \textbf{Lesion} & \textbf{All} \\ 
\midrule

\multirow{5}{*}{DSC$\uparrow$} & Multimodal Knowledge & 75.56 & \textbf{78.46} & \textbf{89.89} & \textbf{87.51} & \textbf{79.44} & \textbf{81.31} & \textbf{83.22} & \textbf{91.77} & 42.72 & \textbf{79.48} \\ 

& Text Knowledge Only & \textbf{75.65} & 78.34 & 87.75 & 84.61 & 78.80 & 79.42 & 82.65 & 89.52 & \textbf{42.92} & 78.5 \\

& MedCPT & 74.02 & 77.33 & 81.32 & 85.22 & 77.82 & 81.28 & 81.98 & 89.84 & 41.95 & 77.94 \\

& CLIP & 74.45 & 77.75 & 83.78 & 85.96 & 75.67 & 78.82 & 81.26 & 91.27 & 41.32 & 77.84 \\

& BERT-Base & 74.59 & 75.48 & 84.08 & 85.96 & 76.10 & 80.97 & 81.90 & 85.60 & 42.65 & 77.52 \\

\midrule

\multirow{5}{*}{NSD$\uparrow$} & Multimodal Knowledge & \textbf{75.50} & \textbf{84.75} & \textbf{91.06} & \textbf{83.19} & \textbf{78.30} & \textbf{71.86} & \textbf{82.73} & \textbf{88.56} & \textbf{38.68} & \textbf{78.35} \\ 

& Text Knowledge Only & 75.29 & 84.67 & 89.22 & 80.29 & 78.06 & 69.77 & 81.89 & 85.88 & 38.68 & 77.33 \\

& MedCPT & 72.13 & 82.80 & 82.54 & 80.64 & 76.10 & 70.35 & 80.01 & 85.49 & 36.45 & 75.70 \\

& CLIP & 73.01 & 83.26 & 85.27 & 81.66 & 74.62 & 68.49 & 79.83 & 87.21 & 36.61 & 75.99 \\

& BERT-Base & 72.86 & 79.99 & 85.36 & 81.74 & 74.06 & 70.08 & 80.04 & 80.98 & 37.33 & 75.18 \\

\bottomrule

\end{tabular}
}
\label{tab:knowledge_data_ablation}
\end{table}

\subsection{Visual Backbone Ablation}
\label{sec:ab_study_vision}

\begin{figure}[t]
    \centering
    \includegraphics[width = 0.9\textwidth]{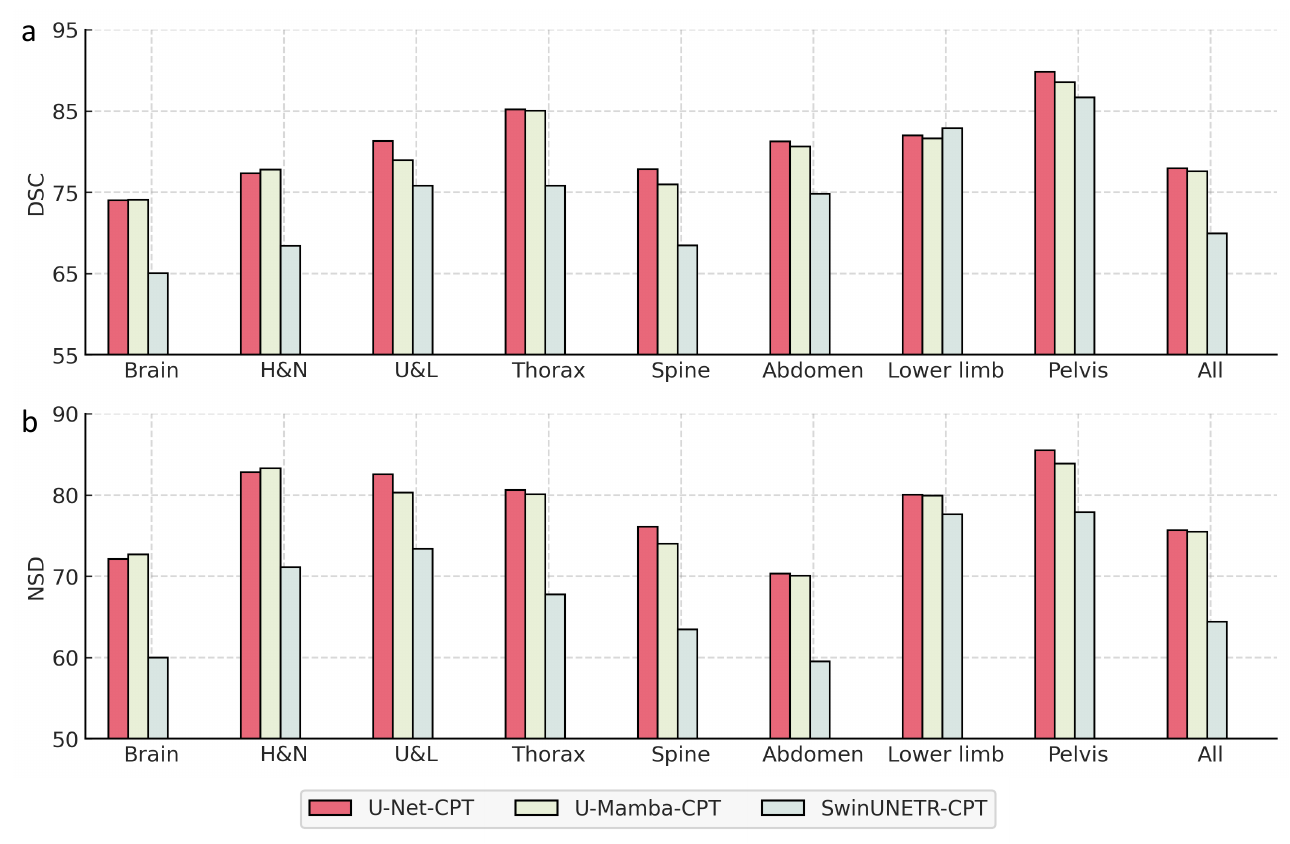}
    \vspace{0.1cm}
    \caption{\textbf{Evaluations on \samdataset{}-Nano variants with different visual backbones.} `All' denote the average scores over all the classes(n=429), including lesion classes. \textbf{a}, DSC comparison; \textbf{b}, NSD comparison. U-Net-CPT and U-Mamba-CPT perform very close, while both surpass SwinUNETR-CPT considerably.}
    \label{fig:vis_ab_hist}
\end{figure}

In this section, we conduct experiments on \textbf{\samdataset{}-Nano} dataset, and discuss the effect of visual backbone.
We investigate three different backbones, 
namely, CNN-based U-Net, SwinUNETR~\cite{hatamizadeh2021swin}, and U-Mamba~\cite{gu2023mamba}. We configure three \textbf{\model{}-Nano} variants with SwinUNETR~(107M), U-Mamba~(114M), and U-Net~(110M) of comparable size: 
For SwinUNETR, we use the same hyperparameters as in the official implementation~\cite{SwinUNetr}. For U-Mamba, we refer to the official implementation~\cite{ma2024u}, preserving the configuration of the U-Net and only adding Mamba layers at the end of the last 3 blocks of the U-Net encoder.

To avoid repeating multimodal knowledge injection for each visual backbone, 
we use MedCPT~\cite{jin2023medcpt} as the text encoder for all these variants, without loss of generality. MedCPT is a text encoder trained on 255 million in-house user click logs from PubMed~\cite{roberts2001pubmed} and shows state-of-the-art performance on various biomedical language tasks, such as medical language retrieval.

Thus, we denote these variants as U-Net-CPT, SwinUNETR-CPT, and U-Mamba-CPT.
We demonstrate the region-wise evaluation results in Supplementary Figure~\ref{fig:vis_ab_hist}. 
U-Net-CPT outperforms U-Mamba-CPT slightly on both DSC (0.35) and NSD (0.22) scores averaged over all classes.
However, both U-Net-CPT and U-Mamba-CPT exceed SwinUNETR-CPT by a significant margin. These observations confirm our choice of U-Net as the visual backbone of \textbf{\model{}}. 
More detailed comparisons on each dataset and class are in Supplementary Tables~\ref{tab:visual_ab_class_results},~\ref{tab:visual_ab_class_results2},~\ref{tab:visual_ab_dataset_results}, and~\ref{tab:visual_ab_dataset_results2}.

\begin{table*}[!tbh]
\renewcommand{\arraystretch}{1.3} 
\center
\small
\caption{Class-wise comparison of three \model{}-Nano variants with different visual backbones on common anatomical structures in abdomen, brain, spine, and pelvis. `U-Net' denotes the \model{}-Nano based on U-Net; `U-Mamba' denotes the variant based on U-Mamba; `SwinUNETR' denotes the variant based on SwinUNETR.}
\vspace{0.3cm}
\resizebox{1.0\textwidth}{!}{\begin{tabular}{lclcccccc}
\toprule
\rowcolor{lightgray} & & & \multicolumn{3}{c}{\textbf{DSC}$\uparrow$} & \multicolumn{3}{c}{\textbf{NSD}$\uparrow$} \\
\rowcolor{lightgray} \multirow{-2}{*}{\textbf{Region}} & \multirow{-2}{*}{\textbf{Modality}} & \multirow{-2}{*}{\textbf{Anatomical Target}} & \textbf{U-Net} & \textbf{U-Mamba} & \textbf{SwinUNETR} & \textbf{U-Net} & \textbf{U-Mamba} & \textbf{SwinUNETR} \\ 
\midrule

\multirow{16}{*}{Abdomen} & CT & Adrenal gland        & 78.54 & 79.03 & 71.97 & 86.94 & 87.41 & 79.02 \\
& CT & Duodenum             & 75.88 & 73.89 & 66.05 & 65.15 & 63.77 & 50.51 \\
& CT & Gallbladder          & 78.48 & 77.9 & 67.31 & 72.19 & 71.51 & 54.17 \\
& CT & Inferior vena cava   & 88.41 & 87.35 & 79.98 & 84.05 & 82.96 & 66.37 \\
& CT & Intestine            & 82.72 & 83.04 & 65.65 & 63.93 & 65.84 & 37.69 \\
& CT & Kidney               & 93.41 & 93.25 & 90.74 & 88.69 & 88.64 & 82.23 \\
& CT & Liver                & 94.97 & 95.42 & 92.35 & 79.47 & 81.66 & 65.55 \\
& CT & Pancreas             & 82.12 & 83.44 & 77.43 & 70.91 & 73.35 & 60.9 \\
& CT & Small bowel          & 78.24 & 71.92 & 57.71 & 69.27 & 62.77 & 41.44 \\
& CT & Spleen               & 93.8 & 94.36 & 90.25 & 88.82 & 90.02 & 76.55 \\
& CT & Stomach              & 87.93 & 87.39 & 79.88 & 71.14 & 71.46 & 48.43 \\
& MRI & Adrenal gland        & 62.99 & 63.86 & 59.31 & 74.7 & 75.85 & 69.1 \\
& MRI & Kidney               & 90.34 & 90.6 & 88.76 & 69.4 & 70.8 & 65.11 \\
& MRI & Liver                & 90.69 & 90.34 & 89.57 & 61.02 & 61.67 & 55.64 \\
& MRI & Pancreas             & 82.78 & 81.32 & 75.76 & 69.41 & 68.49 & 58.7 \\
& MRI & Spleen               & 82.11 & 78.48 & 81.21 & 61.7 & 61.9 & 55.71 \\\hline

\multirow{8}{*}{Brain} & CT & Brainstem            & 85.2 & 84.33 & 82.35 & 69.55 & 68.91 & 61.72 \\
& CT & Hippocampus          & 77.17 & 70.85 & 74.46 & 77.51 & 71.06 & 72.25 \\
& CT & Temporal lobe        & 94.26 & 93.81 & 91.0 & 83.86 & 82.51 & 71.79 \\
& MRI & Brain ventricle      & 75.86 & 76.04 & 65.04 & 83.23 & 83.34 & 70.19 \\
& MRI & Cerebellum           & 88.72 & 89.18 & 80.13 & 80.23 & 82.15 & 57.62 \\
& MRI & Parietal lobe        & 74.47 & 73.52 & 64.93 & 53.83 & 54.02 & 42.72 \\
& MRI & Optic radiation      & 60.41 & 59.83 & 61.32 & 52.84 & 51.7 & 52.9 \\
& MRI & Thalamus             & 84.72 & 83.77 & 79.08 & 83.69 & 81.27 & 72.55 \\\hline

\multirow{8}{*}{Spine} & CT & Lumbar vertebrae     & 73.88 & 73.37 & 67.38 & 72.58 & 72.01 & 61.53 \\
& CT & Sacrum               & 92.81 & 91.31 & 89.01 & 92.75 & 91.47 & 85.08 \\
& CT & Spinal cord          & 79.99 & 82.07 & 63.82 & 82.54 & 86.83 & 61.75 \\
& CT & Thoracic vertebrae   & 79.04 & 75.64 & 62.92 & 78.67 & 75.49 & 58.4 \\
& MRI & Intervertebral discs & 88.21 & 87.76 & 85.97 & 90.92 & 89.81 & 87.49 \\
& MRI & Lumbar vertebrae     & 81.49 & 81.15 & 74.77 & 70.83 & 69.76 & 59.43 \\
& MRI & Sacrum               & 81.58 & 81.45 & 74.59 & 73.0 & 74.14 & 61.25 \\
& MRI & Thoracic vertebrae   & 59.57 & 65.46 & 52.99 & 53.72 & 58.69 & 43.84 \\\hline
         
\multirow{8}{*}{Pelvis} & CT & Gluteus maximus      & 95.39 & 94.64 & 92.54 & 90.19 & 89.45 & 81.87 \\
& CT & Gluteus medius       & 90.19 & 91.09 & 91.19 & 84.74 & 84.93 & 80.59 \\
& CT & Gluteus minimus      & 91.85 & 90.92 & 86.39 & 90.26 & 88.97 & 80.41 \\
& CT & Hip                  & 92.95 & 92.91 & 92.99 & 93.24 & 93.1 & 90.72 \\
& CT & Iliopsoas            & 89.98 & 82.42 & 83.17 & 88.72 & 81.21 & 74.85 \\
& CT & Iliac vena           & 89.17 & 89.31 & 84.71 & 90.69 & 90.6 & 83.26 \\
& CT & Urinary bladder      & 83.71 & 84.94 & 78.64 & 63.38 & 64.83 & 52.45 \\
& MRI & Prostate             & 84.23 & 80.32 & 79.52 & 57.34 & 49.03 & 44.69 \\\bottomrule

\end{tabular}}
\label{tab:visual_ab_class_results}
\end{table*}

\begin{table*}[!htp]
\renewcommand{\arraystretch}{1.3} 
\center
\small
\caption{Class-wise comparison of three \model{}-Nano variants with different visual backbones on common anatomical structures in head and neck, lower limb, upper limb and whole body, and on common lesions. `U-Net' denotes the \model{}-Nano based on U-Net; `U-Mamba' denotes the variant based on U-Mamba; `SwinUNETR' denotes the variant based on SwinUNETR.}
\vspace{0.3cm}
\resizebox{1.0\textwidth}{!}{\begin{tabular}{lclcccccc}
\toprule
\rowcolor{lightgray}                            &                               &                            
& \multicolumn{3}{c}{\textbf{DSC}$\uparrow$} & \multicolumn{3}{c}{\textbf{NSD}$\uparrow$} \\
\rowcolor{lightgray} \multirow{-2}{*}{\textbf{Region}} & \multirow{-2}{*}{\textbf{Modality}} & \multirow{-2}{*}{\textbf{Anatomical Target}} & \textbf{U-Net} & \textbf{U-Mamba} & \textbf{SwinUNETR} & \textbf{U-Net} & \textbf{U-Mamba} & \textbf{SwinUNETR} \\ 
\midrule

\multirow{16}{*}{Head and neck} & CT & Brain                & 96.43 & 91.95 & 95.61 & 91.38 & 88.45 & 82.17 \\
& CT & Carotid artery       & 60.76 & 62.78 & 31.49 & 68.93 & 72.42 & 33.11 \\
& CT & Cervical esophagus   & 64.13 & 59.58 & 41.62 & 68.21 & 59.77 & 41.72 \\
& CT & Cochlea              & 68.99 & 69.07 & 66.73 & 89.94 & 89.45 & 89.35 \\
& CT & Eustachian tube bone & 81.81 & 82.44 & 78.51 & 97.1 & 97.6 & 96.08 \\
& CT & Eyeball              & 84.18 & 84.34 & 79.12 & 85.85 & 86.24 & 76.38 \\
& CT & Lacrimal gland       & 47.32 & 48.27 & 0.0 & 65.35 & 66.95 & 0.0 \\
& CT & Lens                 & 73.9 & 75.35 & 63.77 & 91.87 & 91.69 & 82.36 \\
& CT & Mandible             & 93.26 & 93.58 & 88.38 & 95.28 & 95.52 & 85.75 \\
& CT & Middle ear           & 82.52 & 83.53 & 81.2 & 88.51 & 89.91 & 88.92 \\
& CT & Optic nerve          & 70.52 & 70.77 & 62.75 & 88.25 & 88.33 & 78.39 \\
& CT & Parotid gland        & 83.89 & 84.66 & 80.64 & 73.01 & 74.85 & 63.7 \\
& CT & Submandibular gland  & 78.91 & 79.45 & 75.78 & 75.9 & 76.93 & 68.87 \\
& CT & Tympanic cavity      & 81.4 & 80.87 & 76.85 & 94.7 & 93.91 & 92.3 \\
& CT & Thyroid gland        & 82.79 & 83.96 & 74.85 & 82.09 & 84.53 & 64.63 \\
& MRI & Brain                & 94.72 & 94.92 & 94.41 & 55.57 & 57.45 & 54.3 \\\hline

\multirow{5}{*}{Lower limb} & CT & Head of femur        & 88.21 & 89.21 & 85.96 & 75.11 & 77.22 & 66.86 \\
& MRI & Femur bone           & 95.43 & 95.24 & 94.15 & 87.43 & 85.94 & 81.78 \\
& MRI & Femur cartilage      & 67.85 & 66.98 & 63.02 & 85.83 & 85.94 & 82.27 \\
& MRI & Tibia bone           & 96.25 & 96.02 & 94.26 & 92.24 & 91.13 & 84.68 \\
& MRI & Tibia cartilage      & 65.47 & 67.24 & 63.09 & 87.57 & 89.2 & 85.46 \\\hline
            
\multirow{7}{*}{Thorax} & CT & Autochthon           & 91.97 & 90.26 & 87.01 & 85.79 & 83.89 & 72.14 \\
& CT & Heart atrium         & 90.16 & 87.1 & 86.02 & 76.89 & 72.57 & 65.35 \\
& MRI & Heart ventricle      & 87.84 & 83.78 & 84.54 & 71.1 & 64.83 & 61.81 \\
& CT & Lung                 & 90.05 & 88.22 & 85.26 & 80.81 & 78.71 & 66.04 \\
& CT & Rib                  & 82.54 & 84.08 & 68.12 & 86.6 & 87.62 & 71.14 \\
& CT & Myocardium           & 84.28 & 80.6 & 79.67 & 75.1 & 68.33 & 62.09 \\
& CT & Thoracic cavity      & 94.64 & 94.82 & 93.81 & 70.21 & 69.91 & 68.41 \\ \hline
   
\multirow{3}{*}{Upper limb} & CT & Clavicle             & 82.82 & 78.93 & 86.24 & 84.3 & 80.15 & 85.73 \\
& CT & Humerus              & 80.38 & 77.63 & 65.62 & 80.17 & 78.08 & 58.56 \\
& CT & Scapula              & 80.47 & 80.98 & 75.47 & 83.45 & 83.91 & 77.06 \\\hline

\multirow{3}{*}{Abnormal} & CT & Lung nodule          & 5.14 & 14.9 & 14.17 & 8.01 & 19.03 & 15.67 \\
& MRI & Myocardial edema     & 9.32 & 12.79 & 15.9 & 14.48 & 18.75 & 20.06 \\
& MRI & Stroke               & 49.0 & 49.95 & 47.74 & 46.05 & 47.35 & 43.42 \\\bottomrule

\end{tabular}}
\label{tab:visual_ab_class_results2}
\end{table*}
\begin{table*}[!tbh]
\renewcommand{\arraystretch}{1.3} 
\center
\caption{Dataset-wise Results of three \model{}-Nano variants with different visual backbones on 49 datasets in \samdataset{}-Nano. `U-Net' denotes the \model{}-Nano based on U-Net; `U-Mamba' denotes the variant based on U-Mamba; `SwinUNETR' denotes the variant based on SwinUNETR.}
\vspace{0.3cm}
\label{tab:visual_ab_dataset_results}
\resizebox{1.0\textwidth}{!}{\begin{tabular}{lcccccc}

\toprule
\rowcolor{lightgray} \multicolumn{1}{l}{} & \multicolumn{3}{c}{\textbf{DSC$\uparrow$}} & \multicolumn{3}{c}{\textbf{NSD$\uparrow$}} \\
\rowcolor{lightgray} \multicolumn{1}{l}{\multirow{-2}{*}{\textbf{Dataset}}} & \multicolumn{1}{c}{\textbf{U-Net}} & \multicolumn{1}{c}{\textbf{U-Mamba}} & \multicolumn{1}{c}{\textbf{SwinUNETR}} & \multicolumn{1}{c}{\textbf{U-Net}} & \multicolumn{1}{c}{\textbf{U-Mamba}} & \multicolumn{1}{c}{\textbf{SwinUNETR}} \\
\midrule

AbdomenCT1K~\cite{AbdomenCT1K}                               & 92.49 & 92.67 & 89.88 & 82.49 & 83.6 & 74.1 \\\hline
ACDC~\cite{ACDC}                                             & 85.3 & 83.82 & 82.3 & 66.39 & 63.68 & 59.04 \\\hline
AMOS22 CT~\cite{AMOS22}                                   & 85.81 & 84.71 & 78.05 & 82.25 & 79.65 & 66.03 \\\hline
AMOS22 MRI~\cite{AMOS22}                                 & 80.77 & 80.53 & 75.47 & 75.39 & 75.63 & 66.96 \\\hline
ATLAS~\cite{ATLAS}                                           & 67.9 & 64.5 & 63.82 & 39.22 & 39.33 & 36.07 \\\hline
ATLASR2~\cite{ATLASR2}                                       & 53.8 & 55.96 & 51.66 & 49.68 & 52.06 & 46.06 \\\hline
Brain Atlas~\cite{Brain_Atlas}                               & 74.77 & 75.05 & 64.99 & 72.64 & 73.53 & 59.49 \\\hline
BrainPTM~\cite{BrainPTM}                                     & 66.25 & 65.09 & 64.96 & 51.95 & 51.08 & 49.48 \\\hline
CHAOS MRI~\cite{CHAOS}                                   & 82.59 & 81.14 & 81.5 & 46.68 & 48.2 & 42.96 \\\hline
CMRxMotion~\cite{CMRxMotion}                                 & 87.01 & 86.47 & 85.48 & 68.85 & 67.58 & 63.75 \\\hline
Couinaud~\cite{Couinaud}                         & 81.36 & 80.87 & 75.58 & 54.62 & 54.21 & 44.92 \\\hline
CrossMoDA2021~\cite{CrossMoDA2021}                           & 71.91 & 73.37 & 68.7 & 90.98 & 92.04 & 89.86 \\\hline
CT-ORG~\cite{CTORG}                                         & 89.82 & 88.75 & 87.65 & 75.9 & 75.4 & 70.02 \\\hline
CTPelvic1K~\cite{CTPelvic1K}                                 & 94.76 & 95.24 & 92.58 & 95.91 & 96.29 & 90.89 \\\hline
FeTA2022~\cite{FeTA2022}                                     & 68.98 & 69.84 & 60.57 & 75.31 & 76.54 & 65.0 \\\hline
FLARE22~\cite{FLARE22}                                       & 88.97 & 89.44 & 83.05 & 85.37 & 86.02 & 73.77 \\\hline
FUMPE~\cite{FUMPE}                                           & 36.11 & 34.21 & 26.01 & 31.91 & 32.15 & 24.98 \\\hline
HAN Seg~\cite{HANSeg}                                       & 68.95 & 69.89 & 55.67 & 73.44 & 74.97 & 55.62 \\\hline
Instance22~\cite{INSTANCE}                                 & 61.0 & 58.64 & 36.44 & 51.65 & 46.78 & 26.63 \\\hline
ISLES2022~\cite{ISLES2022}                                   & 44.2 & 43.95 & 43.82 & 42.41 & 42.63 & 40.77 \\\hline
KiPA22~\cite{KiPA22}                                         & 67.52 & 66.34 & 61.49 & 65.89 & 63.84 & 57.35 \\\hline
KiTS23~\cite{KiTS23}                                         & 53.49 & 54.5 & 49.75 & 43.66 & 45.48 & 38.41 \\\hline
LAScarQS2022 Task1~\cite{LAScarQS2022}                     & 65.6 & 67.08 & 64.35 & 73.42 & 75.82 & 69.96 \\\hline
LAScarQS2022 Task2~\cite{LAScarQS2022}                     & 88.78 & 87.52 & 86.06 & 69.87 & 68.91 & 62.89 \\\hline
LNDb~\cite{LNDb}                                             & 5.14 & 14.9 & 14.17 & 8.01 & 19.03 & 15.67 \\\hline
LUNA16~\cite{LUNA16}                                         & 95.92 & 96.37 & 95.28 & 93.02 & 93.39 & 87.52 \\\hline
MM\-WHS CT~\cite{MMWHS}                                   & 88.23 & 85.58 & 85.67 & 69.85 & 64.93 & 62.98 \\\hline
MM\-WHS MRI~\cite{MMWHS}                                 & 85.82 & 85.17 & 82.15 & 66.85 & 66.85 & 59.54 \\\hline
MRSpineSeg~\cite{MRSpineSeg}                                 & 76.21 & 76.0 & 68.33 & 72.45 & 71.54 & 61.29 \\\hline
MyoPS2020~\cite{MyoPS2020}                                   & 59.54 & 60.16 & 59.79 & 40.68 & 41.59 & 40.49 \\\hline
NSCLC~\cite{NSCLC}                                           & 74.17 & 75.28 & 73.66 & 56.99 & 57.81 & 55.93 \\\hline
Pancreas CT~\cite{PancreasCT}                               & 83.79 & 84.52 & 80.61 & 72.39 & 74.27 & 64.5 \\\hline
PARSE2022~\cite{PARSE2022}                                   & 68.56 & 68.98 & 66.84 & 54.16 & 62.3 & 50.61 \\\hline
PDDCA~\cite{PDDCA}                                           & 73.44 & 74.41 & 66.88 & 74.29 & 76.28 & 62.7 \\\hline
PROMISE12~\cite{PROMISE12}                                   & 84.23 & 80.32 & 79.52 & 57.34 & 49.03 & 44.69 \\
\bottomrule
\end{tabular}}
\end{table*}

\begin{table*}[!tbh]
\renewcommand{\arraystretch}{1.3} 
\center
\caption{(Continued) Dataset-wise results of three \model{}-Nano variants with different visual backbones on 49 datasets in \samdataset{}-Nano. `U-Net' denotes the \model{}-Nano based on U-Net; `U-Mamba' denotes the variant based on U-Mamba; `SwinUNETR' denotes the variant based on SwinUNETR.}
\vspace{0.3cm}

\label{tab:visual_ab_dataset_results2}
\resizebox{1.0\textwidth}{!}{\begin{tabular}{lcccccc}

\toprule
\rowcolor{lightgray} \multicolumn{1}{l}{} & \multicolumn{3}{c}{\textbf{DSC$\uparrow$}} & \multicolumn{3}{c}{\textbf{NSD$\uparrow$}} \\
\rowcolor{lightgray} \multicolumn{1}{l}{\multirow{-2}{*}{\textbf{Dataset}}} & \multicolumn{1}{c}{\textbf{U-Net}} & \multicolumn{1}{c}{\textbf{U-Mamba}} & \multicolumn{1}{c}{\textbf{SwinUNETR}} & \multicolumn{1}{c}{\textbf{U-Net}} & \multicolumn{1}{c}{\textbf{U-Mamba}} & \multicolumn{1}{c}{\textbf{SwinUNETR}} \\
\midrule
SEGA~\cite{SEGA}                                             & 77.68 & 78.54 & 77.95 & 63.48 & 66.03 & 65.03 \\\hline
SegRap2023 Task1~\cite{SegRap2023}                     & 85.12 & 84.71 & 80.8 & 89.72 & 89.12 & 83.27 \\\hline
SegRap2023 Task2~\cite{SegRap2023}                     & 64.95 & 65.61 & 60.77 & 43.85 & 44.5 & 38.78 \\\hline
SegTHOR~\cite{SegTHOR}                                       & 84.59 & 84.73 & 79.84 & 67.75 & 69.23 & 59.19 \\\hline
SKI10~\cite{SKI10}                                           & 81.25 & 81.37 & 78.63 & 88.27 & 88.06 & 83.55 \\\hline
SLIVER07~\cite{SLIVER07}                                     & 96.59 & 96.94 & 94.96 & 80.76 & 84.35 & 69.25 \\\hline
TS Cardiac~\cite{Totalsegmentator}     & 86.96 & 83.72 & 81.52 & 83.04 & 79.93 & 72.13 \\\hline
TS Muscles~\cite{Totalsegmentator}     & 87.04 & 85.07 & 84.61 & 85.15 & 83.08 & 78.19 \\\hline
TS Organs~\cite{Totalsegmentator}       & 84.13 & 82.36 & 77.69 & 78.71 & 76.86 & 65.08 \\\hline
TS Ribs~\cite{Totalsegmentator}           & 82.72 & 84.1 & 68.65 & 86.86 & 87.7 & 71.88 \\\hline
TS Vertebrae~\cite{Totalsegmentator} & 83.25 & 82.04 & 70.25 & 83.5 & 82.43 & 66.1 \\\hline
VerSe~\cite{VerSe}                                           & 76.51 & 70.91 & 68.36 & 75.56 & 70.0 & 65.09 \\\hline
WMH~\cite{WMH} & 62.07 & 63.85 & 61.82 & 75.69 & 77.98 & 75.58 \\\hline
WORD~\cite{WORD}                                             & 84.79 & 85.25 & 78.6 & 72.42 & 73.89 & 59.68 \\
\bottomrule

\end{tabular}}
\end{table*}

\clearpage
\newpage

\subsection{Query Decoder Ablation}
\label{sec:ab_study_query_decoder}

{
\color{black}
As illustrated in Equation 8 in the main manuscript, we devise the query decoder to adapt the text prompt to specific image scan,
{\em i.e.}, updating the text embedding of one anatomical target by interacting with the specific multi-scale visual feature for update.
To quantitatively validate its effectiveness, we conducted a comprehensive ablation study across 16 diverse datasets encompassing both CT and MRI modalities, covering 6 human body regions and 119 diverse categories.
We trained two SAT-Nano variants {\em with} and {\em without} the query decoder, while all other parameters and hyperparameters are kept identical. 

As shown in Supplementary Figure~\ref{fig:Query_Decoder}, removing the query decoder causes a consistent performance drop on all datasets.
The average performance drop is 3.40 on DSC score and 3.48 on NSD score.
It validates that the transformer decoder is effective on large-vocabulary segmentation. 
\begin{figure}[htbp]
    \centering
    \includegraphics[width = \textwidth]{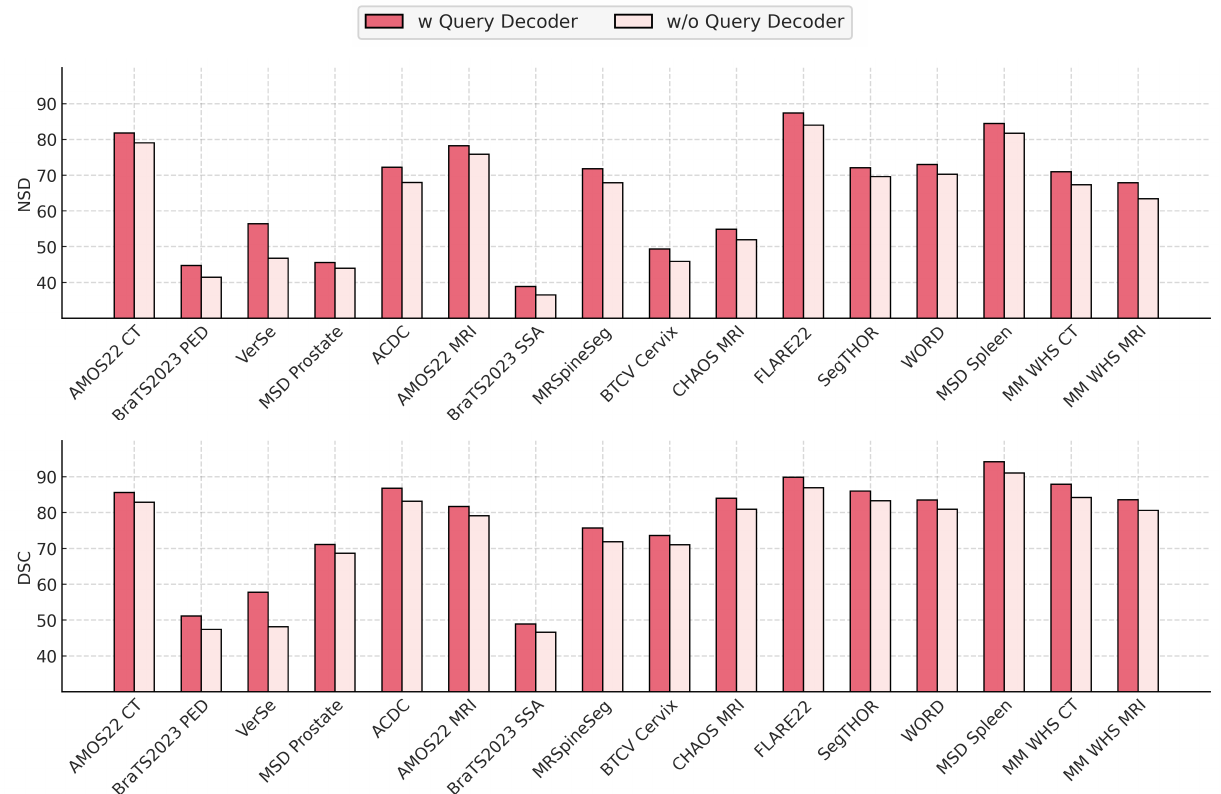}
    \vspace{.2cm}
    \caption{
    \textcolor{black}{
    \textbf{Ablation study on the query decoder.} Two SAT-Nano variants (with/without query decoder) are trained and evaluated on 16 representation datasets.}
    }
    \label{fig:Query_Decoder}
\end{figure}
}

\newpage

\subsection{Text Encoder Ablation Details}

\begin{table*}[!htp]
\renewcommand{\arraystretch}{1.3} 
\center
\small
\caption{Class-wise results of four \model{}-Nano variants with different text encoders on common anatomical structures in thorax, limbs, spine, pelvis, whole body and common lesions. `Ours' denotes the \model{}-Nano prompted with the text encoder pre-trained on our multimodal medical knowledge graph; `MedCPT' denotes the variant with MedCPT as text encoder; `Clip' denotes the variant with Clip text encoder; `BB' denotes the variant with BERT-base as text encoder.}
\vspace{0.3cm}
\resizebox{1.0\textwidth}{!}{\begin{tabular}{lclcccccccc}
\toprule
\rowcolor{lightgray} & & & \multicolumn{4}{c}{\textbf{DSC}$\uparrow$} & \multicolumn{4}{c}{\textbf{NSD}$\uparrow$} \\
\rowcolor{lightgray} \multirow{-2}{*}{\textbf{Region}} & \multirow{-2}{*}{\textbf{Modality}} & \multirow{-2}{*}{\textbf{Anatomical Target}} & \textbf{Ours} & \textbf{MedCPT} & \textbf{Clip} & \textbf{BB} & \textbf{Ours} & \textbf{MedCPT} & \textbf{Clip} & \textbf{BB} \\ 
\midrule

\multirow{7}{*}{Thorax} & CT & Autochthon           & 93.06 & 91.97 & 92.41 & 92.15 & 90.33 & 85.79 & 87.17 & 85.95 \\
& CT & Heart atrium         & 91.13 & 90.16 & 90.21 & 89.29 & 79.94 & 76.89 & 77.48 & 77.68 \\
& MRI & Heart ventricle      & 87.49 & 87.84 & 88.36 & 89.31 & 70.48 & 71.1 & 69.71 & 72.71 \\
& CT & Lung                 & 91.14 & 90.05 & 91.38 & 90.06 & 84.42 & 80.81 & 82.23 & 80.59 \\
& CT & Rib                  & 86.04 & 82.54 & 83.45 & 83.83 & 88.76 & 86.6 & 87.69 & 88.45 \\
& CT & Myocardium           & 84.96 & 84.28 & 84.81 & 86.7 & 74.58 & 75.1 & 72.97 & 75.97 \\
& CT & Thoracic cavity      & 95.35 & 94.64 & 94.97 & 94.56 & 74.4 & 70.21 & 71.36 & 69.38 \\\hline

\multirow{5}{*}{Lower limb} & CT & Head of femur        & 91.86 & 88.21 & 89.02 & 88.17 & 81.87 & 75.11 & 77.6 & 74.59 \\
& MRI & Femur bone           & 95.8 & 95.43 & 95.52 & 95.37 & 88.57 & 87.43 & 87.01 & 86.9 \\
& MRI & Femur cartilage      & 68.8 & 67.85 & 67.45 & 66.77 & 88.03 & 85.83 & 86.62 & 86.15 \\
& MRI & Tibia bone           & 96.53 & 96.25 & 96.42 & 96.31 & 93.83 & 92.24 & 92.47 & 92.48 \\
& MRI & Tibia cartilage      & 67.07 & 65.47 & 66.43 & 65.87 & 89.48 & 87.57 & 88.36 & 87.98 \\\hline

\multirow{8}{*}{Spine} & CT & Lumbar vertebrae     & 77.03 & 73.88 & 77.84 & 73.05 & 76.64 & 72.58 & 77.52 & 71.33 \\
& CT & Sacrum               & 94.15 & 92.81 & 93.39 & 85.87 & 94.82 & 92.75 & 93.49 & 85.59 \\
& CT & Spinal cord          & 82.49 & 79.99 & 82.25 & 76.93 & 87.46 & 82.54 & 87.22 & 77.93 \\
& CT & Thoracic vertebrae   & 80.21 & 79.04 & 76.04 & 78.59 & 80.33 & 78.67 & 75.91 & 78.25 \\
& MRI & Intervertebral discs & 88.67 & 88.21 & 88.75 & 88.05 & 91.43 & 90.92 & 91.56 & 90.42 \\
& MRI & Lumbar vertebrae     & 81.95 & 81.49 & 80.17 & 81.87 & 72.96 & 70.83 & 72.2 & 71.08 \\
& MRI & Sacrum               & 83.71 & 81.58 & 82.11 & 81.4 & 77.75 & 73.0 & 76.27 & 72.88 \\
& MRI & Thoracic vertebrae   & 64.5 & 59.57 & 60.67 & 64.72 & 59.06 & 53.72 & 55.87 & 58.12 \\\hline
   
\multirow{3}{*}{Upper limb} & CT & Clavicle             & 92.11 & 82.82 & 85.13 & 85.65 & 93.24 & 84.3 & 86.71 & 87.04 \\
& CT & Humerus              & 85.91 & 80.38 & 79.3 & 79.68 & 86.43 & 80.17 & 79.2 & 79.62 \\
& CT & Scapula              & 92.54 & 80.47 & 88.45 & 88.3 & 94.7 & 83.45 & 91.47 & 91.46 \\\hline
         
\multirow{8}{*}{Pelvis} & CT & Gluteus maximus      & 94.54 & 95.39 & 95.32 & 94.05 & 90.96 & 90.19 & 90.21 & 88.63 \\
& CT & Gluteus medius       & 91.99 & 90.19 & 94.07 & 92.93 & 88.9 & 84.74 & 89.13 & 87.15 \\
& CT & Gluteus minimus      & 93.76 & 91.85 & 93.01 & 89.71 & 93.36 & 90.26 & 91.67 & 87.83 \\
& CT & Hip                  & 95.48 & 92.95 & 93.93 & 93.89 & 96.14 & 93.24 & 94.24 & 94.13 \\
& CT & Iliopsoas            & 91.32 & 89.98 & 89.95 & 87.99 & 91.52 & 88.72 & 88.95 & 86.64 \\
& CT & Iliac vena           & 90.69 & 89.17 & 90.04 & 85.38 & 92.32 & 90.69 & 91.58 & 86.72 \\
& CT & Urinary bladder      & 86.28 & 83.71 & 86.97 & 82.32 & 69.11 & 63.38 & 67.28 & 61.65 \\
& MRI & Prostate             & 83.24 & 84.23 & 84.5 & 82.67 & 55.04 & 57.34 & 57.83 & 52.78 \\\hline

\multirow{3}{*}{Lesions} & CT & Lung nodule          & 18.91 & 5.14 & 6.92 & 9.96 & 21.34 & 8.01 & 8.15 & 12.73 \\
& MRI & Myocardial edema     & 12.24 & 9.32 & 13.08 & 14.41 & 18.83 & 14.48 & 19.43 & 19.69 \\
& MRI & Stroke               & 48.14 & 49.0 & 49.16 & 47.83 & 47.18 & 46.05 & 46.34 & 45.07 \\\bottomrule

\end{tabular}}
\label{tab:text_ab_class_results2}
\end{table*}

\begin{table*}[!tbh]
\renewcommand{\arraystretch}{1.3} 
\center
\small
\caption{Class-wise results of four \model{}-Nano variants with different text encoders on common anatomical structures in abdomen, brain, head and neck. `Ours' denotes the \model{}-Nano prompted with the text encoder pre-trained on our multimodal medical knowledge graph; `MedCPT' denotes the variant with MedCPT as text encoder; `Clip' denotes the variant with Clip text encoder; `BB' denotes the variant with BERT-base as text encoder.}
\vspace{0.3cm}
\resizebox{1.0\textwidth}{!}{\begin{tabular}{lclcccccccc}
\toprule
\rowcolor{lightgray} & & & \multicolumn{4}{c}{\textbf{DSC}$\uparrow$} & \multicolumn{4}{c}{\textbf{NSD}$\uparrow$} \\
\rowcolor{lightgray} \multirow{-2}{*}{\textbf{Region}} & \multirow{-2}{*}{\textbf{Modality}} & \multirow{-2}{*}{\textbf{Anatomical Target}} & \textbf{Ours} & \textbf{MedCPT} & \textbf{Clip} & \textbf{BB} & \textbf{Ours} & \textbf{MedCPT} & \textbf{Clip} & \textbf{BB} \\ 
\midrule

\multirow{16}{*}{Abdomen} & CT & Adrenal gland        & 80.14 & 78.54 & 77.98 & 78.52 & 88.48 & 86.94 & 86.8 & 86.91 \\
& CT & Duodenum             & 77.23 & 75.88 & 76.65 & 76.58 & 68.65 & 65.15 & 66.17 & 66.11 \\
& CT & Gallbladder          & 81.29 & 78.48 & 79.35 & 79.28 & 76.4 & 72.19 & 72.72 & 72.76 \\
& CT & Inferior vena cava   & 89.26 & 88.41 & 87.79 & 87.81 & 86.06 & 84.05 & 83.08 & 83.26 \\
& CT & Intestine            & 86.21 & 82.72 & 83.41 & 82.81 & 72.69 & 63.93 & 65.32 & 64.77 \\
& CT & Kidney               & 94.08 & 93.41 & 93.85 & 93.34 & 90.37 & 88.69 & 89.12 & 88.39 \\
& CT & Liver                & 95.8 & 94.97 & 95.03 & 95.02 & 83.69 & 79.47 & 79.43 & 79.5 \\
& CT & Pancreas             & 85.13 & 82.12 & 83.72 & 83.61 & 76.41 & 70.91 & 72.91 & 72.52 \\
& CT & Small bowel          & 75.78 & 78.24 & 81.51 & 78.62 & 68.82 & 69.27 & 72.44 & 69.3 \\
& CT & Spleen               & 94.71 & 93.8 & 93.21 & 93.57 & 91.16 & 88.82 & 87.86 & 88.35 \\
& CT & Stomach              & 85.1 & 87.93 & 89.0 & 89.3 & 71.75 & 71.14 & 71.79 & 72.39 \\
& MRI & Adrenal gland        & 61.13 & 62.99 & 59.29 & 60.35 & 73.1 & 74.7 & 70.85 & 71.36 \\
& MRI & Kidney               & 91.93 & 90.34 & 90.94 & 90.85 & 74.61 & 69.4 & 71.5 & 71.03 \\
& MRI & Liver                & 91.29 & 90.69 & 91.63 & 91.6 & 64.15 & 61.02 & 61.52 & 60.69 \\
& MRI & Pancreas             & 81.43 & 82.78 & 81.84 & 82.51 & 69.09 & 69.41 & 68.2 & 69.02 \\
& MRI & Spleen               & 84.53 & 82.11 & 82.22 & 82.63 & 63.45 & 61.7 & 61.49 & 62.31 \\\hline

\multirow{8}{*}{Brain} & CT & Brainstem            & 86.66 & 85.2 & 86.0 & 84.1 & 74.95 & 69.55 & 72.32 & 68.18 \\
& CT & Hippocampus          & 78.36 & 77.17 & 78.84 & 77.02 & 79.7 & 77.51 & 80.48 & 77.55 \\
& CT & Temporal lobe        & 94.52 & 94.26 & 94.9 & 94.23 & 85.6 & 83.86 & 86.67 & 83.86 \\
& MRI & Brain ventricle      & 77.48 & 75.86 & 76.58 & 76.0 & 86.27 & 83.23 & 84.36 & 83.26 \\
& MRI & Cerebellum           & 89.89 & 88.72 & 90.23 & 88.85 & 84.8 & 80.23 & 84.21 & 80.83 \\
& MRI & Parietal lobe        & 75.57 & 74.47 & 74.77 & 73.97 & 57.82 & 53.83 & 53.86 & 53.23 \\
& MRI & Optic radiation      & 61.06 & 60.41 & 62.17 & 60.99 & 55.35 & 52.84 & 55.41 & 53.56 \\
& MRI & Thalamus             & 85.28 & 84.72 & 84.14 & 84.88 & 85.44 & 83.69 & 82.98 & 83.89 \\\hline

\multirow{16}{*}{Head and neck} & CT & Brain                & 89.97 & 96.43 & 97.91 & 97.85 & 86.86 & 91.38 & 94.59 & 93.13 \\
& CT & Carotid artery       & 66.6 & 60.76 & 62.24 & 60.63 & 80.1 & 68.93 & 71.04 & 69.05 \\
& CT & Cervical esophagus   & 61.46 & 64.13 & 65.79 & 61.99 & 62.33 & 68.21 & 67.41 & 63.18 \\
& CT & Cochlea              & 70.75 & 68.99 & 70.08 & 69.65 & 90.59 & 89.94 & 90.29 & 89.88 \\
& CT & Eustachian tube bone & 79.42 & 81.81 & 81.13 & 81.71 & 95.86 & 97.1 & 97.11 & 97.44 \\
& CT & Eyeball              & 85.84 & 84.18 & 84.93 & 84.88 & 88.56 & 85.85 & 86.48 & 86.61 \\
& CT & Lacrimal gland       & 47.62 & 47.32 & 41.86 & 0.0 & 66.76 & 65.35 & 54.39 & 0.0 \\
& CT & Lens                 & 73.89 & 73.9 & 73.15 & 74.37 & 90.48 & 91.87 & 90.47 & 91.9 \\
& CT & Mandible             & 94.3 & 93.26 & 93.74 & 92.79 & 96.69 & 95.28 & 96.23 & 94.34 \\
& CT & Middle ear           & 87.04 & 82.52 & 86.64 & 85.11 & 93.8 & 88.51 & 93.66 & 91.96 \\
& CT & Optic nerve          & 61.67 & 70.52 & 69.7 & 71.01 & 78.05 & 88.25 & 87.85 & 88.48 \\
& CT & Parotid gland        & 85.27 & 83.89 & 84.65 & 84.09 & 76.23 & 73.01 & 74.93 & 72.81 \\
& CT & Submandibular gland  & 79.94 & 78.91 & 78.24 & 78.25 & 77.51 & 75.9 & 75.23 & 74.39 \\
& CT & Tympanic cavity      & 82.53 & 81.4 & 80.53 & 81.0 & 95.28 & 94.7 & 92.93 & 94.48 \\
& CT & Thyroid gland        & 85.61 & 82.79 & 84.98 & 82.49 & 87.85 & 82.09 & 86.55 & 80.97 \\
& MRI & Brain                & 95.21 & 94.72 & 94.83 & 94.36 & 58.25 & 55.57 & 57.14 & 53.53 \\\bottomrule

\end{tabular}}
\label{tab:text_ab_class_results}
\end{table*}
\begin{table*}[!tbh]
\renewcommand{\arraystretch}{1.1} 
\center
\footnotesize
\caption{Dataset-wise results of four \model{}-Nano variants with different text encoders on 49 datasets in \samdataset{}. `Ours' denotes the \model{}-Nano prompted with the text encoder pre-trained on our multimodal medical knowledge graph; `MedCPT' denotes the variant with MedCPT as text encoder; `Clip' denotes the variant with CLIP text encoder; `BB' denotes the variant with BERT-base as text encoder.}
\label{tab:text_ab_dataset_results}
\vspace{0.3cm}
\resizebox{1.0\textwidth}{!}{\begin{tabular}{lcccccccc}

\toprule
\rowcolor{lightgray} \multicolumn{1}{l}{} & \multicolumn{4}{c}{\textbf{DSC$\uparrow$}} & \multicolumn{4}{c}{\textbf{NSD$\uparrow$}} \\
\rowcolor{lightgray} \multicolumn{1}{l}{\multirow{-2}{*}{\textbf{Dataset}}} & \multicolumn{1}{c}{\textbf{Ours}} & \multicolumn{1}{c}{\textbf{MedCPT}} & \multicolumn{1}{c}{\textbf{Clip}} & \multicolumn{1}{c}{\textbf{BB}} & \multicolumn{1}{c}{\textbf{Ours}} & \multicolumn{1}{c}{\textbf{MedCPT}} & \multicolumn{1}{c}{\textbf{Clip}} & \multicolumn{1}{c}{\textbf{BB}} \\
\midrule

AbdomenCT1K~\cite{AbdomenCT1K}                               & 93.14 & 92.49 & 92.35 & 92.41 & 85.31 & 82.49 & 82.44 & 82.23 \\\hline
ACDC~\cite{ACDC}                                             & 86.1 & 85.3 & 85.88 & 85.03 & 69.39 & 66.39 & 68.24 & 65.82 \\\hline
AMOS22 CT~\cite{AMOS22}                                      & 85.81 & 84.48 & 84.47 & 84.31 & 82.25 & 79.07 & 79.28 & 78.84 \\\hline
AMOS22 MRI~\cite{AMOS22}                                     & 80.58 & 80.77 & 80.24 & 75.44 & 76.79 & 75.39 & 74.93 & 70.12 \\\hline
ATLAS~\cite{ATLAS}                                           & 62.62 & 67.9 & 61.37 & 69.7 & 38.47 & 39.22 & 36.11 & 40.53 \\\hline
ATLASR2~\cite{ATLASR2}                                       & 55.15 & 53.8 & 55.2 & 53.39 & 54.49 & 49.68 & 51.41 & 50.18 \\\hline
Brain Atlas~\cite{Brain_Atlas}                               & 76.35 & 74.77 & 75.15 & 75.57 & 76.11 & 72.64 & 73.34 & 73.59 \\\hline
BrainPTM~\cite{BrainPTM}                                     & 65.85 & 66.25 & 66.13 & 65.57 & 53.8 & 51.95 & 53.12 & 51.35 \\\hline
CHAOS MRI~\cite{CHAOS}                                       & 85.06 & 82.59 & 83.81 & 83.67 & 52.41 & 46.68 & 49.08 & 48.21 \\\hline
CMRxMotion~\cite{CMRxMotion}                                 & 88.57 & 87.01 & 87.41 & 86.81 & 75.15 & 68.85 & 70.18 & 67.8 \\\hline
Couinaud Liver~\cite{Couinaud}                               & 79.28 & 81.36 & 72.85 & 80.44 & 53.89 & 54.62 & 49.25 & 53.78 \\\hline
CrossMoDA2021~\cite{CrossMoDA2021}                           & 73.62 & 71.91 & 72.27 & 71.78 & 93.59 & 90.98 & 91.3 & 91.93 \\\hline
CT ORG~\cite{CTORG}                                          & 87.02 & 89.82 & 90.92 & 89.15 & 74.49 & 75.9 & 77.46 & 75.0 \\\hline
CTPelvic1K~\cite{CTPelvic1K}                                 & 95.72 & 94.76 & 95.02 & 94.79 & 97.37 & 95.91 & 96.21 & 95.84 \\\hline
FeTA2022~\cite{FeTA2022}                                     & 72.19 & 68.98 & 70.42 & 68.69 & 79.61 & 75.31 & 77.42 & 75.12 \\\hline
FLARE22~\cite{FLARE22}                                       & 88.42 & 88.97 & 88.54 & 89.19 & 86.26 & 85.37 & 84.69 & 85.57 \\\hline
FUMPE~\cite{FUMPE}                                           & 35.16 & 36.11 & 36.98 & 30.14 & 35.83 & 31.91 & 37.52 & 28.93 \\\hline
HAN Seg~\cite{HANSeg}                                        & 70.63 & 68.95 & 69.64 & 65.37 & 76.56 & 73.44 & 74.13 & 67.88 \\\hline
Instance22~\cite{INSTANCE}                                   & 52.49 & 61.0 & 56.67 & 56.56 & 42.54 & 51.65 & 46.37 & 46.67 \\\hline
ISLES2022~\cite{ISLES2022}                                   & 41.14 & 44.2 & 43.12 & 42.27 & 39.87 & 42.41 & 41.26 & 39.96 \\\hline
KiPA22~\cite{KiPA22}                                         & 63.98 & 67.52 & 65.54 & 67.46 & 62.13 & 65.89 & 62.63 & 66.51 \\\hline
KiTS23~\cite{KiTS23}                                         & 60.09 & 53.49 & 52.9 & 55.22 & 51.63 & 43.66 & 42.84 & 45.42 \\\hline
LAScarQS22 Task1~\cite{LAScarQS2022}                         & 67.8 & 65.6 & 67.2 & 66.44 & 79.54 & 73.42 & 75.94 & 74.5 \\\hline
LAScarQS22 Task2~\cite{LAScarQS2022}                         & 91.66 & 88.78 & 90.14 & 89.09 & 79.28 & 69.87 & 72.84 & 69.73 \\\hline
LNDb~\cite{LNDb}                                             & 18.91 & 5.14 & 6.92 & 9.96 & 21.34 & 8.01 & 8.15 & 12.73 \\\hline
LUNA16~\cite{LUNA16}                                         & 96.52 & 95.92 & 96.23 & 96.02 & 94.85 & 93.02 & 92.89 & 93.25 \\\hline
MM WHS CT~\cite{MMWHS}                                       & 87.79 & 88.23 & 87.75 & 89.08 & 68.79 & 69.85 & 69.05 & 72.82 \\\hline
MM WHS MRI~\cite{MMWHS}                                      & 84.97 & 85.82 & 84.02 & 85.38 & 65.32 & 66.85 & 63.81 & 65.89 \\\hline
MRSpineSeg~\cite{MRSpineSeg}                                 & 77.25 & 76.21 & 76.64 & 76.97 & 74.14 & 72.45 & 73.91 & 72.67 \\\hline
MyoPS2020~\cite{MyoPS2020}                                   & 59.55 & 59.54 & 60.85 & 59.68 & 41.64 & 40.68 & 42.94 & 40.9 \\\hline
NSCLC~\cite{NSCLC}                                           & 75.21 & 74.17 & 75.57 & 75.42 & 60.88 & 56.99 & 59.52 & 57.3 \\\bottomrule
\end{tabular}}
\end{table*}

\begin{table*}[!tbh]
\renewcommand{\arraystretch}{1.3} 
\center
\footnotesize
\caption{(Continued) Dataset-wise results of four \model{}-Nano variants with different text encoders on 49 datasets in \samdataset{}. `Ours' denotes the \model{}-Nano prompted with the text encoder pre-trained on our multimodal medical knowledge graph; `MedCPT' denotes the variant with MedCPT as text encoder; `Clip' denotes the variant with CLIP text encoder; `BB' denotes the variant with BERT-base as text encoder.}
\label{tab:text_ab_dataset_results_2}
\vspace{0.3cm}
\resizebox{1.0\textwidth}{!}{\begin{tabular}{lcccccccc}

\toprule
\rowcolor{lightgray} \multicolumn{1}{l}{} & \multicolumn{4}{c}{\textbf{DSC$\uparrow$}} & \multicolumn{4}{c}{\textbf{NSD$\uparrow$}} \\
\rowcolor{lightgray} \multicolumn{1}{l}{\multirow{-2}{*}{\textbf{Dataset}}} & \multicolumn{1}{c}{\textbf{Ours}} & \multicolumn{1}{c}{\textbf{MedCPT}} & \multicolumn{1}{c}{\textbf{Clip}} & \multicolumn{1}{c}{\textbf{BB}} & \multicolumn{1}{c}{\textbf{Ours}} & \multicolumn{1}{c}{\textbf{MedCPT}} & \multicolumn{1}{c}{\textbf{Clip}} & \multicolumn{1}{c}{\textbf{BB}} \\

\midrule
Pancreas CT~\cite{PancreasCT}                                & 86.47 & 83.79 & 84.47 & 83.89 & 79.27 & 72.39 & 73.61 & 72.31 \\\hline
PARSE2022~\cite{PARSE2022}                                   & 73.73 & 68.56 & 69.29 & 68.04 & 70.03 & 54.16 & 56.97 & 51.62 \\\hline
PDDCA~\cite{PDDCA}                                           & 72.98 & 73.44 & 72.78 & 72.9 & 75.45 & 74.29 & 74.46 & 73.57 \\\hline
PROMISE12~\cite{PROMISE12}                                   & 83.24 & 84.23 & 84.5 & 82.67 & 55.04 & 57.34 & 57.83 & 52.78 \\\hline
SEGA~\cite{SEGA}                                             & 80.08 & 77.68 & 79.02 & 79.32 & 69.56 & 63.48 & 66.88 & 65.51 \\\hline
SegRap2023 Task1~\cite{SegRap2023}                           & 85.22 & 85.12 & 85.7 & 84.93 & 90.29 & 89.72 & 90.94 & 89.54 \\\hline
SegRap2023 Task2~\cite{SegRap2023}                           & 67.35 & 64.95 & 64.56 & 65.69 & 46.51 & 43.85 & 42.52 & 43.86 \\\hline
SegTHOR~\cite{SegTHOR}                                       & 86.17 & 84.59 & 85.79 & 84.8 & 72.19 & 67.75 & 70.74 & 68.62 \\\hline
SKI10~\cite{SKI10}                                           & 82.05 & 81.25 & 81.46 & 81.08 & 89.98 & 88.27 & 88.61 & 88.38 \\\hline
SLIVER07~\cite{SLIVER07}                                     & 97.41 & 96.59 & 96.88 & 96.74 & 86.63 & 80.76 & 82.39 & 81.89 \\\hline
TotalSegmentator Cardiac~\cite{Totalsegmentator}             & 88.99 & 86.96 & 87.96 & 85.93 & 86.64 & 83.04 & 84.38 & 81.99 \\\hline
TotalSegmentator Muscles~\cite{Totalsegmentator}             & 90.32 & 87.04 & 88.35 & 87.93 & 89.49 & 85.15 & 86.65 & 85.95 \\\hline
TotalSegmentator Organs~\cite{Totalsegmentator}              & 87.18 & 84.13 & 86.02 & 85.37 & 83.27 & 78.71 & 80.77 & 79.8 \\\hline
TotalSegmentator Ribs~\cite{Totalsegmentator}                & 86.27 & 82.72 & 83.67 & 84.0 & 89.03 & 86.86 & 87.99 & 88.7 \\\hline
TotalSegmentator Vertebrae~\cite{Totalsegmentator}           & 85.99 & 83.25 & 80.3 & 81.57 & 86.73 & 83.5 & 80.64 & 81.77 \\\hline
VerSe~\cite{VerSe}                                           & 78.19 & 76.51 & 70.62 & 70.77 & 77.7 & 75.56 & 70.32 & 69.69 \\\hline
WMH Segmentation Challenge~\cite{WMH}                        & 64.61 & 62.07 & 62.58 & 61.59 & 78.89 & 75.69 & 76.48 & 75.14 \\\hline
WORD~\cite{WORD}                                             & 86.49 & 84.79 & 85.16 & 84.79 & 76.96 & 72.42 & 73.42 & 72.4 \\\bottomrule
\end{tabular}}
\end{table*}
\clearpage

\begin{table*}[!tbh]
\renewcommand{\arraystretch}{1.3} 
\center
\caption{The configurations of 72 nnU-Nets trained on each dataset. All nnU-Nets are planned under `3d\_fullres' setting. The total size of all the nnU-Nets is around 2247M.}
\vspace{0.3cm}
\label{tab:nnunet_config}
\resizebox{1.0\textwidth}{!}{\begin{tabular}{lclcccccc}

\toprule
\rowcolor{lightgray} \multicolumn{1}{l}{\textbf{Dataset}} & \multicolumn{1}{c}{\textbf{Input Size}} & \multicolumn{1}{c}{\textbf{\#Stage}} & \multicolumn{1}{c}{\textbf{\#Depth }} & \multicolumn{1}{c}{\textbf{\#Width}} & \multicolumn{1}{c}{\textbf{Model Size}} \\ 
\midrule

AbdomenCT1K~\cite{AbdomenCT1K}                     & [96 160 160]  & 6 & [2 2 2 2 2 2]   & [32 64 128 256 320 320]     & 31M \\
ACDC~\cite{ACDC}                                   & [10 256 224]  & 6 & [2 2 2 2 2 2]   & [32 64 128 256 320 320]     & 31M \\
AMOS22 CT~\cite{AMOS22}                            & [64 192 160]  & 6 & [2 2 2 2 2 2]   & [32 64 128 256 320 320]     & 31M \\
AMOS22 MRI~\cite{AMOS22}                           & [64 160 224]  & 6 & [2 2 2 2 2 2]   & [32 64 128 256 320 320]     & 31M \\
ATLASR2~\cite{ATLASR2}                             & [128 128 128] & 6 & [2 2 2 2 2 2]   & [32 64 128 256 320 320]     & 31M \\
ATLAS~\cite{ATLAS}                                 & [48 192 224]  & 6 & [2 2 2 2 2 2]   & [32 64 128 256 320 320]     & 31M \\
autoPET~\cite{autoPET}                             & [80 192 160]  & 6 & [2 2 2 2 2 2]   & [32 64 128 256 320 320]     & 31M \\
Brain Atlas~\cite{Brain_Atlas}                     & [112 128 112] & 5 & [2 2 2 2 2]     & [32 64 128 256 320]         & 17M \\
BrainPTM~\cite{BrainPTM}                           & [112 144 112] & 5 & [2 2 2 2 2]     & [32 64 128 256 320]         & 17M \\
BraTS2023 GLI~\cite{BraTS2023GLI}                  & [128 160 112] & 6 & [2 2 2 2 2 2]   & [32 64 128 256 320 320]     & 31M \\
BraTS2023 MEN~\cite{BraTS2023MEN}                  & [128 160 112] & 6 & [2 2 2 2 2 2]   & [32 64 128 256 320 320]     & 31M \\
BraTS2023 MET~\cite{BraTS2023MET}                  & [128 160 112] & 6 & [2 2 2 2 2 2]   & [32 64 128 256 320 320]     & 31M \\
BraTS2023 PED~\cite{BraTS2023PED}                  & [128 128 128] & 6 & [2 2 2 2 2 2]   & [32 64 128 256 320 320]     & 31M \\
BraTS2023 SSA~\cite{BraTS2023SSA}                  & [128 160 112] & 6 & [2 2 2 2 2 2]   & [32 64 128 256 320 320]     & 31M \\
BTCV Cervix~\cite{BTCV}                      & [64 192 160]  & 6 & [2 2 2 2 2 2]   & [32 64 128 256 320 320]     & 31M \\
BTCV Abdomen~\cite{BTCV}                           & [48 192 192]  & 6 & [2 2 2 2 2 2]   & [32 64 128 256 320 320]     & 31M \\
CHAOS CT~\cite{CHAOS}                              & [48 224 192]  & 6 & [2 2 2 2 2 2]   & [32 64 128 256 320 320]     & 31M \\
CHAOS MRI~\cite{CHAOS}                             & [32 192 288]  & 6 & [2 2 2 2 2 2]   & [32 64 128 256 320 320]     & 31M \\
CMRxMotion~\cite{CMRxMotion}                       & [10 448 384]  & 7 & [2 2 2 2 2 2 2] & [32 64 128 256 320 320 320] & 45M \\
Couinaud~\cite{Couinaud}                           & [64 192 192]  & 6 & [2 2 2 2 2 2]   & [32 64 128 256 320 320]     & 31M \\
COVID-19 CT Seg~\cite{COVID19}                     & [56 192 192]  & 6 & [2 2 2 2 2 2]   & [32 64 128 256 320 320]     & 31M \\
CrossMoDA2021~\cite{CrossMoDA2021}                 & [48 224 192]  & 6 & [2 2 2 2 2 2]   & [32 64 128 256 320 320]     & 31M \\
CT-ORG~\cite{CTORG}                                & [128 128 128] & 6 & [2 2 2 2 2 2]   & [32 64 128 256 320 320]     & 31M \\
CTPelvic1K~\cite{CTPelvic1K}                       & [96 160 160]  & 6 & [2 2 2 2 2 2]   & [32 64 128 256 320 320]     & 31M \\
DAP Atlas~\cite{DAPAtlas}                          & [80 128 128]  & 6 & [2 2 2 2 2 2]   & [32 64 128 256 320 320]     & 31M \\
FeTA2022~\cite{FeTA2022}                           & [96 112 96]   & 5 & [2 2 2 2 2]     & [32 64 128 256 320]         & 17M \\
FLARE22~\cite{FLARE22}                             & [40 224 192]  & 6 & [2 2 2 2 2 2]   & [32 64 128 256 320 320]     & 31M \\
FUMPE~\cite{FUMPE}                                 & [80 192 160]  & 6 & [2 2 2 2 2 2]   & [32 64 128 256 320 320]     & 31M \\
HAN Seg~\cite{HANSeg}                              & [40 224 192]  & 6 & [2 2 2 2 2 2]   & [32 64 128 256 320 320]     & 31M \\
Hecktor2022~\cite{HECTOR2022}                      & [128 128 128] & 6 & [2 2 2 2 2 2]   & [32 64 128 256 320 320]     & 31M \\
Instance22~\cite{INSTANCE}                         & [16 320 320]  & 7 & [2 2 2 2 2 2 2] & [32 64 128 256 320 320 320] & 45M \\
ISLES2022~\cite{ISLES2022}                         & [80 96 80]    & 5 & [2 2 2 2 2]     & [32 64 128 256 320]         & 17M \\
KiPA22~\cite{KiPA22}                               & [160 128 112] & 6 & [2 2 2 2 2 2]   & [32 64 128 256 320 320]     & 31M \\
KiTS23~\cite{KiTS23}                               & [128 128 128] & 6 & [2 2 2 2 2 2]   & [32 64 128 256 320 320]     & 31M \\
LAScarQS22 Task1~\cite{LAScarQS2022}               & [24 256 256]  & 7 & [2 2 2 2 2 2 2] & [32 64 128 256 320 320 320] & 45M \\
LAScarQS22 Task2~\cite{LAScarQS2022}               & [40 256 224]  & 6 & [2 2 2 2 2 2]   & [32 64 128 256 320 320]     & 31M \\
LNDb~\cite{LNDb}                                   & [96 160 160]  & 6 & [2 2 2 2 2 2]   & [32 64 128 256 320 320]     & 31M \\
\bottomrule
\end{tabular}}
\end{table*}

\begin{table*}[!tbh]
\renewcommand{\arraystretch}{1.3} 
\center
\caption{(Continued) The configurations of 72 nnU-Nets trained on each dataset. All nnU-Nets are planned under `3d\_fullres' setting. The total size of all the nnU-Nets is around 2247M.}
\vspace{0.3cm}
\label{tab:nnunet_config2}
\resizebox{1.0\textwidth}{!}{\begin{tabular}{lclcccccc}

\toprule
\rowcolor{lightgray} \multicolumn{1}{l}{\textbf{Dataset}} & \multicolumn{1}{c}{\textbf{Input Size}} & \multicolumn{1}{c}{\textbf{\#Stage}} & \multicolumn{1}{c}{\textbf{\#Depth }} & \multicolumn{1}{c}{\textbf{\#Width}} & \multicolumn{1}{c}{\textbf{Model Size}} \\ 
\midrule

LUNA16~\cite{LUNA16}                               & [80 192 160]  & 6 & [2 2 2 2 2 2]   & [32 64 128 256 320 320]     & 31M \\
MM-WHS CT~\cite{MMWHS}                             & [80 192 160]  & 6 & [2 2 2 2 2 2]   & [32 64 128 256 320 320]     & 31M \\
MM-WHS MR~\cite{MMWHS}                             & [96 160 160]  & 6 & [2 2 2 2 2 2]   & [32 64 128 256 320 320]     & 31M \\
MRSpineSeg~\cite{MRSpineSeg}                       & [8 640 320]   & 7 & [2 2 2 2 2 2 2] & [32 64 128 256 320 320 320] & 43M \\
MSD Colon~\cite{MSD}                               & [56 192 192]  & 6 & [2 2 2 2 2 2]   & [32 64 128 256 320 320]     & 31M \\
MSD Heart~\cite{MSD}                               & [80 192 160]  & 6 & [2 2 2 2 2 2]   & [32 64 128 256 320 320]     & 31M \\
MSD HepaticVessel~\cite{MSD}                       & [64 192 192]  & 6 & [2 2 2 2 2 2]   & [32 64 128 256 320 320]     & 31M \\
MSD Hippocampus~\cite{MSD}                         & [40 56 40]    & 4 & [2 2 2 2]       & [32 64 128 256]             & 6M \\
MSD Liver~\cite{MSD}                               & [128 128 128] & 6 & [2 2 2 2 2 2]   & [32 64 128 256 320 320]     & 31M \\
MSD Lung~\cite{MSD}                                & [80 192 160]  & 6 & [2 2 2 2 2 2]   & [32 64 128 256 320 320]     & 31M \\
MSD Pancreas~\cite{MSD}                            & [40 224 224]  & 6 & [2 2 2 2 2 2]   & [32 64 128 256 320 320]     & 31M \\
MSD Prostate~\cite{MSD}                            & [16 320 320]  & 7 & [2 2 2 2 2 2 2] & [32 64 128 256 320 320 320] & 45M \\
MSD Spleen~\cite{MSD}                              & [64 192 160]  & 6 & [2 2 2 2 2 2]   & [32 64 128 256 320 320]     & 31M \\
MyoPS2020~\cite{MyoPS2020}                         & [48 224 224]  & 6 & [2 2 2 2 2 2]   & [32 64 128 256 320 320]     & 31M \\
NSCLC~\cite{NSCLC}                                 & [48 224 192]  & 6 & [2 2 2 2 2 2]   & [32 64 128 256 320 320]     & 31M \\
Pancreas CT~\cite{PancreasCT}                       & [80 192 160]  & 6 & [2 2 2 2 2 2]   & [32 64 128 256 320 320]     & 31M \\
PARSE2022~\cite{PARSE2022}                         & [96 160 160]  & 6 & [2 2 2 2 2 2]   & [32 64 128 256 320 320]     & 31M \\
PDDCA~\cite{PDDCA}                                 & [48 192 192]  & 6 & [2 2 2 2 2 2]   & [32 64 128 256 320 320]     & 31M \\
PROMISE12~\cite{PROMISE12}                         & [20 320 256]  & 7 & [2 2 2 2 2 2 2] & [32 64 128 256 320 320 320] & 45M \\
SEGA~\cite{SEGA}                                   & [72 160 160]  & 6 & [2 2 2 2 2 2]   & [32 64 128 256 320 320]     & 31M \\
SegRap2023 Task1~\cite{SegRap2023}             & [28 256 224]  & 6 & [2 2 2 2 2 2]   & [32 64 128 256 320 320]     & 31M \\
SegRap2023 Task2~\cite{SegRap2023}             & [28 256 256]  & 7 & [2 2 2 2 2 2 2] & [32 64 128 256 320 320 320] & 45M \\
SegTHOR~\cite{SegTHOR}                             & [64 192 160]  & 6 & [2 2 2 2 2 2]   & [32 64 128 256 320 320]     & 31M \\
SKI10~\cite{SKI10}                                 & [64 192 160]  & 6 & [2 2 2 2 2 2]   & [32 64 128 256 320 320]     & 31M \\
SLIVER07~\cite{SLIVER07}                           & [80 192 160]  & 6 & [2 2 2 2 2 2]   & [32 64 128 256 320 320]     & 31M \\
ToothFairy~\cite{ToothFairy}                       & [80 192 160]  & 6 & [2 2 2 2 2 2]   & [32 64 128 256 320 320]     & 31M \\
TS Heart~\cite{Totalsegmentator}     & [128 128 128] & 6 & [2 2 2 2 2 2]   & [32 64 128 256 320 320]     & 31M \\
TS Muscles~\cite{Totalsegmentator}   & [128 128 128] & 6 & [2 2 2 2 2 2]   & [32 64 128 256 320 320]     & 31M \\
TS Organs~\cite{Totalsegmentator}    & [128 128 128] & 6 & [2 2 2 2 2 2]   & [32 64 128 256 320 320]     & 31M \\
TS Ribs~\cite{Totalsegmentator}      & [128 128 128] & 6 & [2 2 2 2 2 2]   & [32 64 128 256 320 320]     & 31M \\
TS Vertebrae~\cite{Totalsegmentator} & [128 128 128] & 6 & [2 2 2 2 2 2]   & [32 64 128 256 320 320]     & 31M \\
TS V2~\cite{Totalsegmentator}       & [128 128 128] & 6 & [2 2 2 2 2 2]   & [32 64 128 256 320 320]     & 31M \\
VerSe~\cite{VerSe}                                 & [160 128 112] & 6 & [2 2 2 2 2 2]   & [32 64 128 256 320 320]     & 31M \\
WMH~\cite{WMH}                                     & [48 224 192]  & 6 & [2 2 2 2 2 2]   & [32 64 128 256 320 320]     & 31M \\
WORD~\cite{WORD}                                   & [64 192 160]  & 6 & [2 2 2 2 2 2]   & [32 64 128 256 320 320]     & 31M \\
\bottomrule
\end{tabular}}
\end{table*}
\clearpage

\section{All classes in \samdataset{}}
\begin{table*}[!tbh]
\center
\caption{Detailed name list of the 497 classes involved in \samdataset{}.}
\footnotesize
\label{tab:Name_List1}
\vspace{0.3cm}
\resizebox{1.0\textwidth}{!}{\begin{tabular}{p{0.31\textwidth} p{0.31\textwidth}p{0.31\textwidth}}

\toprule
\rowcolor{lightgray} Name List 1 & Name List 2 & Name List 3 \\
\midrule

(ct) abdominal tissue    &    (ct) adrenal gland    &    (ct) anterior eyeball \\
(ct) aorta    &    (ct) arytenoid    &    (ct) autochthon \\
(ct) bone    &    (ct) brachiocephalic trunk    &    (ct) brachiocephalic vein \\
(ct) brain    &    (ct) brainstem    &    (ct) breast \\
(ct) bronchie    &    (ct) buccal mucosa    &    (ct) carotid artery \\
(ct) caudate lobe    &    (ct) celiac trunk    &    (ct) cervical esophagus \\
(ct) cervical vertebrae    &    (ct) cervical vertebrae 1 (c1)    &    (ct) cervical vertebrae 2 (c2) \\
(ct) cervical vertebrae 3 (c3)    &    (ct) cervical vertebrae 4 (c4)    &    (ct) cervical vertebrae 5 (c5) \\
(ct) cervical vertebrae 6 (c6)    &    (ct) cervical vertebrae 7 (c7)    &    (ct) cheek \\
(ct) clavicle    &    (ct) cochlea    &    (ct) colon \\
(ct) colon cancer    &    (ct) common carotid artery    &    (ct) costal cartilage \\
(ct) covid-19 infection    &    (ct) cricopharyngeal inlet    &    (ct) duodenum \\
(ct) esophagus    &    (ct) eustachian tube bone    &    (ct) eyeball \\
(ct) fat    &    (ct) femur    &    (ct) gallbladder \\
(ct) gluteus maximus    &    (ct) gluteus medius    &    (ct) gluteus minimus \\
(ct) gonad    &    (ct) head of femur    &    (ct) head of left femur \\
(ct) head of right femur    &    (ct) heart    &    (ct) heart ascending aorta \\
(ct) heart atrium    &    (ct) heart tissue    &    (ct) heart ventricle \\
(ct) hip    &    (ct) hippocampus    &    (ct) humerus \\
(ct) iliac artery    &    (ct) iliac vena    &    (ct) iliopsoas \\
(ct) inferior alveolar nerve    &    (ct) inferior vena cava    &    (ct) internal auditory canal \\
(ct) internal carotid artery    &    (ct) internal jugular vein    &    (ct) intestine \\
(ct) intracranial hemorrhage    &    (ct) kidney    &    (ct) kidney cyst \\
(ct) kidney tumor    &    (ct) lacrimal gland    &    (ct) larynx \\
(ct) larynx glottis    &    (ct) larynx supraglottis    &    (ct) left adrenal gland \\
(ct) left anterior eyeball    &    (ct) left auricle of heart    &    (ct) left autochthon \\
(ct) left brachiocephalic vein    &    (ct) left breast    &    (ct) left carotid artery \\
(ct) left cheek    &    (ct) left clavicle    &    (ct) left cochlea \\
(ct) left common carotid artery    &    (ct) left eustachian tube bone    &    (ct) left eyeball \\
(ct) left femur    &    (ct) left gluteus maximus    &    (ct) left gluteus medius \\
(ct) left gluteus minimus    &    (ct) left heart atrium    &    (ct) left heart ventricle \\
(ct) left hip    &    (ct) left hippocampus    &    (ct) left humerus \\
(ct) left iliac artery    &    (ct) left iliac vena    &    (ct) left iliopsoas \\
(ct) left internal auditory canal    &    (ct) left internal carotid artery    &    (ct) left internal jugular vein \\
(ct) left kidney    &    (ct) left kidney cyst    &    (ct) left lacrimal gland \\
(ct) left lateral inferior segment of liver    &    (ct) left lateral superior segment of liver    &    (ct) left lens \\
(ct) left lobe of liver    &    (ct) left lung    &    (ct) left lung lower lobe \\
(ct) left lung upper lobe    &    (ct) left mandible    &    (ct) left mastoid process \\
(ct) left medial segment of liver    &    (ct) left middle ear    &    (ct) left optic nerve \\
(ct) left parotid gland    &    (ct) left posterior eyeball    &    (ct) left rib \\
(ct) left rib 1    &    (ct) left rib 10    &    (ct) left rib 11 \\
(ct) left rib 12    &    (ct) left rib 2    &    (ct) left rib 3 \\
(ct) left rib 4    &    (ct) left rib 5    &    (ct) left rib 6 \\
(ct) left rib 7    &    (ct) left rib 8    &    (ct) left rib 9 \\
(ct) left scapula    &    (ct) left subclavian artery    &    (ct) left submandibular gland \\
(ct) left temporal lobe    &    (ct) left temporomandibular joint    &    (ct) left thyroid \\
(ct) left tympanic cavity    &    (ct) left vestibule semicircular canal    &    (ct) lens \\
(ct) lips    &    (ct) liver    &    (ct) liver tumor \\
(ct) liver vessel    &    (ct) lumbar vertebrae    &    (ct) lumbar vertebrae 1 (l1) \\
(ct) lumbar vertebrae 2 (l2)    &    (ct) lumbar vertebrae 3 (l3)    &    (ct) lumbar vertebrae 4 (l4) \\
(ct) lumbar vertebrae 5 (l5)    &    (ct) lumbar vertebrae 6 (l6)    &    (ct) lung \\
(ct) lung effusion    &    (ct) lung lower lobe    &    (ct) lung nodule \\
\bottomrule
\end{tabular}}
\end{table*}

\begin{table*}[!tbh]
\center
\footnotesize
\caption{(Continued) Detailed name list of the 497 classes involved in \samdataset{}.}
\label{tab:Name_List2}
\vspace{0.3cm}
\resizebox{1.0\textwidth}{!}{\begin{tabular}{p{0.31\textwidth} p{0.31\textwidth}p{0.31\textwidth}}

\toprule
\rowcolor{lightgray} Name List 1 & Name List 2 & Name List 3 \\
\midrule

(ct) lung tumor    &    (ct) lung upper lobe    &    (ct) mandible \\
(ct) manubrium of sternum    &    (ct) mastoid process    &    (ct) mediastinal tissue \\
(ct) middle ear    &    (ct) muscle    &    (ct) myocardium \\
(ct) nasal cavity    &    (ct) nasopharyngeal lymph node    &    (ct) nasopharyngeal tumor \\
(ct) optic chiasm    &    (ct) optic nerve    &    (ct) oral cavity \\
(ct) pancreas    &    (ct) pancreas tumor    &    (ct) parotid gland \\
(ct) pharynx constrictor muscle    &    (ct) pituitary gland    &    (ct) portal vein and splenic vein \\
(ct) posterior eyeball    &    (ct) prostate    &    (ct) pulmonary artery \\
(ct) pulmonary embolism    &    (ct) pulmonary vein    &    (ct) rectum \\
(ct) renal artery    &    (ct) renal vein    &    (ct) rib \\
(ct) rib 1    &    (ct) rib 10    &    (ct) rib 11 \\
(ct) rib 12    &    (ct) rib 2    &    (ct) rib 3 \\
(ct) rib 4    &    (ct) rib 5    &    (ct) rib 6 \\
(ct) rib 7    &    (ct) rib 8    &    (ct) rib 9 \\
(ct) rib cartilage    &    (ct) right adrenal gland    &    (ct) right anterior eyeball \\
(ct) right anterior inferior segment of liver    &    (ct) right anterior superior segment of liver    &    (ct) right autochthon \\
(ct) right brachiocephalic vein    &    (ct) right breast    &    (ct) right carotid artery \\
(ct) right cheek    &    (ct) right clavicle    &    (ct) right cochlea \\
(ct) right common carotid artery    &    (ct) right eustachian tube bone    &    (ct) right eyeball \\
(ct) right femur    &    (ct) right gluteus maximus    &    (ct) right gluteus medius \\
(ct) right gluteus minimus    &    (ct) right heart atrium    &    (ct) right heart ventricle \\
(ct) right hip    &    (ct) right hippocampus    &    (ct) right humerus \\
(ct) right iliac artery    &    (ct) right iliac vena    &    (ct) right iliopsoas \\
(ct) right internal auditory canal    &    (ct) right internal carotid artery    &    (ct) right internal jugular vein \\
(ct) right kidney    &    (ct) right kidney cyst    &    (ct) right lacrimal gland \\
(ct) right lens    &    (ct) right lobe of liver    &    (ct) right lung \\
(ct) right lung lower lobe    &    (ct) right lung middle lobe    &    (ct) right lung upper lobe \\
(ct) right mandible    &    (ct) right mastoid process    &    (ct) right middle ear \\
(ct) right optic nerve    &    (ct) right parotid gland    &    (ct) right posterior eyeball \\
(ct) right posterior inferior segment of liver    &    (ct) right posterior superior segment of liver    &    (ct) right rib \\
(ct) right rib 1    &    (ct) right rib 10    &    (ct) right rib 11 \\
(ct) right rib 12    &    (ct) right rib 2    &    (ct) right rib 3 \\
(ct) right rib 4    &    (ct) right rib 5    &    (ct) right rib 6 \\
(ct) right rib 7    &    (ct) right rib 8    &    (ct) right rib 9 \\
(ct) right scapula    &    (ct) right subclavian artery    &    (ct) right submandibular gland \\
(ct) right temporal lobe    &    (ct) right temporomandibular joint    &    (ct) right thyroid \\
(ct) right tympanic cavity    &    (ct) right vestibule semicircular canal    &    (ct) sacral vertebrae 1 (s1) \\
(ct) sacrum    &    (ct) scapula    &    (ct) skin \\
(ct) skull    &    (ct) small bowel    &    (ct) spinal canal \\
(ct) spinal cord    &    (ct) spleen    &    (ct) sternum \\
(ct) stomach    &    (ct) subclavian artery    &    (ct) submandibular gland \\
(ct) superior vena cava    &    (ct) temporal lobe    &    (ct) temporomandibular joint \\
(ct) thoracic cavity    &    (ct) thoracic vertebrae    &    (ct) thoracic vertebrae 1 (t1) \\
(ct) thoracic vertebrae 10 (t10)    &    (ct) thoracic vertebrae 11 (t11)    &    (ct) thoracic vertebrae 12 (t12) \\
(ct) thoracic vertebrae 2 (t2)    &    (ct) thoracic vertebrae 3 (t3)    &    (ct) thoracic vertebrae 4 (t4) \\
(ct) thoracic vertebrae 5 (t5)    &    (ct) thoracic vertebrae 6 (t6)    &    (ct) thoracic vertebrae 7 (t7) \\
(ct) thoracic vertebrae 8 (t8)    &    (ct) thoracic vertebrae 9 (t9)    &    (ct) thymus \\
(ct) thyroid gland    &    (ct) trachea    &    (ct) tympanic cavity \\
(ct) urinary bladder    &    (ct) uterocervix    &    (ct) uterus \\
(ct) vertebrae    &    (ct) vestibule semicircular canal    &    (mri) adrenal gland \\
(mri) amygdala    &    (mri) anterior hippocampus    &    (mri) aorta \\
\bottomrule
\end{tabular}}
\end{table*}

\begin{table*}[!tbh]
\center
\footnotesize
\caption{(Continued) Detailed name list of the 497 classes involved in \samdataset{}.}
\label{tab:Name_List3}
\vspace{0.3cm}
\resizebox{1.0\textwidth}{!}{\begin{tabular}{p{0.31\textwidth}p{0.31\textwidth}p{0.31\textwidth}}

\toprule
\rowcolor{lightgray} Name List 1 & Name List 2 & Name List 3 \\
\midrule

(mri) basal ganglia    &    (mri) brain    &    (mri) brain edema \\
(mri) brain tumor    &    (mri) brain ventricle    &    (mri) brainstem \\
(mri) brainstem excluding substantia nigra    &    (mri) cerebellum    &    (mri) cingulate gyrus \\
(mri) cochlea    &    (mri) corpus callosum    &    (mri) corticospinal tract \\
(mri) deep grey matter    &    (mri) duodenum    &    (mri) enhancing brain tumor \\
(mri) esophagus    &    (mri) external cerebrospinal fluid    &    (mri) femur bone \\
(mri) femur cartilage    &    (mri) frontal lobe    &    (mri) gallbladder \\
(mri) grey matter    &    (mri) heart ascending aorta    &    (mri) heart atrium \\
(mri) heart ventricle    &    (mri) hippocampus    &    (mri) inferior vena cava \\
(mri) insula    &    (mri) intervertebral disc between lumbar vertebrae 1 (l1) and lumbar vertebrae 2 (l2)    &    (mri) intervertebral disc between lumbar vertebrae 2 (l2) and lumbar vertebrae 3 (l3) \\
(mri) intervertebral disc between lumbar vertebrae 3 (l3) and lumbar vertebrae 4 (l4)    &    (mri) intervertebral disc between lumbar vertebrae 4 (l4) and lumbar vertebrae 5 (l5)    &    (mri) intervertebral disc between lumbar vertebrae 5 (l5) and sacrum \\
(mri) intervertebral disc between thoracic vertebrae 10 (t10) and thoracic vertebrae 11 (t11)    &    (mri) intervertebral disc between thoracic vertebrae 11 (t11) and thoracic vertebrae 12 (t12)    &    (mri) intervertebral disc between thoracic vertebrae 12 (t12) and lumbar vertebrae 1 (l1) \\
(mri) intervertebral disc between thoracic vertebrae 9 (t9) and thoracic vertebrae 10 (t10)    &    (mri) intervertebral discs    &    (mri) kidney \\
(mri) lateral ventricle    &    (mri) left adrenal gland    &    (mri) left amygdala \\
(mri) left angular gyrus    &    (mri) left anterior cingulate gyrus    &    (mri) left anterior orbital gyrus \\
(mri) left anterior temporal lobe lateral part    &    (mri) left anterior temporal lobe medial part    &    (mri) left caudate nucleus \\
(mri) left cerebellum    &    (mri) left corticospinal tract    &    (mri) left cuneus \\
(mri) left fusiform gyrus    &    (mri) left heart atrium    &    (mri) left heart atrium scar \\
(mri) left heart ventricle    &    (mri) left hippocampus    &    (mri) left inferior frontal gyrus \\
(mri) left insula anterior inferior cortex    &    (mri) left insula anterior long gyrus    &    (mri) left insula anterior short gyrus \\
(mri) left insula middle short gyrus    &    (mri) left insula posterior long gyrus    &    (mri) left insula posterior short gyrus \\
(mri) left kidney    &    (mri) left lateral orbital gyrus    &    (mri) left lateral remainder occipital lobe \\
(mri) left lateral ventricle excluding temporal horn    &    (mri) left lateral ventricle temporal horn    &    (mri) left lingual gyrus \\
(mri) left medial orbital gyrus    &    (mri) left middle and inferior temporal gyrus    &    (mri) left middle frontal gyrus \\
(mri) left nucleus accumbens    &    (mri) left optic radiation    &    (mri) left pallidum \\
(mri) left parahippocampal and ambient gyrus    &    (mri) left postcentral gyrus    &    (mri) left posterior cingulate gyrus \\
(mri) left posterior orbital gyrus    &    (mri) left posterior temporal lobe    &    (mri) left pre-subgenual frontal cortex \\
(mri) left precentral gyrus    &    (mri) left putamen    &    (mri) left straight gyrus \\
(mri) left subcallosal area    &    (mri) left subgenual frontal cortex    &    (mri) left substantia nigra \\
(mri) left superior frontal gyrus    &    (mri) left superior parietal gyrus    &    (mri) left superior temporal gyrus anterior part \\
(mri) left superior temporal gyrus middle part    &    (mri) left supramarginal gyrus    &    (mri) left thalamus \\
(mri) liver    &    (mri) liver tumor    &    (mri) lumbar vertebrae \\
(mri) lumbar vertebrae 1 (l1)    &    (mri) lumbar vertebrae 2 (l2)    &    (mri) lumbar vertebrae 3 (l3) \\
(mri) lumbar vertebrae 4 (l4)    &    (mri) lumbar vertebrae 5 (l5)    &    (mri) myocardial edema \\
(mri) myocardial scar    &    (mri) myocardium    &    (mri) necrotic brain tumor core \\
(mri) occipital lobe    &    (mri) optic radiation    &    (mri) pancreas \\
(mri) parietal lobe    &    (mri) peripheral zone of prostate    &    (mri) posterior hippocampus \\
\bottomrule
\end{tabular}}
\end{table*}

\begin{table*}[!tbh]
\center
\footnotesize
\caption{(Continued) Detailed name list of the 497 classes involved in \samdataset{}.}
\label{tab:Name_List4}
\vspace{0.3cm}
\resizebox{1.0\textwidth}{!}{\begin{tabular}{p{0.31\textwidth}
p{0.31\textwidth}p{0.31\textwidth}}

\toprule
\rowcolor{lightgray} Name List 1 & Name List 2 & Name List 3 \\
\midrule

(mri) prostate    &    (mri) pulmonary artery    &    (mri) right adrenal gland \\
(mri) right amygdala    &    (mri) right angular gyrus    &    (mri) right anterior cingulate gyrus \\
(mri) right anterior orbital gyrus    &    (mri) right anterior temporal lobe lateral part    &    (mri) right anterior temporal lobe medial part \\
(mri) right caudate nucleus    &    (mri) right cerebellum    &    (mri) right corticospinal tract \\
(mri) right cuneus    &    (mri) right fusiform gyrus    &    (mri) right heart atrium \\
(mri) right heart ventricle    &    (mri) right hippocampus    &    (mri) right inferior frontal gyrus \\
(mri) right insula anterior inferior cortex    &    (mri) right insula anterior long gyrus    &    (mri) right insula anterior short gyrus \\
(mri) right insula middle short gyrus    &    (mri) right insula posterior long gyrus    &    (mri) right insula posterior short gyrus \\
(mri) right kidney    &    (mri) right lateral orbital gyrus    &    (mri) right lateral remainder occipital lobe \\ 
(mri) right lateral ventricle excluding temporal horn    &    (mri) right lateral ventricle temporal horn    &    (mri) right lingual gyrus \\
(mri) right medial orbital gyrus    &    (mri) right middle and inferior temporal gyrus    &    (mri) right middle frontal gyrus \\
(mri) right nucleus accumbens    &    (mri) right optic radiation    &    (mri) right pallidum \\
(mri) right parahippocampal and ambient gyrus    &    (mri) right postcentral gyrus    &    (mri) right posterior cingulate gyrus \\
(mri) right posterior orbital gyrus    &    (mri) right posterior temporal lobe    &    (mri) right pre-subgenual frontal cortex \\
(mri) right precentral gyrus    &    (mri) right putamen    &    (mri) right straight gyrus \\
(mri) right subcallosal area    &    (mri) right subgenual frontal cortex    &    (mri) right substantia nigra \\
(mri) right superior frontal gyrus    &    (mri) right superior parietal gyrus    &    (mri) right superior temporal gyrus anterior part \\
(mri) right superior temporal gyrus middle part    &    (mri) right supramarginal gyrus    &    (mri) right thalamus \\
(mri) sacrum    &    (mri) spleen    &    (mri) stomach \\
(mri) stroke    &    (mri) temporal lobe    &    (mri) thalamus \\
(mri) third ventricle    &    (mri) thoracic vertebrae    &    (mri) thoracic vertebrae 10 (t10) \\
(mri) thoracic vertebrae 11 (t11)    &    (mri) thoracic vertebrae 12 (t12)    &    (mri) thoracic vertebrae 9 (t9) \\
(mri) tibia bone    &    (mri) tibia cartilage    &    (mri) transition zone of prostate \\
(mri) urinary bladder    &    (mri) vertebrae    &    (mri) vestibular schwannoma \\
(mri) white matter    &    (mri) white matter hyperintensities    &    (pet) head and neck tumor \\
(pet) lymph node    &    (pet) tumor    & \\
\bottomrule
\end{tabular}}
\end{table*}
\clearpage

\section{Dataset Details of SAT-DS}
\begin{table*}[!tbh]
\center
\scriptsize
\caption{The 72 datasets we collect to build up \samdataset{}. Dataset not marked with  \textasteriskcentered~ are included in \samdataset{}-Nano.}
\label{tab:dataset_details}
\vspace{0.3cm}
\resizebox{1.0\textwidth}{!}{\begin{tabular}{lclcccccc}

\toprule
\rowcolor{lightgray} \textbf{Dataset Name} & \textbf{\#Scans} & \textbf{\#Classes} & \textbf{\#Annotations}  & \textbf{Region} \\
\midrule

\rowcolor{lightgray!50} \textbf{CT Data}           &              &        &        &               \\
AbdomenCT1K~\cite{AbdomenCT1K} & 988      & 4      &3,950    & Abdomen       \\
ACDC~\cite{ACDC}           & 300          & 4      &1,200        & Thorax        \\
AMOS22 CT~\cite{AMOS22}           & 300          & 16      & 4,765       & Abdomen        \\
BTCV Abdomen \textasteriskcentered ~\cite{BTCV}           & 30          & 15      & 448       & Abdomen        \\
BTCV Cervix \textasteriskcentered ~\cite{BTCV}           & 30          & 4      & 118       & Abdomen        \\
CHAOS CT \textasteriskcentered ~\cite{CHAOS}    & 20           & 1      &20        & Abdomen       \\
Couinaud~\cite{Couinaud} & 161 & 10   &1,599        & Abdomen       \\
COVID-19 CT Seg \textasteriskcentered ~\cite{COVID19} & 20       & 4      &80        & Thorax        \\
CrossMoDA2021~\cite{CrossMoDA2021}           & 105          & 2      & 210       & Head and Neck        \\
CT-ORG~\cite{CTORG}           & 140          & 6      & 680       & Whole Body        \\
CTPelvic1K~\cite{CTPelvic1K}           & 117          & 5      & 585       & Lower Limb        \\
DAP Atlas \textasteriskcentered ~\cite{DAPAtlas}           & 533          & 179      & 93072       & Whole Body        \\
FLARE22~\cite{FLARE22}     & 50           & 15     & 750        & Abdomen       \\
FUMPE~\cite{FUMPE}           & 35          & 1      & 33       & Thorax        \\
HAN Seg~\cite{HANSeg}      & 41           & 41     &1,681        & Head and Neck \\
INSTANCE~\cite{INSTANCE}           & 100          & 1      & 100       & Brain        \\
KiPA22~\cite{KiPA22}           & 70          & 4      & 280       & Abdomen        \\
KiTS23~\cite{KiTS23}           & 489          & 3      & 1226       & Abdomen        \\
LNDb~\cite{LNDb}           & 236          & 1      & 206       & Thorax        \\
LUNA16~\cite{LUNA16}       & 888          & 4      & 3,551        & Thorax        \\
MM-WHS CT~\cite{MMWHS}           & 40          & 9      & 180       & Thorax        \\
MSD Colon \textasteriskcentered ~\cite{MSD}       & 126          & 1      &126        & Abdomen       \\
MSD HepaticVessel \textasteriskcentered ~\cite{MSD} & 303        & 2      &606        & Abdomen       \\
MSD Liver \textasteriskcentered ~\cite{MSD}       & 131          & 2      &249        & Abdomen       \\
MSD Lung \textasteriskcentered ~\cite{MSD}        & 63           & 1      &63        & Thorax        \\
MSD Pancreas \textasteriskcentered ~\cite{MSD}    & 281          & 2      &562        & Abdomen       \\
MSD Spleen \textasteriskcentered ~\cite{MSD}      & 41           & 1      &41        & Abdomen       \\
NSCLC~\cite{NSCLC}         & 85           & 2      &162        & Thorax        \\
Pancreas CT~\cite{PancreasCT}           & 80          & 1      & 80       & Abdomen        \\
Parse2022~\cite{PARSE2022}           & 100          & 1      & 100       & Thorax        \\
PDDCA~\cite{PDDCA}           & 48          & 12      & 543       & Head and Neck        \\
SEGA~\cite{SEGA}           & 56          & 1      & 56       & Thorax       \\
SegRap2023 Task 1~\cite{SegRap2023}           & 120          & 61      & 7320       & Head and Neck        \\
SegRap2023 Task 2~\cite{SegRap2023}           & 120          & 2      & 240       & Head and Neck        \\
SegTHOR~\cite{SegTHOR}           & 40          & 4      & 160       & Thorax        \\
SLIVER07~\cite{SLIVER07}   & 20           & 1      &20        & Abdomen       \\
\bottomrule
\end{tabular}}
\end{table*}

\begin{table*}[!tbh]
\center
\scriptsize
\caption{(Continued) The 72 datasets we collect to build up \samdataset{}. Dataset not marked with  \textasteriskcentered~ are included in \samdataset{}-Nano.}
\label{tab:dataset_details2}
\vspace{0.3cm}
\resizebox{1.0\textwidth}{!}{\begin{tabular}{lclcccccc}

\toprule
\rowcolor{lightgray} \textbf{Dataset Name} & \textbf{\#Scans} & \textbf{\#Classes} & \textbf{\#Annotations}  & \textbf{Region} \\
\midrule
ToothFairy \textasteriskcentered ~\cite{ToothFairy}           & 153          & 1      & 153       & Head and Neck        \\
TotalSegmentor Cardiac~\cite{Totalsegmentator} & 1,202 & 17 &13,264       & Whole Body    \\
TotalSegmentor Muscles~\cite{Totalsegmentator} & 1,202 & 31 &21,510       & Whole Body    \\
TotalSegmentor Organs~\cite{Totalsegmentator} & 1,202 & 24 &20,361       & Whole Body    \\
TotalSegmentor Ribs~\cite{Totalsegmentator} & 1,202 & 39 &32,666       & Whole Body    \\
TotalSegmentor Vertebrae~\cite{Totalsegmentator} & 1,202 & 29 &19,503       & Whole Body    \\
TotalSegmentor V2 \textasteriskcentered ~\cite{Totalsegmentator} & 1,202 & 24 &15,729       & Whole Body    \\
VerSe~\cite{VerSe} & 96 & 29 & 1,295      & Spine    \\
WORD~\cite{WORD}           & 150          & 18     &2,700        & Abdomen       \\

\rowcolor{lightgray!50} \textbf{MRI Data}          &              &        &        &               \\
AMOS22 MRI~\cite{AMOS22}           & 60          & 16      & 896       & Abdomen        \\
ATLAS~\cite{ATLAS}           & 60          & 2      & 120       & Abdomen        \\
ATLASR2~\cite{ATLASR2}           & 654          & 1      & 652       & Brain        \\
Brain Atlas~\cite{Brain_Atlas} & 30       & 108    &3,240        & Brain         \\
BrainPTM~\cite{BrainPTM}   & 60           & 7      &408        & Brain         \\
BraTS2023 GLI \textasteriskcentered ~\cite{BraTS2023GLI}           & 5004          & 4      & 19680       & Brain        \\
BraTS2023 MEN \textasteriskcentered ~\cite{BraTS2023MEN}           & 4000          & 4      & 11420       & Brain        \\
BraTS2023 MET \textasteriskcentered ~\cite{BraTS2023MET}           & 951          & 4      & 3476       & Brain        \\
BraTS2023 PED \textasteriskcentered ~\cite{BraTS2023PED}           & 396          & 4      & 1328       & Brain        \\
BraTS2023 SSA \textasteriskcentered ~\cite{BraTS2023SSA}           & 240          & 4      & 940       & Brain        \\
CHAOS MRI~\cite{CHAOS}         & 60           & 5      &300        & Abdomen       \\
CMRxMotion~\cite{CMRxMotion}           & 138          &   4    &   536     & Thorax        \\
FeTA2022~\cite{FeTA2022}           & 80          &   7   &   560  & Brain        \\
ISLES2022~\cite{ISLES2022} & 500          & 1      &492        & Brain         \\
LAScarQS2022 Task 1~\cite{LAScarQS2022}           & 60          &   2   &  120   & Thorax        \\
LAScarQS2022 Task 2~\cite{LAScarQS2022}           & 130          &  1    &  130   & Thorax        \\
MM-WHS MRI~\cite{MMWHS}           & 40          &    9  &  180  & Thorax        \\
MRSpineSeg~\cite{MRSpineSeg} & 91         & 23     &1,783        & Spine         \\
MSD Cardiac \textasteriskcentered ~\cite{MSD} & 20          & 1      & 20       & Thorax         \\
MSD Hippocampus \textasteriskcentered ~\cite{MSD} & 260          & 3      &780        & Brain         \\
MSD Prostate \textasteriskcentered ~\cite{MSD}    & 64           & 2      &124        & Pelvis        \\
MyoPS2020~\cite{MyoPS2020}           & 135          &   6   &  450   & Thorax        \\
PROMISE12~\cite{PROMISE12} & 50           & 1      &50        & Pelvis        \\
SKI10~\cite{SKI10}         & 99           & 4      &396        & Upper Limb    \\
WMH~\cite{WMH} & 170 & 1    &170        & Brain         \\ 

\midrule
\rowcolor{lightgray!50} \textbf{PET Data}          &              &        &        &               \\
autoPET \textasteriskcentered ~\cite{autoPET}           & 501          &   1   &  501   & Whole Body        \\
HECKTOR2022 \textasteriskcentered ~\cite{HECTOR2022}           & 524          &   2   &  972   & Head and Neck        \\

\rowcolor{lightgray!50} \textbf{Summary}           &  22,186      & 497     & 302,033 & /             \\
\bottomrule

\end{tabular}}
\end{table*}
\clearpage

\section{Examples From the Knowledge Tree}
\begin{table*}[htbp]
\renewcommand{\arraystretch}{1.3} 
\center
\footnotesize
\caption{Textual knowledge examples from the constructed knowledge tree. \textit{Def} indicates the definition of the concept; \textit{Rel} indicates the relationship with other concept.}
\label{tab:knowledge_example3}
\vspace{0.3cm}
\resizebox{1.0\textwidth}{!}{
\begin{tabular}{lp{14cm}} 

\toprule
\rowcolor{lightgray} Concept & Knowledge \\
\midrule

Brain & \textit{Rel}: Situated above, cerebellum \\\hline
Caudate lobe & \textit{Rel}: Connected to (via caudate process), the right lobe \\\hline
Cingulate gyrus & \textit{Rel}: Connected to, parahippocampal gyrus \\\hline
Femur & \textit{Rel}: Surrounded by, strong ligaments \\\hline
Heart atrium & \textit{Rel}: Spatially related to, Right Ventricle (for the Right Atrium) \\\hline
Iliac artery & \textit{Rel}: Runs posterior to, inguinal ligament \\\hline
Lateral ventricle & \textit{Rel}: Forms a frame around, thalamus \\\hline
Left femur & \textit{Rel}: Articulates with, patella \\\hline
Left heart ventricle & \textit{Rel}: Located in, bottom left portion of the heart \\\hline
Left lens & \textit{Rel}: Located within, left eye \\\hline
Left rib 6 & \textit{Rel}: Articulates posteriorly with, vertebral column \\\hline
Lung & \textit{Rel}: Posteriorly related to, vertebra \\\hline
Optic chiasm & \textit{Rel}: Closely associated with, anterior cerebral artery \\\hline
Optic nerve & \textit{Rel}: Passes through, posterior orbit \\\hline
Oral cavity & \textit{Rel}: Bounded externally by, lips and cheek mucosa \\\hline
Pulmonary artery & \textit{Rel}: Carries, deoxygenated blood to the lungs \\\hline
Pulmonary artery & \textit{Rel}: Lies anterior to, the right mainstem bronchus \\\hline
Pulmonary embolism & \textit{Rel}: Affects, main pulmonary artery (for larger obstructions) \\\hline
Right amygdala & \textit{Rel}: Part of, limbic system \\\hline
Right clavicle & \textit{Rel}: Extends between, acromion of the scapula \\\hline
Right hippocampus & \textit{Rel}: Involved in, storage of long-term memory \\\hline
Right kidney & \textit{Rel}: Close proximity to, liver, intestines, diaphragm \\\hline
Spinal cord & \textit{Rel}: Extends from, C1 vertebra \\\hline
Temporomandibular joint & \textit{Rel}: Coated with, fibrocartilage (articulating surfaces) \\\hline
Urinary bladder & \textit{Rel}: Located posterior to, symphysis pubis \\\hline
Thoracic vertebrae 3 (T3) & \textit{Rel}: Articulates with, heads of the ribs (via demi-facets) \\\hline
Bronchie & \textit{Def}: These structures are crucial airways leading into the lungs, forming a part of the lower respiratory system and facilitating the movement of air to and from the bronchopulmonary segments. The walls are composed of respiratory mucosa, which includes mucous-secreting cells, alongside cartilage plates that provide structural support. Furthermore, smooth muscle fibers are present to regulate the airway's diameter, with an outer layer, known as the adventitia, anchoring these airways to the surrounding lung tissues. The trachea, situated at the neck's front, divides into the right and left primary branches of these airways. Each branch serves as a conduit for air into the smaller subdivisions within the lungs. The structural components are quite similar to those of the trachea, with modifications in the form of cartilage plates instead of complete rings, ensuring flexibility and support. In conclusion, these air passages are pivotal for respiratory function, distinguished by their structural composition and their role in air delivery to lung segments. Originating from the trachea, they extend into the lungs, demonstrating significant anatomical relationships with nearby structures, ensuring efficient lung ventilation. \\\bottomrule

\end{tabular}}
\end{table*}

\begin{table*}[htbp]
\renewcommand{\arraystretch}{1.3} 
\center
\footnotesize
\caption{(Continued) Examples of textual knowledge from the constructed knowledge tree. \textit{Def} indicates the definition of the concept; \textit{Rel} indicates the relationship with other concept.}
\label{tab:knowledge_example1}
\vspace{0.3cm}
\resizebox{1.0\textwidth}{!}{
\begin{tabular}{lp{14cm}} 

\toprule
\rowcolor{lightgray} Concept & Knowledge \\
\midrule

Eyeball & \textit{Def}: This spheroidal structure, present in all vertebrates, functions similarly to a simple camera by housing the retina - a layer rich in photoreceptors necessary for vision. It resides within the skull's orbit, bordered by various facial bones and cushioned by fat. This bilateral and spherical organ features a tough exterior known as the sclera, which is overlaid anteriorly by the conjunctiva to prevent drying. Comprising three layers, its fibrous exterior encompasses the sclera and cornea for shape and support, while the vascular and inner layers include crucial components for blood supply and neural functions, respectively. It connects to the brain via the optic nerve, facilitating the transfer of visual information, and houses the lens and vitreous cavity, crucial for focusing light and maintaining shape, filled with vitreous humor. \\\hline

Head of right femur & \textit{Def}: This structure is found at the proximal end of the femur, engaging with the pelvis's acetabulum to establish the hip joint. It adopts a nearly spherical form, positioned superomedially and extending anteriorly from the femur's neck. Its smooth convex surface is interrupted by a depression, the fovea for the ligament, on its posteroinferior aspect. Its surface is mostly smooth and coated with articular cartilage, excluding the fovea where the ligamentum teres finds its attachment. This organ interacts with the acetabulum to create the hip joint and is linked to the neck of the femur, which connects its shaft and head at an angle conducive to efficient walking. Additionally, the ligamentum teres femoris bridges the gap between the acetabulum and a pit located on this structure. \\\hline

Lateral ventricle & \textit{Def}: Situated within the cerebral hemispheres, these C-shaped structures span the cerebrum, extending throughout the occipital, frontal, and parietal lobes. They consist of a central part and three horns extending anteriorly, posteriorly, and inferiorly. The central section of this structure is elongated in an anteroposterior direction and showcases a triangular cross-section. It is topped by the corpus callosum's trunk. Composed of five sections: the frontal horn, the body, the atrium, the occipital horn, and the temporal horn, they are surrounded by significant anatomical structures including the putamen, globus pallidus, thalamus, fornix, septum pellucidum, hippocampus, amygdala, and deep cerebral white matter. \\\hline

Left lacrimal gland & \textit{Def}: This structure, recognized as the tear gland, finds its position atop the eyeball, nestled in the lacrimal fossa of the orbit's anterior upper outer quadrant - a niche carved out by the orbital plate of the frontal bone. Compact, stretching about 2cm, it comprises two interlinked sections: the predominant orbital portion and the diminutive palpebral section. The former claims the territory above the lateral edge of the levator palpebrae superioris muscle, mirroring the dimensions and form of an almond, peculiarly resembling a J-shaped serous gland. Tasked with a continuous flow, this organ secretes a fluid that bathes the eye, simultaneously cleansing and safeguarding its surface while providing essential lubrication. \\\hline

Liver & \textit{Def}: A large organ found in the upper right quadrant of the abdomen, it stands as the largest gland within the human body, with a weight of about 1.5 kilograms. This structure exhibits a reddish-brown hue and is cone or wedge-shaped, having the smaller end positioned near the spleen and stomach while the larger end is over the small intestine. Except for its bare area where it meets the diaphragm, this organ is entirely enveloped by visceral peritoneum. It is comprised of several lobes, namely the right lobe, left lobe, caudate lobe, and quadrate lobe, while essentially being divided into two main lobes which include eight segments housing 1,000 lobules each. These lobules are linked to small ducts, merging into larger ones to eventually form the common hepatic duct. This duct is crucial for transporting bile produced by the organ's cells to the gallbladder and duodenum via the common bile duct. Positioned beneath the diaphragm and resting atop the stomach, right kidney, and intestines, this organ is predominantly intraperitoneal, stretching from the fifth intercostal space in the midclavicular line to the right costal margin. Its superior posterior portion presents a bare area where it interfaces directly with the diaphragm and the inferior vena cava. In terms of structure, the organ is made up of hepatocytes arranged in hexagonally shaped lobules, with each lobule centering around a central vein. Between these hepatocyte cords are vascular spaces known as sinusoids, characterized by their thin fenestrated endothelium and a discontinuous membrane. \\

Lumbar vertebrae & \textit{Def}: Positioned in the lower back, this structure forms part of the spine, comprising five cylindrical bones labeled L1 to L5. These segments are the largest in the vertebral column and are designed for weight-bearing, with their kidney-shaped bodies growing in size from top to bottom. Their pedicles and laminae are robust, supporting the attachment of strong back muscles through short, sturdy spinous processes. Compared to other spinal vertebrae, these bones are distinct because they do not possess features like transverse foramina or bifid spinous processes. Their spinous processes are compact and do not extend beneath their body level. Facing medially, the superior articular facets, and laterally, the inferior articular facets, provide unique spatial relationships, including the presence of accessory and mammillary processes. The fifth bone is notably the largest and plays a pivotal role in transferring the torso's weight to the base of the sacrum, where the spinal cord ends around the L1/L2 level. This organ is essential for weight support while standing and is designed for a high degree of extension due to the size of its corresponding intervertebral discs. Furthermore, its ability to facilitate needle access for procedures like epidural anesthesia highlights its clinical significance. \\\bottomrule

\end{tabular}}
\end{table*}

\begin{table*}[htbp]
\renewcommand{\arraystretch}{1.3} 
\center
\footnotesize
\caption{(Continued) Textual knowledge examples from the constructed knowledge tree. \textit{Def} indicates the definition of the concept; \textit{Rel} indicates the relationship with other concept.}
\label{tab:knowledge_example2}
\vspace{0.3cm}
\resizebox{1.0\textwidth}{!}{
\begin{tabular}{lp{14cm}} 

\toprule
\rowcolor{lightgray} Concept & Knowledge \\
\midrule

Occipital lobe & \textit{Def}: Positioned at the brain's rear, right behind the temporal and parietal regions, and beneath the skull's occipital bone, this organ is the brain's smallest by area, representing roughly 12\% of the cerebral cortex. Its boundaries, somewhat arbitrarily determined, give it a triangular shape, situating it posterior to the parietal and temporal regions. It hosts a series of folds, including gyri and sulci, and features different functional areas. The primary and secondary visual cortices, alongside specific other areas, contribute to the ongoing visual representation based on retinal inputs. Separated from the temporal region by an imaginary line aligning with the parietooccipital sulcus, its extent reaches from its pole to the mentioned sulcus. This structure finds itself beneath the parietal region, above the temporal one towards the brain's back, sitting on the tentorium cerebelli which itself segregates it from the cerebellum below. It stands as the visual processing center, deciphering color, form, and movement. \\\hline


Scapula & \textit{Def}: Located in the upper thoracic region on the dorsal surface of the rib cage, this triangular-shaped, flat bone is found. It sits adjacent to the posterior surface of ribs 2-7 and forms joints with the humerus and clavicle, contributing to the shoulder joint's structure. On its anterior aspect, a smooth, concave area - primarily the subscapular fossa - is present, from which the subscapularis muscle takes its origin. The coracoid process, resembling a beak, protrudes anterolaterally from the superior border here. Meanwhile, the posterior aspect is characterized by a convex, ridge-divided area that separates into the superior supraspinous fossa and the larger inferior infraspinous fossa. Noteworthy are two processes: the coracoid and the acromion, with the latter articulating with the clavicle. This structure bridges the upper limb to the trunk, offering attachment points for a multitude of muscles such as the levator scapulae, teres major, and the muscles of the rotator cuff, thereby playing an essential role in shoulder joint stability and movement. \\\hline

Spleen & \textit{Def}: Located in the left hypochondriac region, specifically in the left upper quadrant, this organ is found posterior to the stomach and anterior to the left hemidiaphragm, nestled at the level of the 9th to 11th ribs. Medially, it is close to the left kidney, and below it directly contacts the left colic flexure. This structure, resembling the size of a clenched fist, possesses a smooth, convex surface facing the diaphragm and is distinguished by a ridge that divides it into an anterior gastric portion and a posterior renal portion. The anterior surface is broad, concave, and aims forward and upwards, while the posterior is rounded, targeting upwards and backwards. Its spongy nature and reddish-purple appearance are attributed to intense vascularization. The structure is comprised of both red and white pulp, each playing crucial roles in filtering blood, immune response, recycling iron, storing blood, and extramedullary hematopoiesis. It is linked to the stomach and kidney through parts of the greater omentum, namely the gastrosplenic and splenorenal ligaments, and its principal venous drainage is provided by the splenic vein, which travels behind the pancreas before joining. \\\hline 

Spinal cord & \textit{Def}: Originating at the lower part of the brainstem, specifically the medulla oblongata, this structure extends down to the lower back, terminating in a cone-shaped end known as the conus medullaris. It is positioned from the topmost neck bone, the C1 vertebra, down to roughly the L1 vertebra at the upper portion of the lower back, just beneath the ribcage. Measuring around 18 inches (45 centimeters) in length, its cylindrical form encompasses a collection of nerve fibers safeguarded within the vertebral column. Segmented with pairs of roots (dorsal and ventral), these join to form the spinal nerves, while its composition includes an external white matter layer surrounding an internal gray matter core. This organ is encased by the central nervous system's three protective membranes: the dura mater, arachnoid, and pia mater, further shielded by the vertebral column's bony architecture. As an essential component of the central nervous system, it serves as the primary conduit for signaling between the brain and the body, as well as the site for initiating reflexes and processing sensory information, closely associated with the spinal nerves emerging at each of its segments. \\\hline

Stomach & \textit{Def}: Positioned to the left of the midline and centrally in the upper abdomen area, this J-shaped organ exhibits a lesser and greater curvature, with its anterior and posterior surfaces smoothly coated by peritoneum. Residing primarily in the epigastric and umbilical regions, its size, shape, and position exhibit variability among individuals and change with position and respiration. This hollow, muscular structure can significantly vary in its capacity, designed to store food temporarily. It comprises four main regions: the cardia, fundus, body, and pylorus, with the fundus being the rounded area adjacent to the cardia and the body constituting the largest portion. It can expand or contract, featuring numerous folds (rugae) when empty that smooth out when distended. A dense layer of small gastric glands within the mucous-membrane lining secretes enzymes and hydrochloric acid for the partial digestion of proteins and fats. It consists of layers including mucosa, submucosa, muscularis externa, and serosa, adapting its shape and size based on its contents. Located between the esophagus and the duodenum, and resting below the diaphragm as part of the gastrointestinal (GI) tract, its positions are explored through radiology and endoscopy, revealing its spatial differences and general appearance. This organ is instrumental in the initial stages of digestion, its structure finely tuned for effective performance. \\\bottomrule

\end{tabular}}
\end{table*}

\end{document}